%% file: article.tex
\documentclass[iop, twocolappendix, numberedappendix, revtex4]{emulateapj}
\usepackage{natbib}
\usepackage{amsmath}
\usepackage{graphicx}
\usepackage{sidecap}
\usepackage{xcolor}
\usepackage{verbatim}
\usepackage{mathtools}
\usepackage{amssymb}
\usepackage[caption=false]{subfig}
\usepackage{amssymb}
\usepackage{bm}
\usepackage{CJK}
\usepackage{ulem}

\def\min{{\rm min}}
\def\max{{\rm max}}

\shortauthors{Herman, Zhu, \& Wu}

\begin{document}
\begin{CJK*}{UTF8}{gbsn}

\title{Revisiting the Long-Period Transiting Planets from {\it Kepler}}

\author{Miranda K. Herman$^{1}$, Wei Zhu (祝伟)$^{2}$, and Yanqin Wu (武延庆)$^{1}$}

\affil{$^{1}$Astronomy \& Astrophysics, University of Toronto, 50 St. George St., Toronto, ON M5S 3H4, Canada;  miranda.herman@utoronto.ca}
\affil{$^{2}$Canadian Institute for Theoretical Astrophysics, University of Toronto, 60 St. George St., Toronto, ON M5S 3H8, Canada}

%%%%%%%%%%%%%%%%%%%%%%%%%%%%%%%%%%%%%%%%%%%%%%%%%%%%%%%

\begin{abstract}

Currently, we have only limited means to probe the presence of planets at large orbital separations. Foreman-Mackey et al. searched for long-period transiting planets in the {\it Kepler} light curves using an automated pipeline. Here, we apply their pipeline, with minor modifications, to a larger sample and use updated stellar parameters from {\it Gaia} DR2. The latter boosts the stellar radii for most of the planet candidates found by FM16, invalidating a number of them as false positives. We identify 15 candidates, including two new ones. All have sizes from $0.3$ to $1~R_{\rm J}$, and all but two have periods from 2 to 10 yr. We report two main findings based on this sample. First, the planet occurrence rate for the above size and period ranges is $0.70^{+0.40}_{-0.20}$ planets per Sun-like star, with the frequency of cold Jupiters agreeing with that from radial-velocity surveys.  Planet occurrence rises with decreasing planet size, roughly describable as $dN/d\log R \propto R^{\alpha}$ with $\alpha = -1.6^{+1.0}_{-0.9}$, i.e., Neptune-sized planets are some four times more common than Jupiter-sized ones. Second, five out of our 15 candidates orbit stars with known transiting planets at shorter periods, including one with five inner planets. We interpret this high incidence rate to mean: (1) almost all our candidates should be genuine; (2) across a large orbital range (from $\sim 0.05$ to a few astronomical units), mutual inclinations in these systems are at most a few degrees; and (3) large outer planets exist almost exclusively in systems with small inner planets.

\end{abstract}


%%%%%%%%%%%%%%%%%%%%%%%%%%%%%%%%%%%%%%%%%%%%%%%%%%%%%%%

\section{Introduction}\label{sec:Intro} 

Currently, we have limited means to probe planet populations at orbital separations on the order of an astronomical unit or above. These planets are difficult to identify in transit surveys like the {\it Kepler} mission as they only transit once or twice within the observational baseline. Furthermore, unless they have Jovian masses, their radial velocity (RV) signals are hard to detect. Though planet microlensing studies have yielded an interesting sample, including planets with Neptune masses, such a sample is small and is mostly composed of planets orbiting M-dwarfs \citep{Gould10}. The elusive nature of these long-period planets is a roadblock on our path toward a complete census of planet populations and a successful theory of planet formation. It has also recently become apparent that long-period planets may be correlated with short-period ones \citep{Bryan19,ZhuWu18}, and may influence the dynamical evolution of the latter. Given this, it seems relevant and worthwhile to expand our knowledge of such planets. In this work, we pursue this task by searching for transiting long-period planets in stellar light curves obtained by the {\it Kepler} mission.

The {\it Kepler} mission \citep{Borucki10} has been responsible for the discovery of thousands of transiting exoplanets and planetary candidates \citep[e.g.,][]{Thompson18}. Relative to other transit surveys such as the {\it Transiting Exoplanet Survey Satellite (TESS)} \citep[][]{Ricker14} or even the upcoming {\it PLATO} mission \citep{Rauer14}, {\it Kepler}'s long observational baseline makes it uniquely capable of probing planets at a wide range of orbital separations. However, as the transit probability decreases rapidly with separation ($\sim 10^{-3}$ at 5 au), many recent exoplanet population studies have restricted themselves to the distribution and occurrence rate of relatively short-period planets \citep[e.g.,][]{Fressin13, Petigura13, Dressing15}.

Long-period transit events are overlooked by standard search procedures, which require three or more transits to be observed. \citet{Yee08} first introduced the idea of searching for single-transit events in the {\it Kepler} data, and a handful of studies have since sought to increase the sample of known long-period planets through visual inspection of individual light curves \citep[e.g.,][]{Wang15, Uehara16}. However this method has its own shortcomings; it is difficult to determine the detection efficiency of such a search procedure, which is critical for studying the underlying planet population.

\citet[][]{FM16} (hereafter FM16) were the first to perform an automated search for long-period transiting planets. FM16 sought long-period, high signal-to-noise ratio (S/N) transit signals in the light curves of 39,036 bright main sequence GK stars. They identified 16 long-period planet candidates (two of which they label as likely false positives). By injecting artificial transits into the light curves, they obtained the search completeness and detection efficiency for their pipeline, allowing them to infer the occurrence rate of planets in the outer region.  FM16 reported a rate of $0.42\pm0.16$ planets per star within a radius range of $0.4~R_{\rm J}<R_{\rm p}<1~R_{\rm J}$ and a period range of 2 yr~$<P<~$25 yr. Their work constitutes significant progress toward understanding the population of outer planets.

However, a major change has occurred since FM16 that warrants a new study. FM16 adopted stellar parameters from the {\it Kepler} Input Catalog (KIC) \citep{Brown11,Huber14}, but the recent second data release of the {\it Gaia} mission \citep{Gaia16, Gaia18b} has updated the stellar radii for a significant fraction of the stars in FM16's target catalog \citep[e.g.,][]{Berger18}. In particular, many of the stars identified as GK dwarfs in the {\it Kepler} catalog have larger radii than initially thought, meaning that the sizes of many of FM16's planet candidates may have been underestimated. A number of candidates are now too large to be compatible with sub-stellar objects, and many small planets are now Jovian in size. This upgrade necessarily requires FM16's planet occurrence rate to be adjusted.

FM16 adopted the philosophy of open-source software and have made their entire pipeline publicly available. This affords us great ease in our work. Here, we repeat their exercise but with a number of modifications. First, we apply the search procedure of FM16 to more than 61,000 {\it Kepler} light curves to identify long-period planet candidates, using the revised stellar properties of \citet{Berger18} to determine their transit characteristics. We also identified a missing factor of $\pi^{1/3}$ in the transit probability calculation in FM16's code. This alone reduces FM16's occurrence rate by a factor of $1.46$. We adopt a procedure different from that in the FM16 code to convert transit depth to planet radius, after a discrepancy in our earlier draft was identified (K. Masuda 2019, private communication). We further employ a procedure to flatten stellar variability to increase the transit S/N. We also provide an appended explanation on why so few double-transit systems are found relative to single-transit ones, a puzzle posed but not well-explored by FM16.

After performing injection and recovery tests to compute the completeness of the search procedure, we use our new sample of planet candidates to provide a new estimate for the planet occurrence rate at long periods. We compare this to earlier results from RV and microlensing studies. Moreover, we carry out a number of detailed analyses to clarify the significance of our results. One intriguing finding of FM16 is that a large fraction of transiting outer planets also have inner transiting companions, including one with five planets. Given the rarity of transiting systems observed by {\it Kepler}, this is not expected unless the inner and outer planetary systems are correlated in occurrence rate and are inclined similarly. We investigate this behavior in detail. Our sample also affords us an opportunity to study the size distribution of outer planets, which provides important input for theories of planet formation.

%%%%%%%%%%%%%%%%%%%%%%%%%%%%%%%%%%%%%%%%%%%%%%%%%%%%%%%%%%%%

\section{Searching for Candidates} \label{sec:Method}

\subsection{Target Selection}

Our candidate search method makes use of the open-source code \texttt{peerless}\footnote{https://github.com/dfm/peerless} from FM16, closely following their procedure with a few modifications. We refer the reader to FM16 for a detailed description of their original search and vetting procedure, and we discuss the changes we implement in Appendix \ref{sec:peerless_mods}. Below, we describe the sample of light curves to which this modified procedure is applied.

\begin{figure}
	\centering
    \includegraphics[width=0.48\textwidth]{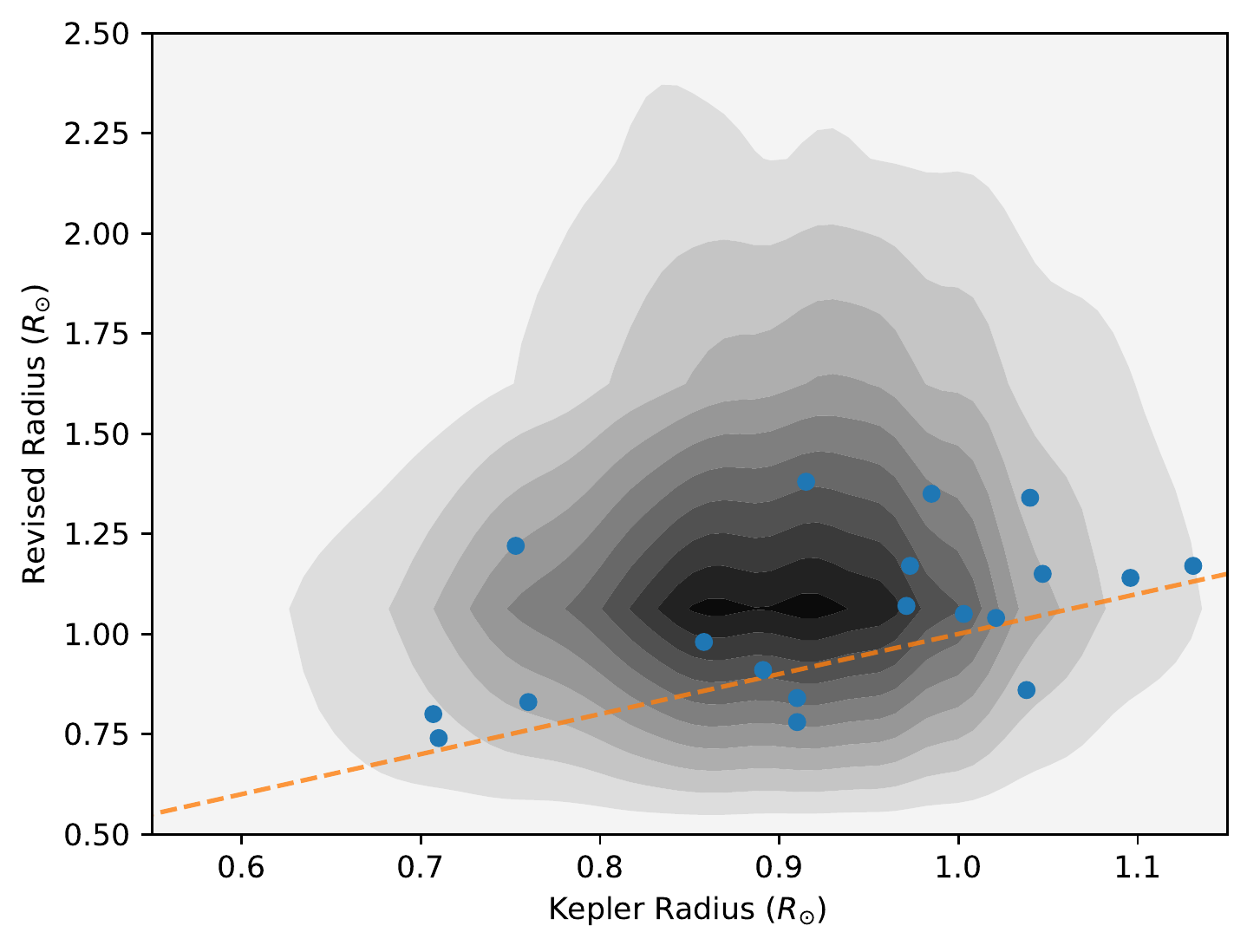}
    \caption{Comparison of the radii reported by {\it Kepler} and their revised values from \citet{Berger18} for all stars in our initial target sample. The orange dashed line denotes a one-to-one ratio, and the blue points are the host stars of our planet candidates. Stars with radii smaller than $0.7 ~R_{\odot}$ or larger than $1.5 ~R_{\odot}$ are excluded from our sample because they no longer qualify, based on our selection criteria.
    }  
    \label{fig:gaia-kepler}
\end{figure}

\begin{figure}
	\centering
   \includegraphics[width=0.48\textwidth]{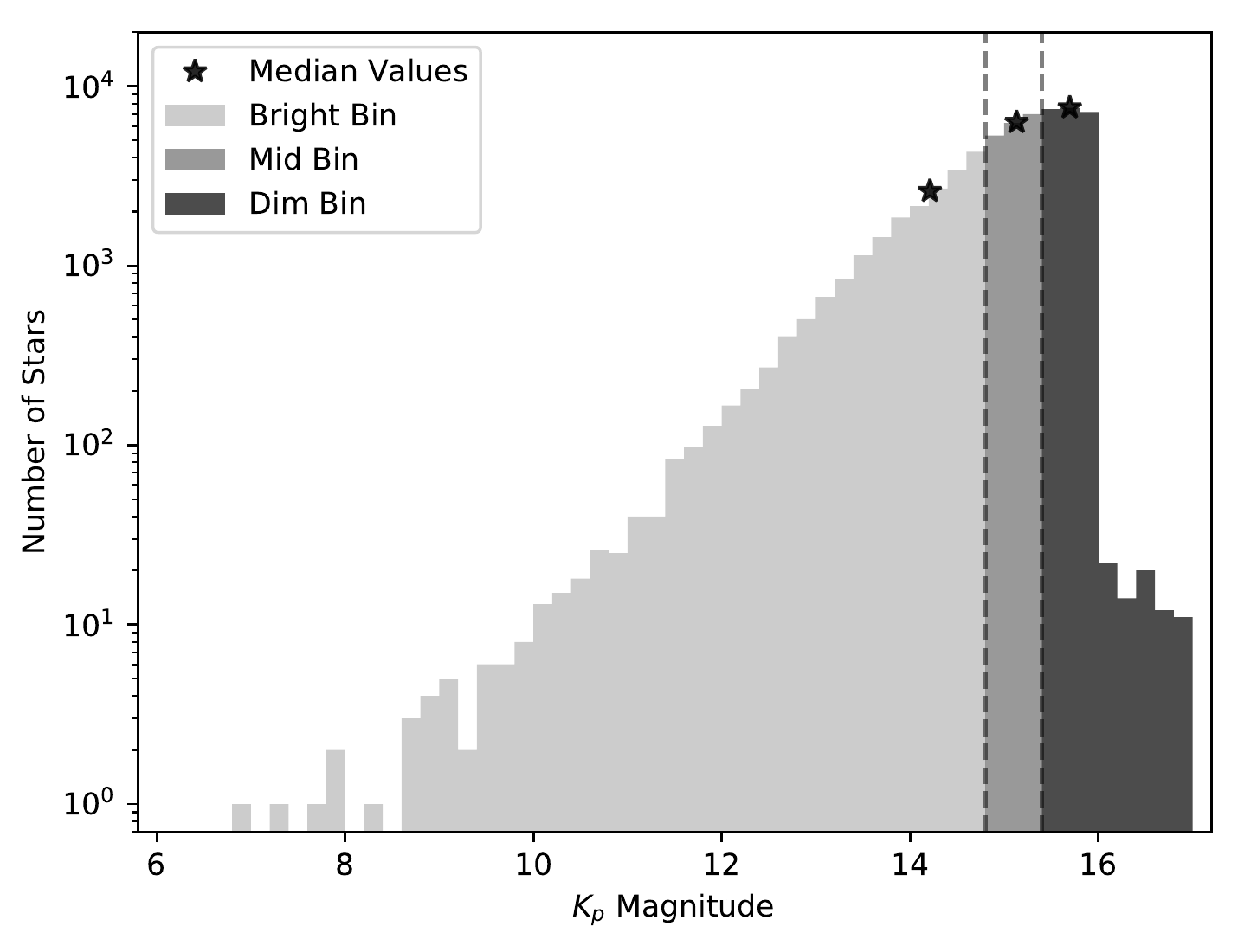}
    \caption{The magnitude distribution of the 61,418 stars in our final target sample. The sample is split into three bins with roughly an equal number of stars per bin. The denoted median values are $14.2$, $15.1$, and $15.7$, respectively. A numerical fit for $dN/d\log{\rm flux} \propto {\rm flux}^{\gamma}$ gives $\gamma \approx -1.4$.}  
    \label{fig:mag-hist}
\end{figure}

Our initial stellar sample of quiet Sun-like stars is selected so as to match that of FM16, with an additional 50,817 stars dimmer than their magnitude limit of 15 (up to $K_p = 17$ mag). Our initial selection criteria are:

\begin{itemize}
    \setlength\itemsep{0.05em}
    \item 4200 K $\leq T_{\rm eff} \leq$ 6100 K,
    \item $R_{*} \leq 1.15 ~R_{\odot}$,
    \item data span $\geq$ 2 yr,
    \item duty cycle $\geq$ 0.6,
    \item $K_{p} \leq 17$ mag, and
    \item CDPP$_{7.5 \rm hr} \leq$ 1000 ppm.
\end{itemize}

However, we further restrict our sample based on the updated stellar radii provided by \citet{Berger18}, who combine precise parallax measurements from {\it Gaia} DR2 \citep{Gaia16, Gaia18b} with the stellar parameters of the {\it Kepler} DR25 catalog to produce revised stellar radii for a large fraction of stars in the {\it Kepler} field. \citet{Berger18} claim that their parallax-derived radii have an uncertainty of $\sim$8\%, 4--5 times more precise than those provided by the latest {\it Kepler} catalog of stellar properties. For stars with poorly constrained parallax/distance measurements or missing photometry, no updated radius is reported. From their catalog, we find revised radii for $\sim$95\% of our targets, and we discard the remaining targets for which no revised radius is available. For a comparison of all stellar radii reported by {\it Kepler} and \citet{Berger18} in our target sample, see Figure \ref{fig:gaia-kepler}.

For a considerable fraction of our target sample, the new radii push the stars to sizes inconsistent with a Sun-like sample. Keeping these stars in our sample would make it difficult to constrain the occurrence rate of long-period transiting planets without taking the differences in stellar type into account; for instance, many of the assumed GK dwarfs in our sample should now be considered giants based on their revised radii. The properties of such stars may have an effect on planet occurrence, and we therefore choose to limit our sample to targets with $0.7 ~R_{\odot} \leq R_{*} \leq 1.5 ~R_{\odot}$ while keeping all other parameters the same. This reduces our sample by $\sim$ 28\%, resulting in a final selection of 61,418 Sun-like stars. While this is by no means the most straightforward method of cutting our stellar sample, the general conclusions we draw should have little dependence on the selection method, given the large number of stars we consider. A histogram displaying the magnitude distribution of our sample is shown in Figure \ref{fig:mag-hist}.

Following our target selection, we apply our modified search and vetting procedure to the pre-search data conditioning (PDC) light curves. A complete description of this process can be found in FM16. This yields $19$ transit candidates, out of $61,418$ targets. To constrain the transit parameters of each candidate, we adopt an MCMC fit (described below), rather than the output from \texttt{peerless}. \footnote{This revision was prompted by Kento Masuda, who notified us that the planet radii in our earlier draft, using \texttt{peerless}, did not match those expected based on the transit depths and stellar radii. Surprisingly, results in FM16 did not share this problem. The cause for this discrepancy is unclear.} Finally, we  use \texttt{peerless} to determine the false-positive probability for each event.

\begin{figure*}[!ht]
    \begin{center}
    \include{lcfig}
    \end{center}
    \caption{Plots of each of the transit candidates not common to FM16. The PDC light curve is shown in black and the posterior-median transit model is shown in blue. In the bottom panel of each plot, the residuals between the data and model are shown in parts per thousand (ppt). The candidates with two observed transits are folded on their derived orbital periods.
    }
    \label{fig:light-curves}
\end{figure*}

\begin{figure*}[!ht]
	\centering
    \includegraphics{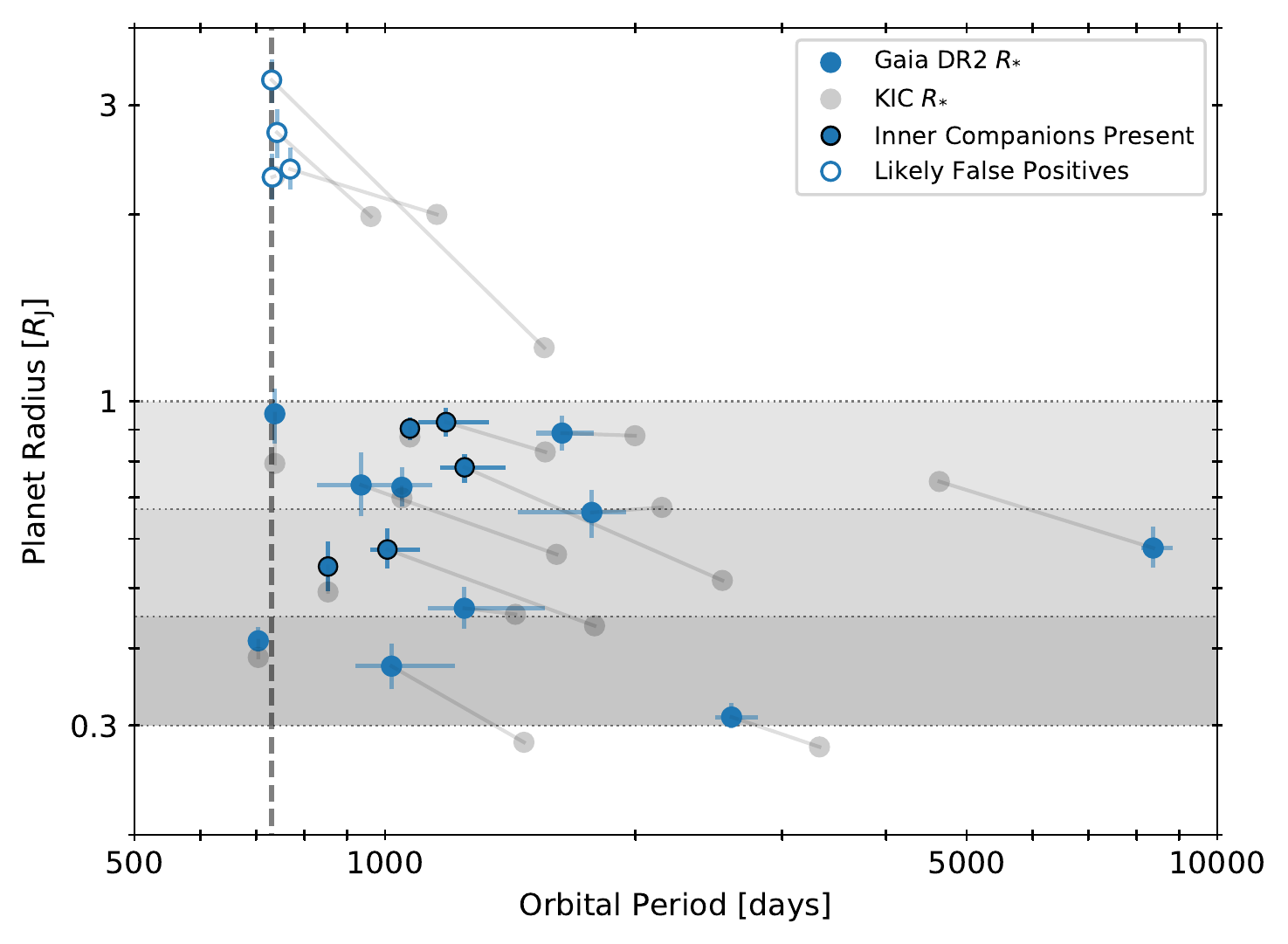}
    \caption{The results of our candidate search. We recover all 12 of the candidates identified in FM16 that are contained within our stellar sample, and identify seven additional long-period transit candidates. Candidate parameters obtained using the revised stellar parameters \citep{Berger18} are plotted in blue, while gray points (connected to the blue ones by thin lines) indicate their parameters if using the older {\it Kepler} input catalog values. Most candidates have moved up in radius after the revision. The candidates identified as likely false positives in Table \ref{tab:catalog} and in Section \ref{sec:false} are denoted by open circles, and those with known inner transiting companions are outlined in black.
    The vertical dashed line denotes the maximum possible period to exhibit at least three transits in the baseline of {\it Kepler} observations. Planets inward of this line should be picked up by the {\it Kepler} pipeline.
    The horizontal bands illustrate the radius bins into which we divide our sample when computing occurrence rates. The largest bin includes Jupiter/Saturn-sized planets; the smallest bin includes Neptune/Uranus-sized; and the middle bin has no corresponding planets in our own solar system.
    }
    \label{fig:candidates}
\end{figure*}

%%%%%%%%%%%%%%%%%%%%%%%%%%%%%%%%%%%%%%%%%%%%%%%%%%%%

\subsection{MCMC Fit to the Transit Light Curves} \label{sec:Fitting}

The physical parameters of these 19 candidates are determined by performing an MCMC fit to the light curves. We use the \texttt{batman} package \citep{batman} to generate the transit
light curve for each candidate in isolation, assuming a quadratic limb darkening law and circular orbit, and use the \texttt{emcee} package \citep{Foreman2013} to constrain the transit parameters. 

We first detrend the light curve, using a third-order spline, after masking the transit based on its duration, which is estimated in the \texttt{peerless} search output. The flux errors are determined from the standard deviation of the out-of-transit light curve. The fitting parameters are the time of mid-transit ($t_0$), logarithmic period ($\log(P)$), planet-star radius ratio ($R_{\rm p}/R_{*}$), impact parameter ($b$), and the quadratic limb darkening coefficients ($u_1$ and $u_2$). The likelihood for the MCMC sampling $\mathcal{L}$ is calculated as $\mathcal{L} \propto \exp(-\chi^2/2)$, where $\chi^2$ is the sum of the standard chi-squared for each transit.

For candidates with two transits, we fit each transit separately, then set the period as the difference between their respective $t_0$ values. We then phase-fold the transits on this period and fit the resulting phase-folded transit. For candidates with only one transit, we first use the \texttt{isocrones} package \citep{isocrones} to determine the mean stellar density and its uncertainty from the {\it Gaia} DR2 parallax \citep{Gaia18b}, corrected for zero-point offset with 0.03 mas, and the KIC broadband photometry. We then use these values as a prior in our transit fits to constrain the planet's orbital period \citep[see][]{Seager03}.\footnote{This fails for eclipsing binaries where both components contribute to the gravitational mass. This explains the periods of the largest candidates in our catalog, which we are not concerned with as they are likely eclipsing binaries.}We also place a constraint on the period such that no additional transits can occur within the {\it Kepler} baseline. As stated earlier, we assume a circular orbit when fitting each transit. Introducing a non-zero eccentricity only increases the uncertainty on the inferred orbital period while having little influence on the median \citep[for an illustration, see Section 3.3 of][]{Villanueva19}.

%%%%%%%%%%%%%%%%%%%%%%%%%%%%%%%%%%%%%%%%%%%%%%%%%%%%

\subsection{Results of the Candidate Search} \label{sec:Results}

The results of our MCMC fits to the candidate transits are given in Table \ref{tab:catalog}. Twelve of these candidates are common to FM16, and we provide new constraints on their transit parameters based on the revised stellar radii of \citet{Berger18}. We note that, by restricting our target list based on the revised stellar radii, we lose four of the candidates identified in FM16 (KIC 8426957, KIC 8738735, KIC 10287723, and KIC 10321319) as their stellar parameters are no longer consistent with those of Sun-like stars. The remaining seven transit light curves not common to FM16's sample are shown in Figure \ref{fig:light-curves}.

We also consider the overlap between our sample and that of both \citet{Wang15} and \citet{Uehara16}. Our list of target light curves contains ten candidates identified by \citet{Wang15}, four of which we recover (see Table \ref{tab:catalog}). The remainder either do not meet our S/N threshold (KIC 5536555, KIC 9413313, KIC 9662267, KIC 11716643, and KIC 12454613) or are rejected for close proximity to a data gap (KIC 6191521). Seven of the candidates found in \citet{Uehara16} are also in our target sample; of these, we recover five candidates. The other two are rejected for having a grazing impact parameter (KIC 3230491) and for close proximity to a data gap \citep[KIC 6191521, also found in][]{Wang15}. We note that these exclusions are automatically included in our completeness calculation and so do not affect our determination of the occurrence rate described below. The radii and orbital periods of all planet candidates are plotted in Figure \ref{fig:candidates}.

Of the seven candidates displayed in Figure \ref{fig:light-curves}, two have two observed transits within the baseline of the {\it Kepler} observations (KIC 7906827 and KIC 10525077), the latter of which was previously discovered by \citet[][]{Wang15} ({\it Kepler}-459b). Two additional single-transit candidates, KIC 9704149 and KIC 11342550, were reported in \citet{Wang15} and \citet[][]{Uehara16} (KOI-1421), respectively. Thus, four of the long-period transit candidates in our sample have never been published before. However, based on our false-alarm probability (FAP) estimates (see Table \ref{tab:catalog}), it is very likely that two of these candidates are not transiting planets but are instead eclipsing binary stars. In Section \ref{sec:false}, we further discuss potential astrophysical false positives among our candidates.

Interestingly, five of the total 19 transit candidates have known inner companions. This means that over a quarter of our long-period candidates are found in systems with at least one shorter-period planet. Considering that only 1745 of the targets in our sample of 61,418 stars contain previously discovered planets and/or planet candidates, this abundance of long-period planets in multiple-planet systems is intriguing. Could these be special systems for which transits are more detectable? We find this to be unlikely. Photometric uncertainties in the host-star light curves are similar to those without inner systems. The S/N ratio of these candidates are also comparable to the others. In Section \ref{sec:Discussion}, we discuss the implications of this abundance of inner planets in terms of the mutual inclination of bodies in multiple-planet systems and the architecture of such systems.

\begin{deluxetable*}{ccccccccccc}
\tabletypesize{\footnotesize}
\tablecaption{System Parameters of the Long-period Transiting Exoplanet Candidates \label{tab:catalog}}
\include{param_tab}
\tablenotetext{a}{The KOI number and target {\it Kepler} number, as available. KIC 8505215 (also known as KOI-99) and KIC 11342550 (also known as KOI-1421) are misidentified by the {\it Kepler} pipeline as shorter period candidates. Visual inspection shows that these putative candidates only transit once and instead have much longer periods. Here, their periods are determined by the transit cord.}
\tablenotetext{b}{Included in \citet{Wang15}.}
\tablenotetext{c}{Included in \citet{Uehara16}.}
\tablenotetext{d}{Candidate has two observed transits. In these cases, the mean S/N is given.}
\tablecomments{The first 12 candidates are common to \citet{FM16}. The values and uncertainties indicate the $16^{\mathrm{th}}$, $50^{\mathrm{th}}$, and $84^{\mathrm{th}}$ percentiles of the posterior samples for each parameter.}
\end{deluxetable*}

%%%%%%%%%%%%%%%%%%%%%%%%%%%%%%%%%%%%%%%%%%%%%%%%%%%%%%%%

\subsection{Search Completeness and Detection Efficiency} \label{sec:Completeness}

To constrain the transit detection efficiency for our target sample, we first characterize the completeness of the search procedure. We follow the commonly used method of injecting artificial transits with known parameters into the {\it Kepler} light curves, and then measure the recovery rate of the procedure \citep[e.g.,][and many others]{Petigura13, Christiansen15, Dressing15, FM16}. To achieve this, we first randomly choose a star from the target sample and load its PDC light curve and stellar properties. We then sample planet properties from the distributions given in Table \ref{tab:injs}, and the transit is calculated and multiplied into the light curve. The search procedure is run on this transit-injected light curve, and if at least one transit within one transit duration passes all steps of the vetting process, the candidate is considered recovered. Note that, depending on the injected period and time of mid-transit, a candidate can display up to two transits in a single light curve. A more detailed analysis of the recovery rate for double- versus single transit events is described in Appendix \ref{sec:dvs}.

\begin{figure*}[!ht]
\begin{center}
    \includegraphics[width=0.49\textwidth]{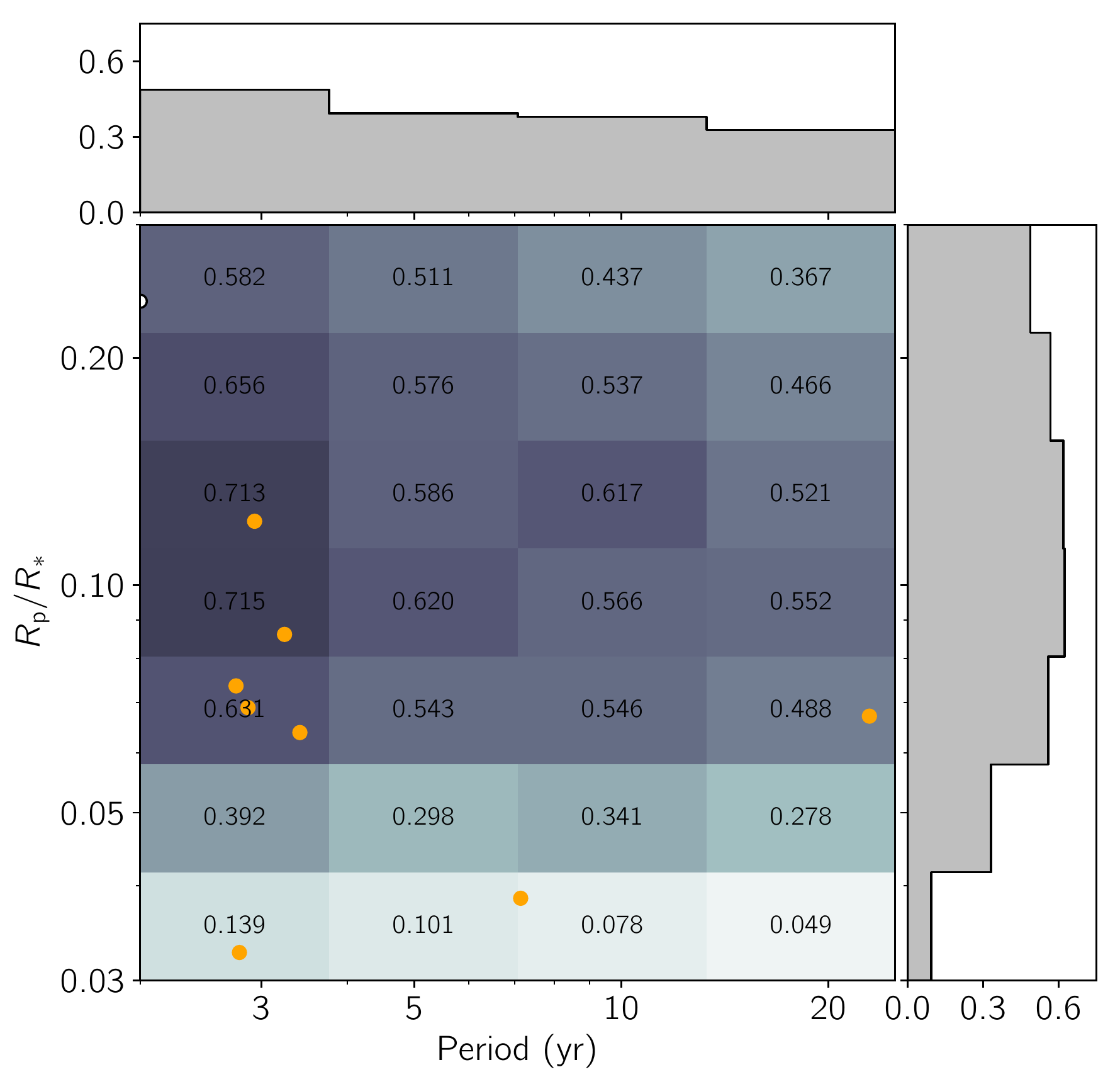}
    \includegraphics[width=0.49\textwidth]{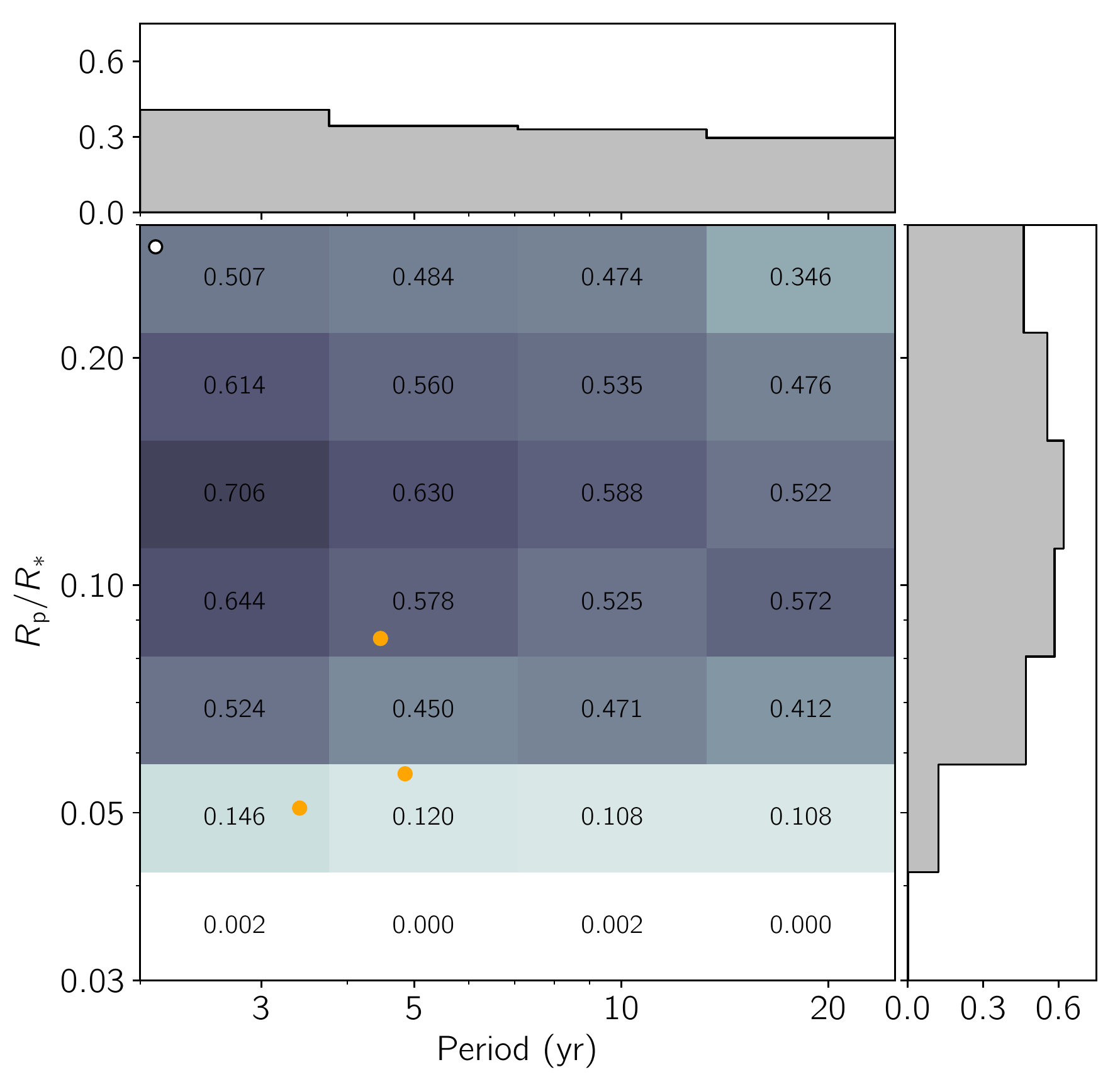} 
    \includegraphics[width=0.49\textwidth]{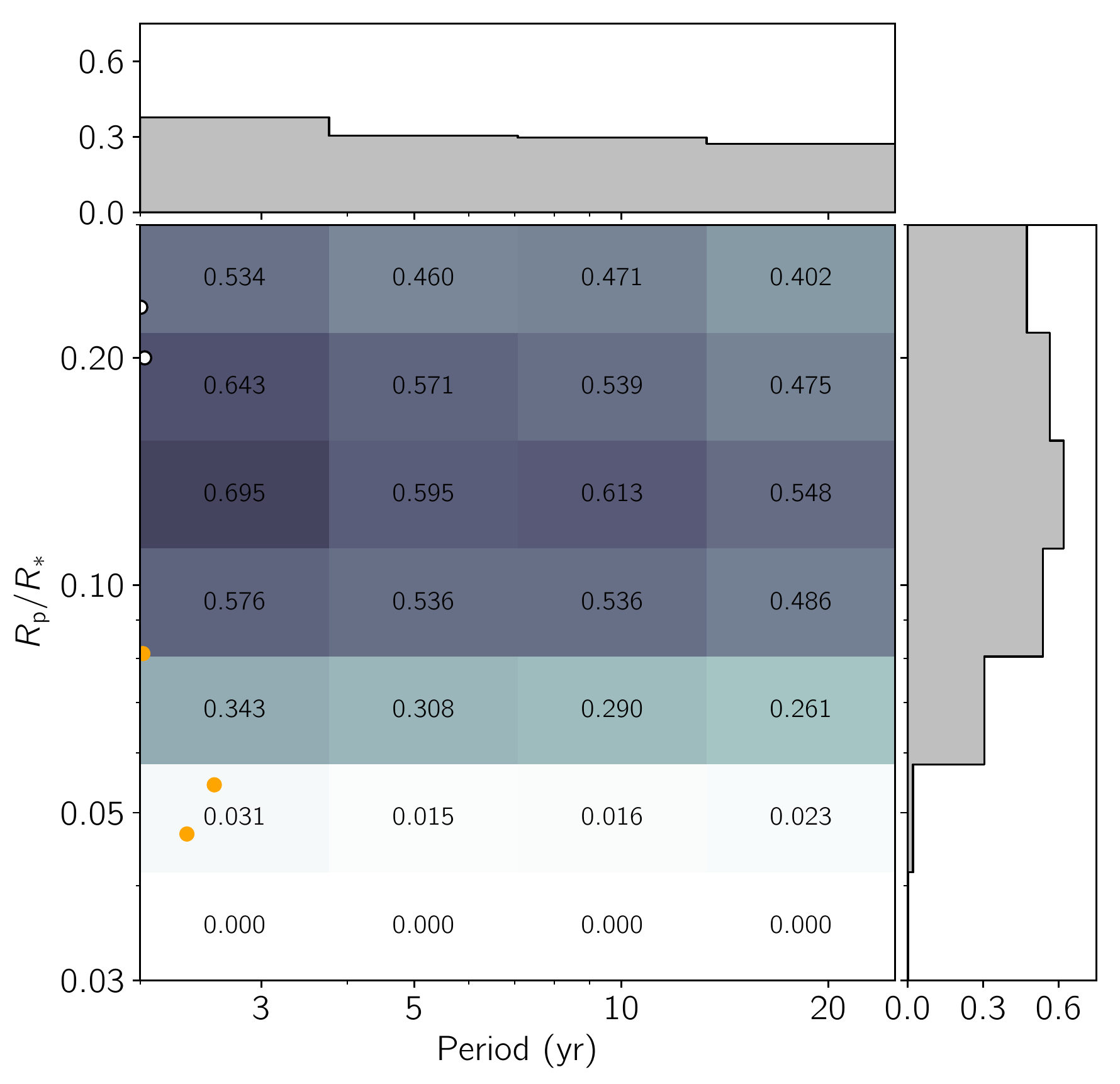}
    \includegraphics[width=0.49\textwidth]{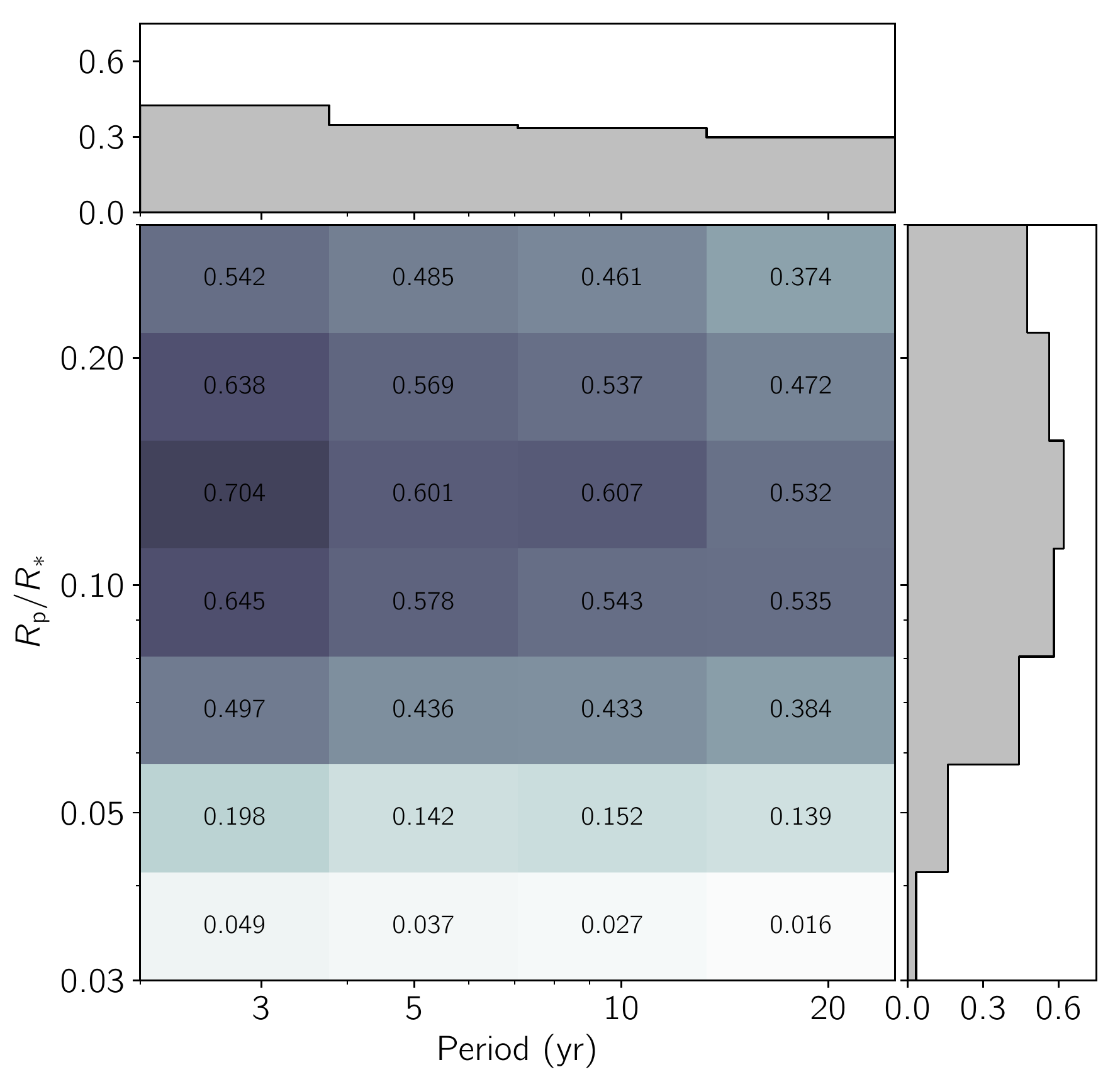}
\end{center}
\caption{The search completeness divided into magnitude bins, as a function of fractional planet radius and orbital period. {\it Top left}: $6 \leq K_p < 14.8$, {\it Top right}: $14.8 \leq K_p < 15.4$, {\it Bottom left}: $15.4 \leq K_p \leq 17$, {\it Bottom right}: The full magnitude range, $6 \leq K_p \leq 17$. The completeness for each shaded bin is determined by calculating the fraction of simulated transits that are recovered by the automated search procedure. The adjacent histograms display the integrated completeness in terms of orbital period (top histograms) and fractional planet radius (right histograms). Note the decrease in the fraction of recovered injections, particularly at smaller radius ratios, as the stellar magnitude increases between the first three plots. We plot each of our planet candidates in orange and the false positives in white, in the panels corresponding to their host star magnitudes (Table \ref{tab:catalog}).
}  
\label{fig:completeness}
\end{figure*}

\begin{deluxetable}{lcc}
\tabletypesize{\footnotesize}
\tablecaption{Parameter Distributions for Simulated Transits
\label{tab:injs}}
\tablehead{%
    \colhead{Name} &\colhead{Parameter} & \colhead{Distribution}}
\startdata
Period & $\log ~P$ & $\mathcal{U}(\log ~2~\mathrm{yr},~\log ~25~\mathrm{ yr})$ \\
Radius Ratio & $\log ~R_\mathrm{p}/R_\star$ & $\mathcal{U}(\log ~0.02,~\log ~0.3)$ \\
Impact Parameter & $b$ & $\mathcal{U}(0,~1+R_\mathrm{p}/R_\star)$ \\
Eccentricity & $e$ & $\beta(0.867,~3.03)$\tablenotemark{a} \\
    & $\omega$ & $\mathcal{U}(0,~2\pi)$\\
Limb Darkening & $q_1$ & $\mathcal{U}(0,~1)$ \\
    & $q_2$ & $\mathcal{U}(0,~1)$ \\
\enddata
\tablenotetext{a}{\citet{Kipping13b}}
\end{deluxetable}

The completeness or the fraction of recovered injections as a function of radius and orbital period for our stellar sample is shown in Figure \ref{fig:completeness}. We report this completeness in three separate magnitude bins, each of which contains roughly one-third of our full target sample ($6 \leq K_p < 14.8$; $14.8 \leq K_p < 15.4$; $15.4 \leq K_p \leq 17$). The magnitude dependence of the completeness is most obvious at small fractional planet radii, where the percentage of recovered injections is essentially zero for the dimmest stars. Note that the radius is reported in terms of radius ratio between the simulated planet and its host star, $R_{\rm p}/R_*$. We deem this more illuminating than reporting the radius in terms of $R_{\rm J}$, as our stellar sample spans a range of radii; a $1 ~R_{\rm J}$ planet around a $0.75 ~R_\odot$ star would be easier to recover, for instance, compared to the same planet around a $1.5 ~R_\odot$ star if all other factors are kept the same. The decrease in the completeness toward the largest $R_{\rm p}/R_*$ values is due in part to our choice of injected planetary properties, chosen to maximize the recovery of signals with smaller radii (FM16). For instance, signals with $R_{\rm p}/R_*$ values above $\sim 0.1$ are more likely to be rejected during the vetting process of the search pipeline for having a grazing impact parameter. This parameter is influenced by the radius both in its injected distribution ($b \leq 1 + R_{\rm p}/R_*$) (Table \ref{tab:injs}) and acceptance criteria ($b \leq 1 - R_{\rm p}/R_*$) (FM16).

This completeness provides an estimate of the probability of detecting a transit with some set of planetary parameters, given that the planet transits its host star within the observing period. We denote this probability as $p_{\mathrm{rec}}$. The total detection efficiency is then determined by combining this with the geometric transit probability and the window function -- in other words, the probability that a planet will transit along our line of sight given its physical parameters, and the probability that it transits within the baseline of our observations. The geometric transit probability for a given system is expressed as

\begin{equation}
    p_{\mathrm{geo}} = \frac{R_* + R_{\rm p}}{a} \frac{1+e \sin \omega}{1 - e^2}
\end{equation}
\citep{Winn10} where $R_*$ is the stellar radius, $R_{\rm p}$ is the planet radius, $a$ is the semi-major axis, $e$ is the orbital eccentricity, and $\omega$ is the argument of periastron. This equation can be rewritten using Kepler's third law such that

\begin{equation} \label{eqn:pgeo}
    p_{\mathrm{geo}} = \left(\frac{4 \pi^2}{G M_*}\right)^{1/3} \left(\frac{1+e \sin \omega}{1 - e^2}\right) \left(R_* + R_{\rm p}\right) P^{-2/3}
\end{equation}
where $M_*$ is the stellar mass and $P$ is the orbital period of the planet.

Like \citet{Burke14} and FM16, we use the binomial probability of observing a minimum of one transit to approximate the window function

\begin{equation} \label{eqn:pwin}
    p_{\mathrm{win}} = \left\{
        \begin{array}{ll}
            1 - \left(1 - f_{\rm duty}\right)^{T/P} & \quad P \leq T \\
            Tf_{\rm duty}/P & \quad P > T
        \end{array}
    \right.
\end{equation}
where $f_{\rm duty}$ is the duty cycle and $T$ is the baseline of observations for the star being considered. The total detection efficiency is then

\begin{equation} \label{eqn:pdet}
    p_{\mathrm{det}} = p_{\mathrm{rec}} ~p_{\mathrm{geo}} ~p_{\mathrm{win}} ~.
\end{equation}
Because our initial completeness is calculated for three separate ranges in magnitude, this detection efficiency is also magnitude-dependent. We take this into account when computing the planet occurrence rate in the following section.

%%%%%%%%%%%%%%%%%%%%%%%%%%%%%%%%%%%%%%%%%%%%%%%%%%%%%%%%

\subsection{False Positives} \label{sec:false}

Before we proceed to analyze our results, we discuss the possibility of false positives within our catalog of 19 planet candidates. 

The general increase in stellar radii following the second data release of {\it Gaia} (see Figure \ref{fig:gaia-kepler}) plays an important role when identifying false positives in our sample. The resulting jump in the size of our planet candidates necessarily changes their physical association; while some candidates simply move from Neptune-sized into the Jovian range, others exceed our expectations for maximum planet radii and thus are unlikely to be planets. Some of the candidates identified by FM16 belong to the latter class.

Among our catalog of 19 candidates (Table \ref{tab:catalog}), the four largest ones have radii larger than $2 ~R_{\rm J}$. The \texttt{peerless} package assigns them a low planet probability (FAP $\geq 0.9$). They are more likely eclipsing binaries blended with foreground stars or with physically associated stellar companions \citep{FM16, Morton16}. We reject these four from the outset. Two candidates, with respective periods of $704$ days and $8375$ days, fall outside the period range ($2-10$ yr) we are considering. In the following, we consider the remaining 13 candidates.

Five of the remaining $13$ have known inner transiting companions. As the inner and the outer companions should be mutually inclined and do not necessarily transit simultaneously, this surprising fact itself suggests that most of these 13 candidates should be real planets (we discuss this in more detail below).

\begin{figure}
	\centering
    \includegraphics[width=0.48\textwidth]{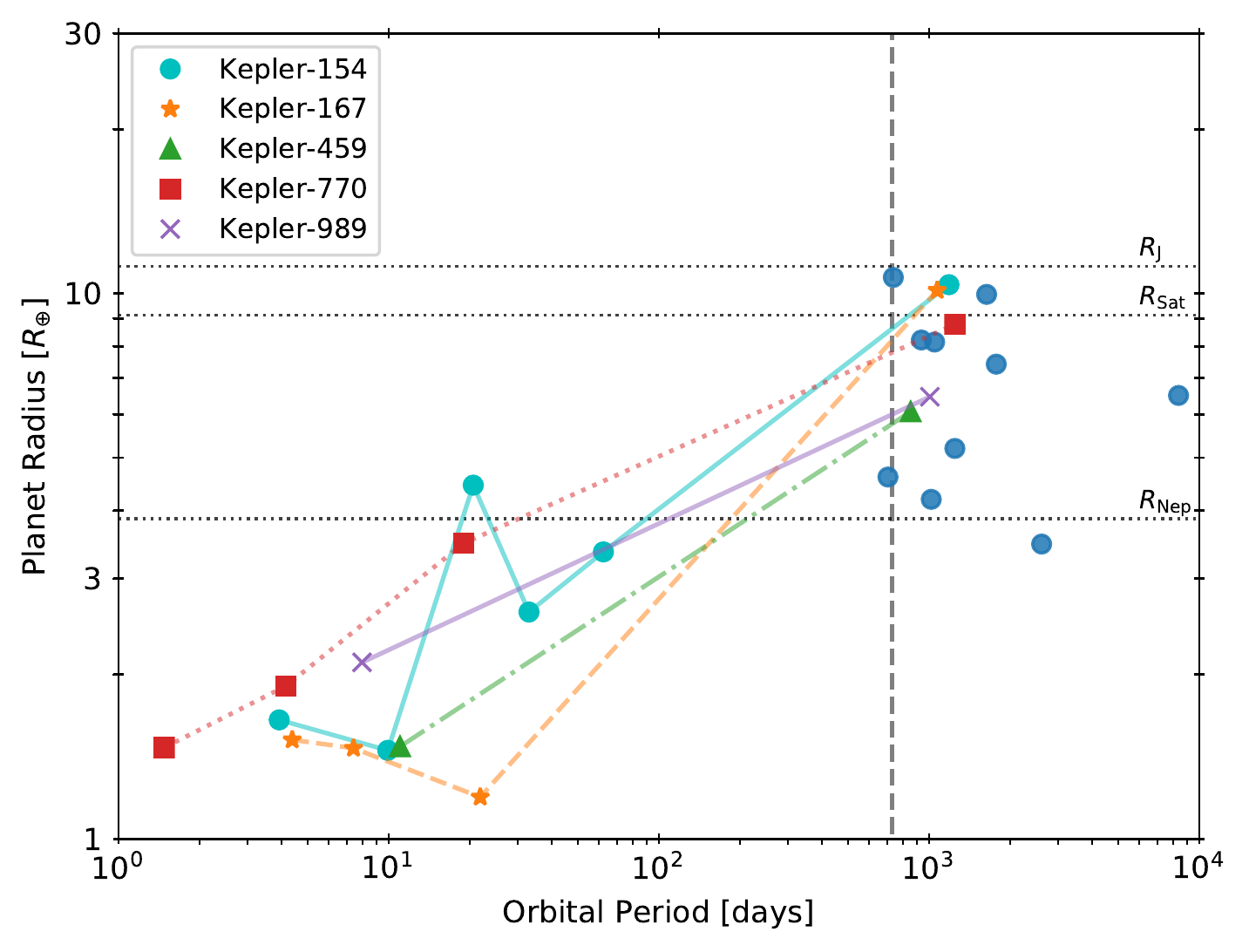}
s    \caption{
    The inner-outer planet connection. Here, multiple-planet systems are connected by colored lines, and dark blue symbols represent the single long-period candidates, as in Figure \ref{fig:candidates}. We use the updated planet radii reported by \citet{Berger18} for all inner planets. The same vertical line from Figure \ref{fig:candidates} is shown, and the radii of Jupiter, Saturn, and Neptune are denoted for comparison.
    }  
    \label{fig:inner+outer}
\end{figure}

Four of our candidates are larger than Saturn. \citet{Santerne12} reported that such large bodies have higher false-positive rates than do smaller planets, if their periods are within a couple hundred days. Currently there is no study on whether such a trend also extends to colder planets. Two of these four giants have inner transiting companions (see Figures \ref{fig:candidates} and \ref{fig:inner+outer}), boosting their credibility as genuine planets \citep{Lissauer12}, but are the other two likely real? To make an educated guess, we review known sources of astrophysical confusion and their prevalence:

\begin{itemize}

\item Brown Dwarfs. Because their sizes are comparable to that of giant planets, brown dwarfs may initially appear to be an important source of confusion. However, brown dwarfs are exceedingly uncommon companions for solar-type stars, particularly at periods $< 2000$ days, with less than $1\%$ of companions being brown dwarfs in mass, far below the rate of planetary companions \citep{Grether06}. This phenomenon is commonly referred to as the brown dwarf desert \citep[][and references therein]{Murphy18, Grether06}. 

\item M-dwarfs. With masses above $0.1 ~M_\odot$, M-dwarfs can have radii as low as that of Jupiter. By analyzing Figures 11 and 16 of \citet{Raghavan10}, we find a binary fraction of just $0.6\%$ for companions with size below $0.15 ~R_\odot = 1.5 R_{\rm J}$ and with period between 2 and 10 yr.

\end{itemize}

So together, brown dwarf or low-mass M-dwarf companions may masquerade as Jovian planets, but with a total occurrence rate well below $2\%$. This is much lower than the occurrence rate of cold Jupiters obtained by RV studies ($\sim 10\%$). We therefore argue that at most one, if any, of the large candidates could be a false-positive.

We have little information regarding the false positive rate for the smaller candidates in our sample. In the following analysis, we assume that they are all genuine planets. This is supported by their low FAP values, and the fact that three out of nine have inner systems.

\subsection{Possible Flux Dilution}

We have been prompted by the referee to consider the issue of flux dilution, which has been shown in the past to interfere with the {\it Kepler} planet sample. Many of our planet candidates are around faint stars, and multiple stars of comparable magnitudes may cohabit the same {\it Kepler} CCD pixel ($\sim 4\arcsec$ across). These stars can be either physically associated or unassociated with the host star, and their light can dilute the signal of a deeper transit into that of a shallower, planet-like, transit. Cohabiting stars are not rare events -- \citet{Furlan17} found that $\sim 30$ \% of {\it Kepler} planet host stars have at least one neighbor lying within $4\arcsec$.

Querying ExoFOP\footnote{https://exofop.ipac.caltech.edu/} revealed that eight of our $15$  candidates have AO follow-up imaging \citep{Law14, Hirsch17, Furlan17, Ziegler18V}, and among these, three have stellar companions with distances from $1\arcsec$ to $4\arcsec$. However, among the $13$ candidates on which we base our analysis, only KIC 3239945 has a known companion resolved by AO, with a $K$-magnitude difference of $3.6$ between the two stars and a separation of {$2.2\arcsec$}. Fortunately, a thorough vetting of this system, which also hosts three inner planets, has been performed by \citet{Kipping16} and the planetary nature of the outer planet (Kepler-167e) is firmly established.

We also query \textit{Gaia} DR2 for companions to the 13 stars in our final candidate sample, using their corresponding {\it Gaia} DR2 IDs \citep[Table 1 of][]{Berger18}. {\it Gaia} DR2 has an effective angular resolution of $\sim0.4\arcsec$ \citep{Gaia18b}, and can completely resolve any stellar companion beyond $1\arcsec$ \citep{Ziegler18V}. We find companions within $4\arcsec$ of three of our targets: KIC 3239945 (also listed in ExoFOP, see above), KIC 4754460, and KIC 8505215. In all cases, the point sources have \textit{Gaia} $G$ magnitude differences greater than 5. The large magnitude differences make it unlikely that our candidates are stellar transits diluted by light from a companion. If the transit signal comes from the brighter source, its depth will be little affected by the companion light; if instead the transit signal arises from the dimmer object, the transit would have to be intrinsically deep ($47\%$ depth for KIC 4754460 and $19\%$ for KIC 8505215) and V-shaped, accompanied by secondary transits. None of these features are compatible with observations.

What about star pairs that cannot be resolved by AO and \textit{Gaia}? This is increasingly rare for those that are not physically associated, but more likely for physically bound ones. Given our sample, the only possible scenario for false-positives is that in which a solar-type star is orbited by two M-dwarfs that transit each other. This would, however, create a very different transit shape, which should be identified by the {\texttt peerless} vetting procedure and FAP analysis. As such, we consider this unlikely.

%%%%%%%%%%%%%%%%%%%%%%%%%%%%%%%%%%%%%%%%%%%%%%%%%%%%%%%%

\section{Discussion} \label{sec:Discussion}

Here, we discuss the implications of our work on three issues: the occurrence rate of long-period planets, the size distribution of such planets, and the inner-outer planet correlation.

\subsection{Occurrence Rate for Outer Planets} \label{sec:occurrence}

Having justified that the majority of our planet candidates are likely true planets, we proceed to constrain the planet occurrence rate. We perform this calculation, using the detection efficiency as determined in Section \ref{sec:Completeness}, for planets from 0.3 to 1 $R_{\rm J}$.

As alluded to earlier, the  occurrence estimate by FM16 has to be revised using the updated stellar radii from \citet{Berger18}. In fact, many of their candidates are now shifted to sizes larger than Jupiter. Furthermore, while the equation describing the geometric transit probability given in FM16 is correct, its implementation in \texttt{peerless} misses a factor of $\pi^{1/3}$. This decreases their reported occurrence rate by a factor of 1.46. We report a new occurrence rate, taking into account the revised radii, this minor calculation error, and the seven additional transit candidates found in our sample.

To estimate this occurrence rate, we assume that the uncertainties on the planet properties, such as radius and orbital period, are negligible. This is a permissible simplification because the mean occurrence rate is calculated in bins considerably larger than the uncertainties. Additionally, like FM16 we assume that no other transits occur in a data gap for those candidates with a single-transit event, i.e.,the orbital period we determine is assumed correct. Each candidate is also assumed to orbit a star accurately described by the {\it Kepler} catalog of \citet{Huber14}, with updated radii from \citet{Berger18}, rather than a companion or background star. We also exclude the likely false positives contained in our sample when computing the occurrence rate, and make the simplifying assumption that all of our planet candidates have zero eccentricity; the latter has a negligible effect on the outcome of our results. We further assume that the distribution of planets in the parameter space of interest follows the relation
\begin{equation}
    \frac{d^2 N}{d\log{P}d\log{R_{\rm p}}} = A \left(\frac{R_{\rm p}}{R_{\rm J}}\right)^\alpha \left(\frac{P}{\rm year}\right)^\beta\ .
    \label{eq:defalpha}
\end{equation}
 In the following, we set $\beta= 0$. This assumed period insensitivity is justified both because we are in a relatively narrow range ($2-10$ yr), and because RV studies have not found any strong period dependence for giant planets in this range \citep{Cumming08,Bryan16}. The total expected number of detected planets, for $N^{(i)}_*$ stars in magnitude bin $i$, is
\begin{equation}
    \bar{N}_{\rm lp} = \sum_i N_\star^{(i)} \int \displaylimits_{2{\rm yrs}}^{10 {\rm yrs}} \int \displaylimits_{R_{p,\min}}^{R_{p,\max}} A \left( {{R_{\rm p}}\over{R_J}}\right)^\alpha p_{\rm det} d\log{P}d\log{R_{\rm p}}\ ,
\end{equation}
where the detection efficiency $p_{\rm det}$ is given by equation (\ref{eqn:pdet}).

We divide the radius range covered by our candidates ($0.3-1.0~R_{\rm J}$) into three equal logarithmic bins. Within each narrow radius bin, we can adopt $\alpha=0$, but we let $A$ vary across the bins.

\begin{figure*}[!ht]
	\centering
    \includegraphics[width=0.32\textwidth]{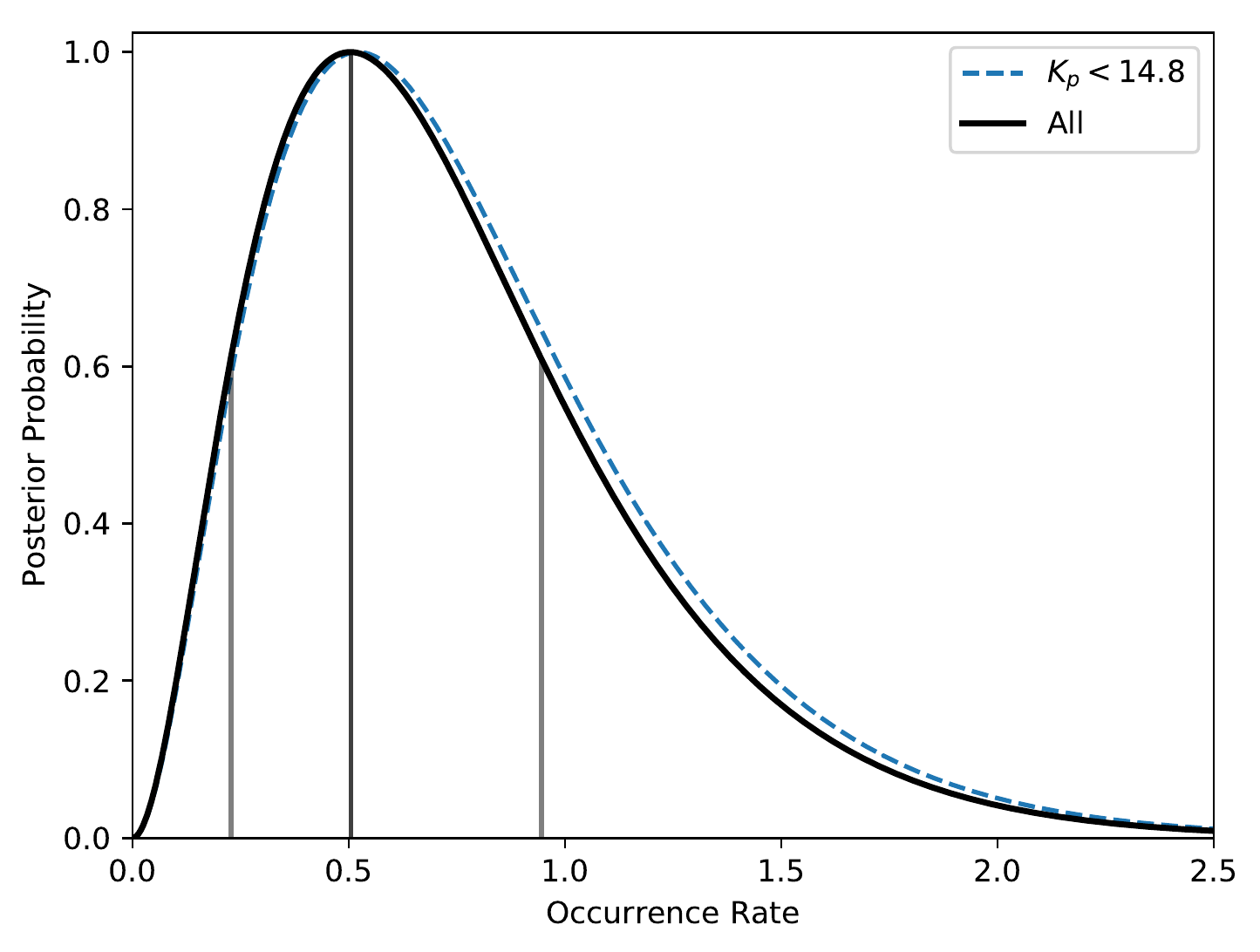}
    \includegraphics[width=0.32\textwidth]{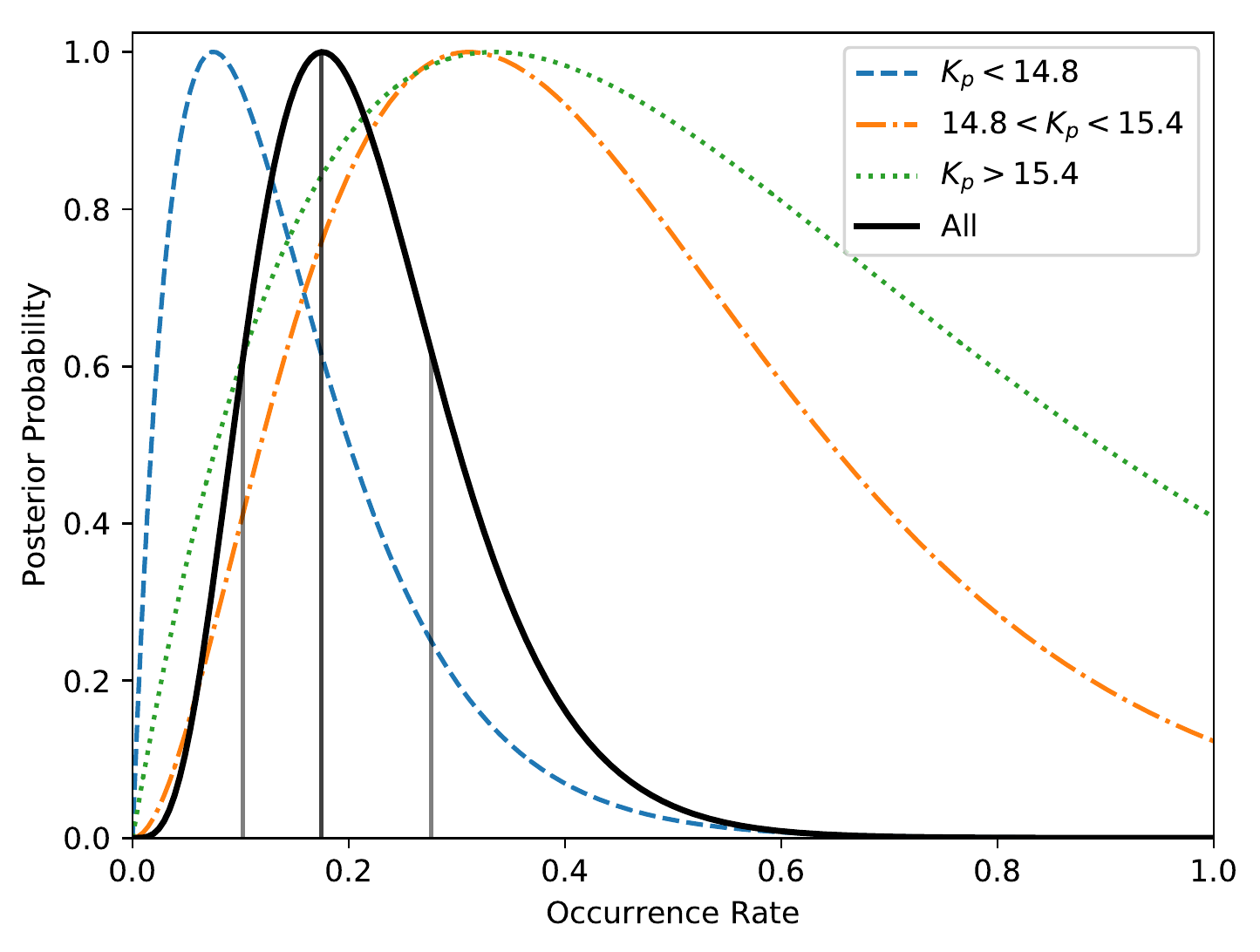}
    \includegraphics[width=0.32\textwidth]{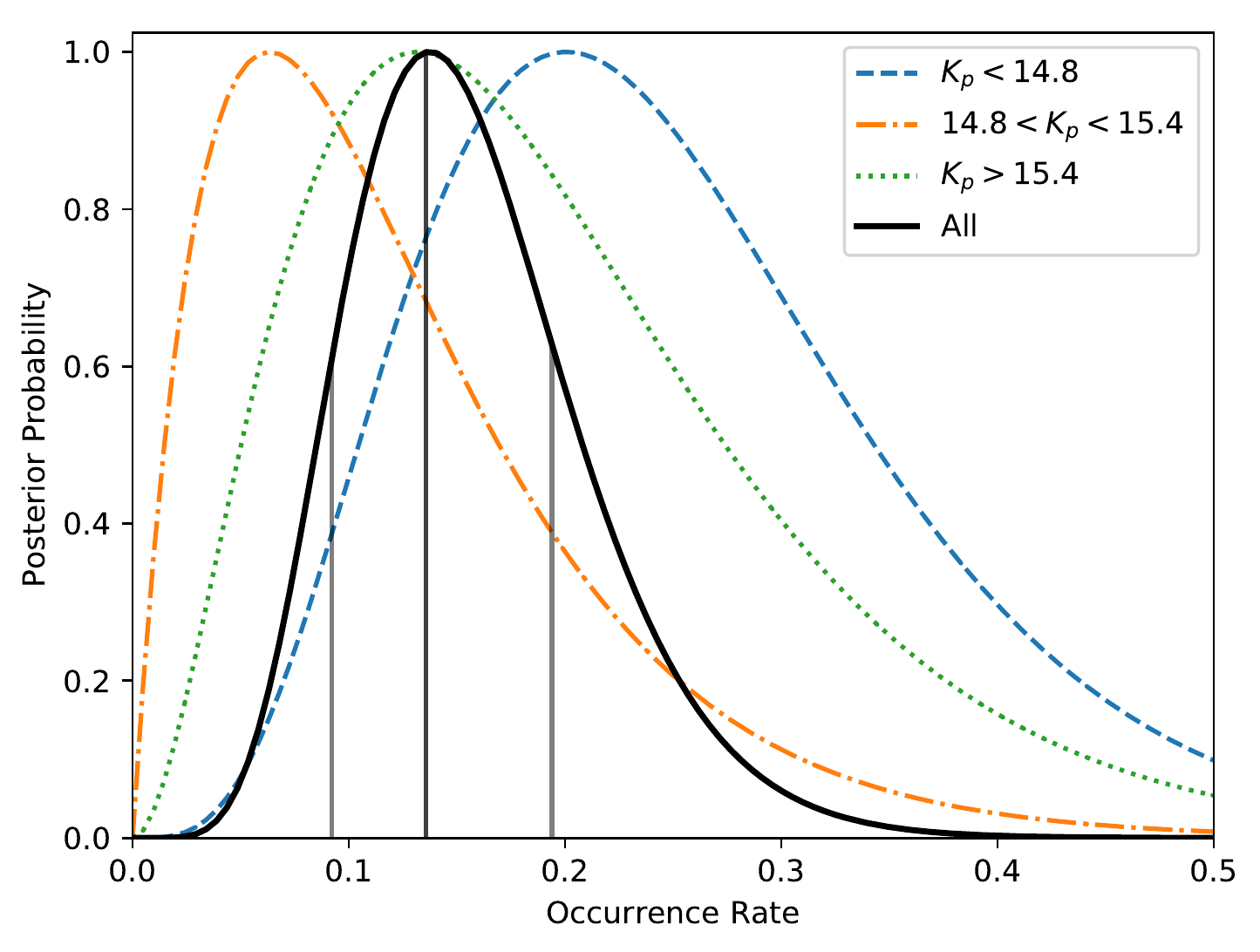}
    \caption{Posterior probability distributions of the occurrence rates. {\it Left}: Small outer planets. Here, we choose $2 ~\rm{yr} < P < 10 ~\rm{yr}$ and $0.30 ~ R_{\rm J} < R_{\rm p} < 0.45 ~ R_{\rm J}$, corresponding to the first bin in Table \ref{tab:occurrence} and Figure \ref{fig:occur_hist_peaks}. The peak and $\exp(-0.5)$ values are indicated for the overall occurrence rate. No planet candidates were detected in the two dimmer magnitude bins for this radius range, so these curves are not shown. {\it Center}: Intermediate-sized planets, where $0.45 ~ R_{\rm J} < R_{\rm p} < 0.67 ~ R_{\rm J}$, corresponding to the second radius bin. {\it Right}: Large outer planets, where $0.67 ~ R_{\rm J} < R_{\rm p} < 1.00 ~ R_{\rm J}$, corresponding to the third radius bin. Note the difference in x-axis ranges between the plots.}  
    \label{fig:posteriors}
\end{figure*}

The probability of getting the actual number of planets we detect, $N_{\rm lp}$, given its expectation value $\bar{N}_{\rm lp}$, is
\begin{equation}
    P(N_{\rm lp}|n_{\rm lp}) = \frac{\bar{N}_{\rm lp}^{N_{\rm lp}} \exp{(-\bar{N}_{\rm lp})}}{N_{\rm lp}!}\ .
\end{equation}
According to Bayes' theorem, the posterior distribution of $n_{\rm lp}$ is given by
\begin{equation}
    P(n_{\rm lp}|N_{\rm lp}) = \frac{P(N_{\rm lp}|n_{\rm lp}) P(n_{\rm lp})}{P(N_{\rm lp})} \propto P(N_{\rm lp}|n_{\rm lp})\ ,
\end{equation}
because both $P(n_{\rm lp})$ and $P(N_{\rm lp})$ are essentially constant. The posterior probability distributions for the three radius bins are presented in Figure \ref{fig:posteriors}. The separation by magnitude bins is a precaution observed because the detection completeness is dependent on the stellar magnitude. In practice, we find such a precaution unnecessary -- it makes little difference to the results whether we separate the computation by magnitude bin and combine the results to determine the total occurrence rate (black curves in Figure \ref{fig:posteriors}), or use the search completeness computed for all magnitudes (the bottom right panel of Figure \ref{fig:completeness}) to determine the total rate.

The derived occurrence rates for different radius bins and separate magnitude bins are presented in Table \ref{tab:occurrence}. The center values are taken to be the peak of the posterior distribution, and the $1-\sigma$ upper and lower bounds are taken to be where the probabilities have fallen by a factor $\exp{(-0.5)}$ from the peak.\footnote{Although the standard procedure is to report the 68\% range centered on the median as the 1-$\sigma$ values, we would like to be consistent with our later approach of using maximum likelihood to constrain the values of $A$ and $\alpha$.} Our results show that planet occurrence rises with decreasing planet sizes, from $0.14$ for the largest size bin (Jupiter-sized) to $0.50$ for the lowest bin (Neptune-sized). There are more small planets than large giants -- the number of detections are smaller for the smaller planets, but their detection efficiency drops even faster, demanding a larger underlying occurrence rate. This result is not sensitive to our choice of bin size -- the trend persists when we increase the number of bins to four or five.

To account for this size dependency in our calculation of the total occurrence rate, we need to introduce a nonzero $\alpha$ into our analysis. This is detailed in Section \ref{subsec:size}. The final result is $0.70^{+0.4}_{-0.2}$ planets per Sun-like star, with sizes between 0.3 and 1 $R_{\rm J}$, and periods between 2 and 10 yr.

\begin{deluxetable}{cccc}
\tabletypesize{\footnotesize}
\tablecaption{Occurrence Rates of Outer Planets
\label{tab:occurrence}}
\include{occur_tab}
\tablenotetext{a}{The occurrence rate is expressed as the number of planets per Sun-like star for the given radius bin, and integrated over $2$ to $10$ yr. The reported values are taken from the peak of the posterior distribution, and the uncertainties are taken as the values at $\exp(-0.5)$ from the peaks.}
\tablenotetext{b}{The total occurrence rate (bottom row) is derived in Section \ref{subsec:size} for $\alpha \ne 0$. It is close, but not identical, to the sum of the three size bins.}
\end{deluxetable}

Here, we attempt to compare our results  to those from other studies. Long-term RV surveys have monitored a large number of stars and have obtained occurrence rates for cold giant planets.  \citet{Mayor11} concluded a planet occurrence rate of $0.139 \pm 0.017$ for planet mass $m\sin i > 50~M_{\oplus}$ and period $P <$ 10 yr.  A more convenient comparison is against that of \citet{Cumming08}, where they expressed  the planet occurrence rate as a function  of planet mass and period. Evaluating their parameterized form for the mass range $M_p \geq M_{\rm Saturn}$ and the period range $2-10$ yr, we obtain an occurrence rate of $0.05 - 0.1$ planets per FGK star. To compare this against our results, we assume that planets with masses $M_p \geq M_{\rm Saturn}$ are also larger than Saturn. For the size range  $R_{\rm Saturn} - R_{\rm J}$, we find an occurrence rate of $0.07_{-0.03}^{+0.04}$ planets per Sun-like star, in good agreement with \citet{Cumming08}.

%%%%%%%%%%%%%%%%%%%%%%%%%%%%%%%%%%%%%%%%%%%%%%%%%%%%

\subsection{Size Distribution of Outer Planets}
\label{subsec:size}

A unique outcome of our study is the size distribution for outer planets, something hitherto unexplored.  In the following, we investigate the observed size distribution and discuss its implications.

The likelihood of producing the observed radius distribution from the underlying distribution (Equation \ref{eq:defalpha}) is given by Poisson statistics \citep[e.g.,][]{Gould10,Suzuki16}:
\begin{equation}
    \mathcal{L} = A \log{(5)} e^{-\bar{N}_{\rm lp}} \prod_i^{N_{\rm lp}} \left(\frac{R_{{\rm p},i}}{R_{\rm J}}\right)^{\alpha} p_{{\rm det},i}\ ,
\end{equation}
where $R_{{\rm p},i}$ is the radius of the $i^{\rm th}$ planet and $p_{{\rm det},i}$ its detection efficiency (a function of radius and period).

\begin{figure}
    \centering
    \includegraphics[width=0.52\textwidth]{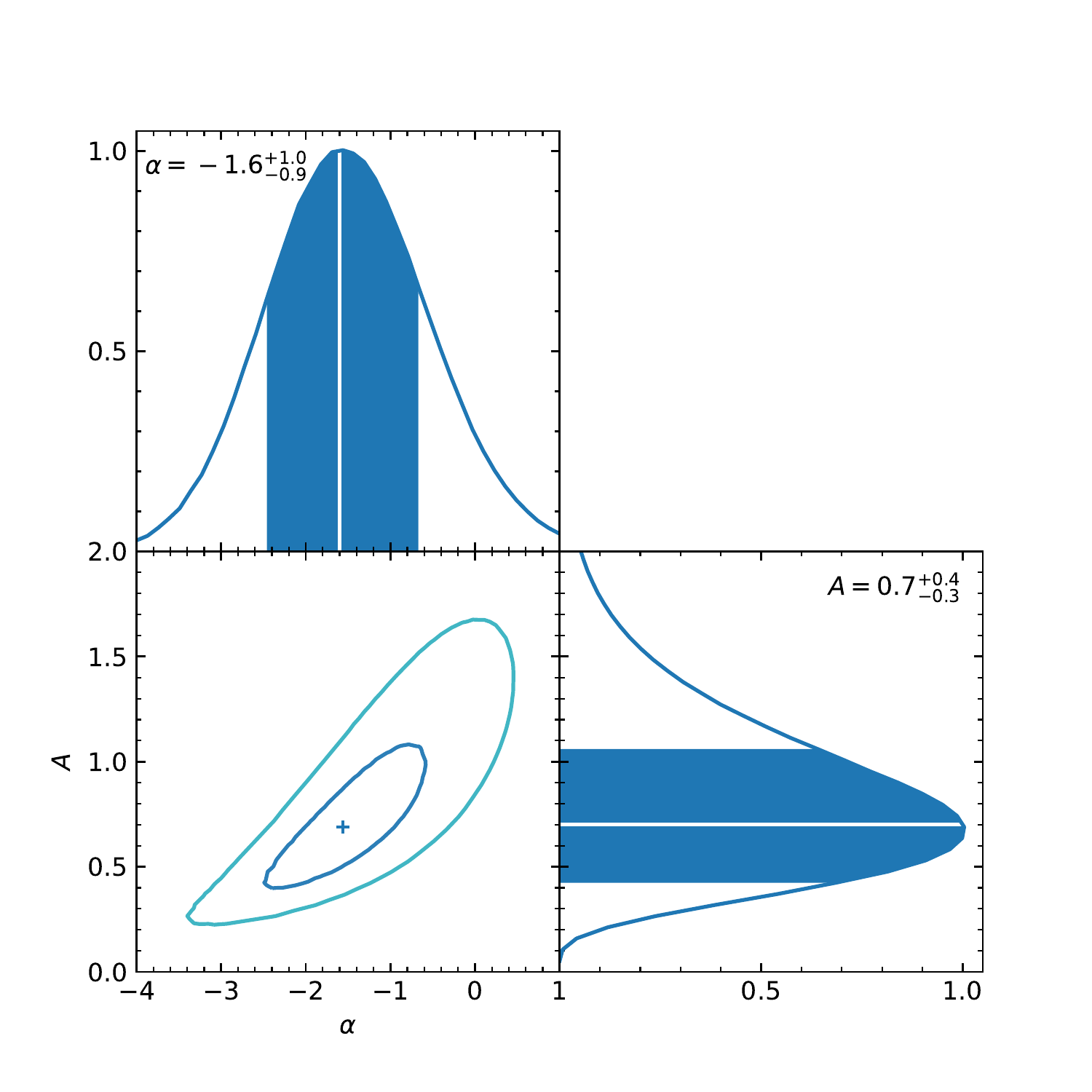}
    \caption{Constraints on the parameters $\alpha$ and $A$. The former is the slope of the planet radius distribution, and the latter its normalization. In the lower left panel, the cross marks the best-fit model, and the contours indicate the 1-$\sigma$ (defined by $\exp{(-0.5)}\mathcal{L}_{\max}$) and 2-$\sigma$ (defined by $\exp{(-2)}\mathcal{L}_{\max}$) boundaries. The marginalized distributions are shown in the top and right panels, with the 1-$\sigma$ ranges shaded.
    }
    \label{fig:contours}
\end{figure}

\begin{figure}
    \centering
    \includegraphics[width=0.48\textwidth]{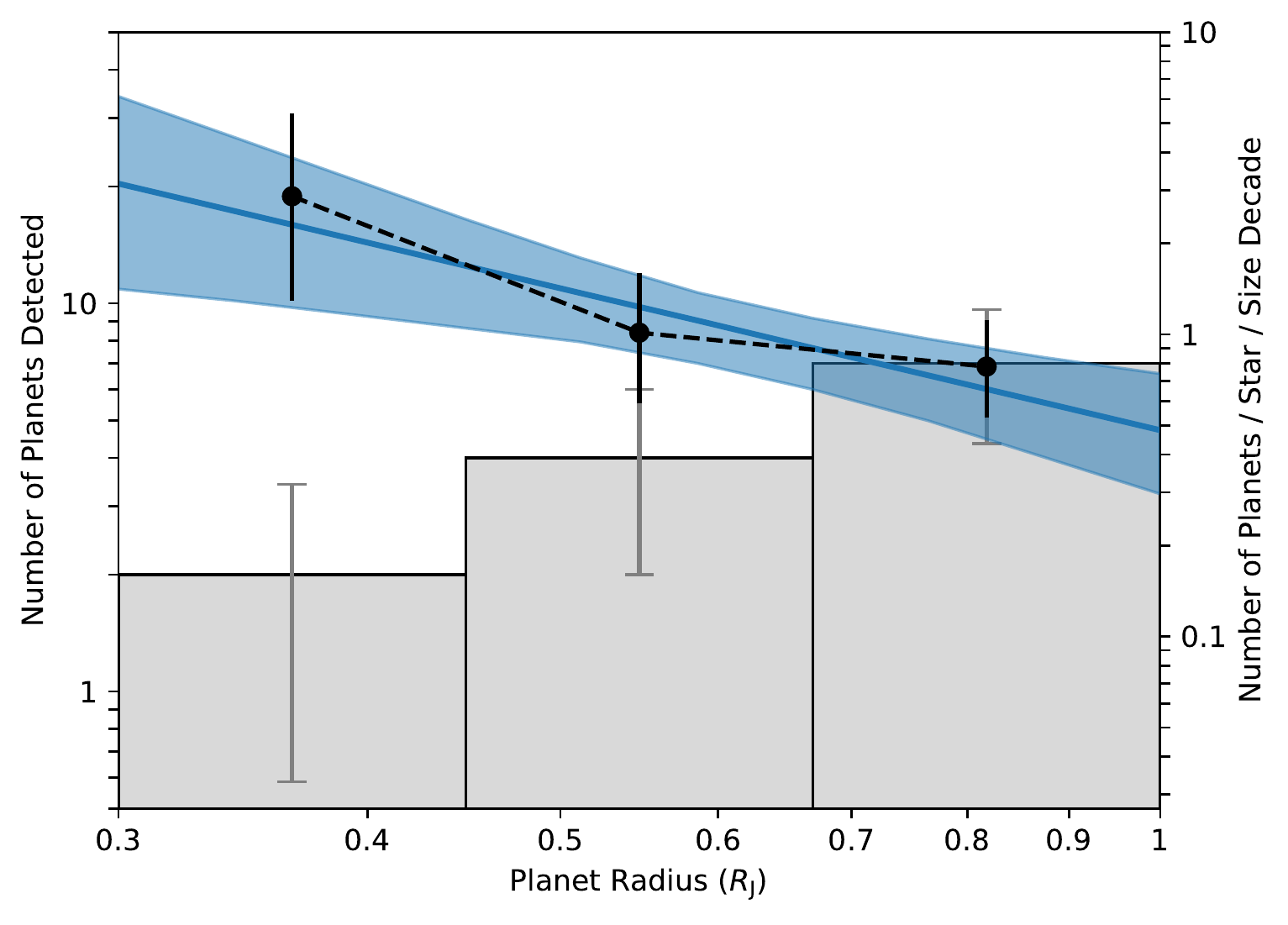}
    \caption{The gray histogram (left axis) shows the observed size distribution of our detected planets ($R_{\rm p} \in [0.3, 1] R_{\rm J}$, and $P \in [2,10]$ yr). Error bars depict Poisson noise. Planet occurrence rates per logarithmic size decade (right axis) are obtained using two approaches.  The blue line (best fit) and shaded region (its $1-\sigma$ range) are results from a maximum likelihood analysis where $\alpha$ is allowed to vary.    The black circles reflect the histogram values, after correcting for detection completeness and  assuming $\alpha=0$ within each size bin. To relate the values here to those in Table \ref{tab:occurrence}, which are integrated over a given size range ($[R_{\rm min},R_{\rm max}]$), multiply the values here by a factor of $\log (R_{\rm max}/R_{\rm min})$.
    }
    \label{fig:occur_hist_peaks}
\end{figure}

We evaluate the likelihood values over a range of $\alpha$ and $A$ and present the results in Figure~\ref{fig:contours}. Both parameters are constrained reasonably well. In particular, the power-law slope $\alpha$ is determined to be $\alpha=-1.6_{-0.9}^{+1.0}$, meaning the occurrence rate rises toward smaller planets. Figure \ref{fig:occur_hist_peaks} shows the best-fit size distribution for our sample, as well as the 1-$\sigma$ allowed range. The occurrence rates derived from the three radius bins separately (assuming $\alpha=0$ in each bin) are also overplotted. These two different approaches obtain consistent results.

Using this derived $\alpha$ value, we can obtain an overall occurrence rate of $0.70^{+0.40}_{-0.20}$, integrated over the size range and period range we defined previously. This is consistent with, though somewhat lower than, what one would obtain by a simple summation of the three radius bins in Table \ref{tab:occurrence}.

How does our result compare to those from microlensing studies, which probe the mass function (but not radius distribution) of outer planets? Planet microlensing is sensitive to planets with a planet-star mass ratio down to Neptune (at $5\times 10^{-5}$ that of the Sun), orbiting at a distance of a few AUs. Most of the planet hosts in these studies are M-dwarfs (median mass $0.6 ~M_\odot$), due to the latter's prevalence in the Galaxy. \citet{Suzuki16} analyzed 30 planetary microlensing events and concluded that the planet occurrence is $0.43$ planets per star, with mass ratio $\geq 3\times 10^{-5}$ and the projected planet-star separation within $[0.45,2.22]$ Einstein radii \citep[see also][]{Gould10,Cassan12}. These separations are similar in physical dimensions to those of our planets, but span a larger dynamic range (a factor of $2.35$ wider). For a crude comparison, we scale down their occurrence rate accordingly to obtain $0.18$. This is some four times lower than our value ($0.70^{+0.40}_{-0.20}$),  suggesting that M-dwarfs may have fewer outer planets than do solar-type stars. On the other hand, the two occurrence rates are marginally consistent with each other, given the large error bars from both surveys and the uncertainties involved in mass-size conversion. More data is required. 

The size distribution we obtain in this study is qualitatively similar to the mass distribution inferred by \citet{Suzuki16}, both indicating that Neptune-massed planets are a few times more abundant than Jovian-massed ones.

What does our size distribution imply for theories of planet structure and formation? First, it is interesting to notice the absence of candidates between $1$ and $2 R_{\rm J}$ in size, despite the high detection completeness for bodies at these sizes. This is what one would expect from calculations of planetary structures: independent of mass and composition, all cold planets should have sizes below $1 R_{\rm J}$ \citep[e.g.,][]{Fortney07}.

Second, while the rising size function toward smaller planets is not surprising, the fact that it is continuous across the size range is unexpected. Planets with sizes between Neptune and Jupiter (our middle size bin, $0.45-0.67 R_{\rm J}$) are not present in our solar system. Lacking appropriate analogs, one may call them super-Neptunes or sub-Saturns. We find that they are at least as common as Jovian-sized ones, if not more so. 

In contrast, population synthesis models based on core-accretion theory \citep[see review by][and references therein]{Benz14}, predict a deficit of such planets. If these planets have core-masses of, say, $25 ~M_\oplus$, their sizes would require hydrogen envelopes with masses from $5$ to $30~ M_\oplus$ \citep[see Figure 8 of][]{Fortney07}. These latter values straddle the range over which runaway gas accretion is predicted to occur, and models predict that the envelopes will grow rapidly (both in mass and size), transforming the planets from Neptunes to Jupiters. As such, the abundance of these intermediate planets is puzzling.

Interestingly, a similar contradiction is noticed by \citet{Suzuki18}. They argued that the smooth planet mass function observed by microlensing surveys is incompatible with, and challenges, our current theory of core-accretion. Our results here provide another piece of evidence in this debate.

%%%%%%%%%%%%%%%%%%%%%%%%%%%%%%%%%%%%%%%%%%%%%%%%%%%%

\subsection{Correlation between Inner and Outer Systems} \label{sec:inner}

Among our list of planet candidates, five out of 13 have inner transiting companions (Figure \ref{fig:inner+outer}). This high incidence is useful both for ascertaining the planetary nature of our candidates (Section \ref{sec:false}) and for studying the correlation of (a) orbital orientation and (b) occurrence between the inner and outer planetary systems. We discuss the latter point here.

The fraction of detected long-period planets with inner transiting companions depends only on the detectability of the inner planets, when the outer one is found to be transiting. Mathematically, this is
\begin{equation} \label{eqn:Nin/Np}
    \frac{N_{\rm lp}^{\rm in}}{N_{\rm lp}} = P({\rm SE}|{\rm Outer}) p_{\rm geo}^{\rm in} p_{\rm rec}^{\rm in}\ .
\end{equation}
Here, $P({\rm SE}|{\rm Outer})$ quantifies the correlation between the inner and outer systems (i.e.,the probability of the inner system being present when the star is known to have an outer planet), and $p_{\rm geo}^{\rm in}$ and $p_{\rm rec}^{\rm in}$ are the geometric transit probability and the recovery rate for the inner planets, respectively. In this case, we are only interested in those planets detectable by {\it Kepler}, for which reliable statistical results are available. We therefore limit ourselves to inner ones with orbital periods shorter than 400 days and radii $\gtrsim 1 ~R_\oplus$ (i.e., super-Earths). The planet search pipeline developed by the {\it Kepler} team \citep{Bryson13} has a recovery rate of nearly 100\% for such planets, \ so we simply adopt $p_{\rm rec}^{\rm in}=1$.

Given the presence of a known outer planet, the geometric transit probability for an inner planet will depend on the inclination dispersion of the system $i_0$ such that
\begin{equation} \label{eqn:fgeo_in}
    p_{\rm geo} \approx \frac{R_*/a_{\rm in}}{\sin{i_0}}\ . 
  \end{equation}
where $a_{\rm in}$ is the orbital separation of the inner one. Below, we adopt $a_{\rm in} \sim 30 R_*$, as is representative for the inner planets here. With this and $p_{\rm rec}^{\rm in}=1$, we may now express equation~(\ref{eqn:Nin/Np}) as 
\begin{equation}
    P({\rm SE}|{\rm Outer}) = 89\% \left(\frac{N_{\rm lp}^{\rm in}/N_{\rm lp}}{5/13}\right) \left(\frac{0.03}{R_*/a_{\rm in}}\right) \left(\frac{i_0}{4^\circ}\right)\ .
\label{eq:psecj}
\end{equation}

Before discussing the implications of the above results, we first consider evidences for the value of $i_0$.

\begin{figure}
	\centering
   \includegraphics[width=0.48\textwidth]{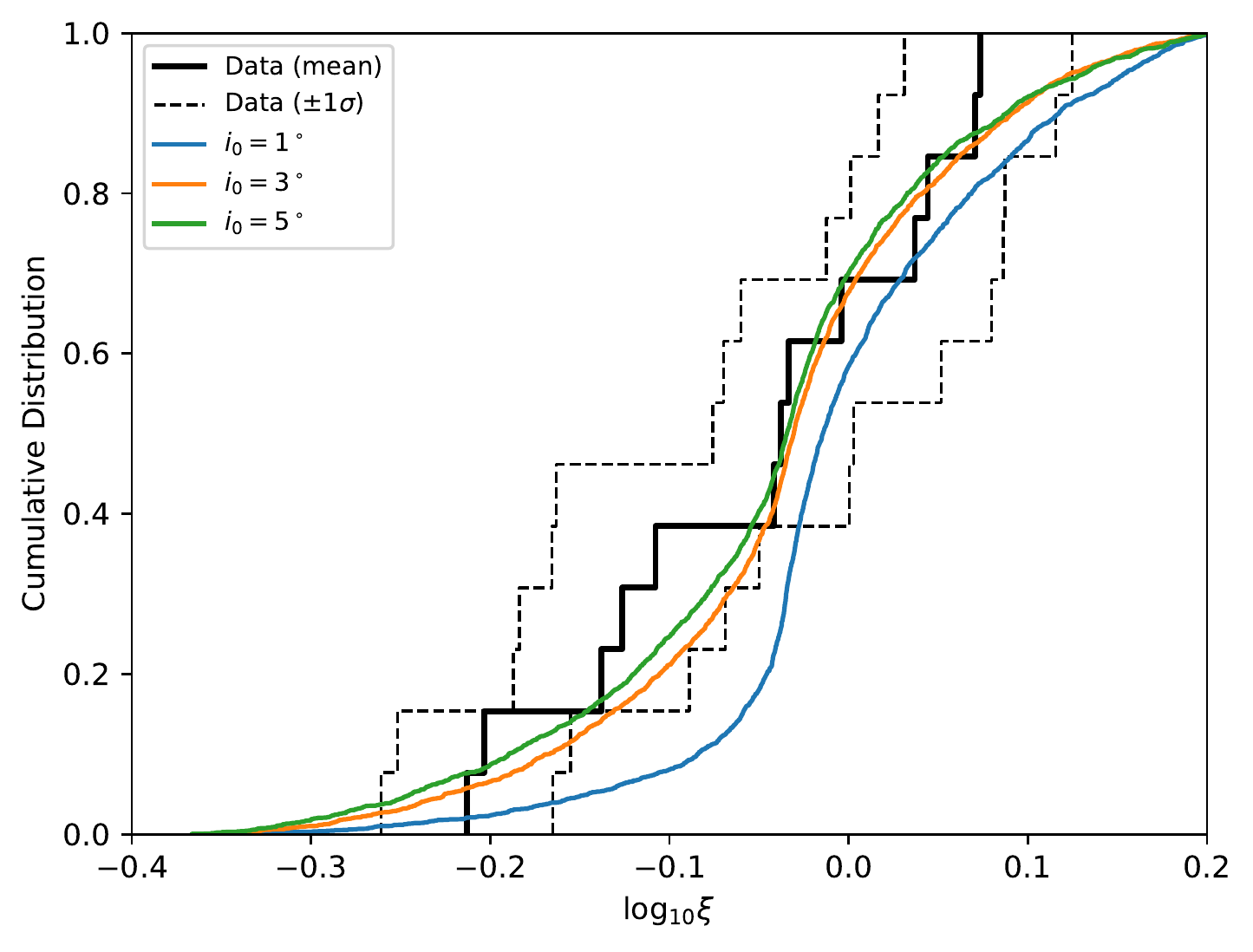}
    \caption{The cumulative distributions of the $\xi$ parameter (Equation~\ref{eqn:xi}) as derived from the data (black) and models with different mutual inclinations. We also show the $1\sigma$ range of the data distribution (dashed lines).
    }  
    \label{fig:xi_cdf}
\end{figure}

\begin{itemize} 

\item  In our solar system, the four terrestrial planets are inclined by $\sim 2-6^\circ$ from the invariable plane, i.e., the plane that is mostly determined by the orbits of Jupiter and Saturn. 

\item From analyzing {\it Kepler} statistics for the close-in {\it Kepler} planets, \citet{Zhu18} concluded that $i_0 \sim 0.8^\circ (k/5)^\beta$, where $k$ is the intrinsic multiplicity of the (inner) system and the power-law index $\beta \in [-4,-2]$. They also found that typically $k \sim 3$ within $400$ days. This yields an average inclination of $i_0 \sim 4^\circ$.

\item A direct constraint on the mutual inclination $i_0$ can be derived from the ratio of transit durations $T_{\rm dur}$ of planet pairs weighted by their respective orbital velocities ($\propto P^{-1/3}$; \citealt{Steffen10})
\begin{equation} \label{eqn:xi}
    \xi \equiv \frac{T_{\rm dur,in}/P_{\rm in}^{1/3}}{T_{\rm dur,out}/P_{\rm out}^{1/3}}\ .
\end{equation}
Taking the nominal periods for the outer planets,\footnote{We use the orbital periods derived from our transit fits assuming zero eccentricity (Table \ref{tab:catalog}). This is justified, given that \citet{Fabrycky14} have shown that the parameter $\xi$ is insensitive to orbital eccentricities.} we plot the resulting distribution of $\xi$ for the $5$ systems  in Figure \ref{fig:xi_cdf}, where the $\pm 1 \sigma$ uncertainties of individual $\xi$ measurements are also shown. We also compute the $\xi$ distributions for models with three different mutual inclinations ($1^\circ$, $3^\circ$, and $5^\circ$), following the method of \citet{Fabrycky14} (see also \citealt{Zhu18}). As Figure~\ref{fig:xi_cdf} shows, models with larger mutual inclinations better resemble the data. Even though we cannot distinguish between the models with $i_0=3^\circ$ and $5^\circ$, this test is robust in excluding the small mutual inclination case ($i_0=1^\circ$). Therefore, it is not unreasonable to use $i_0=4^\circ$ for the mutual inclination between the inner and outer planetary systems.

\end{itemize}

We conclude from the above that $i_0 = 4^\circ$ may be a reasonable value to describe the inner-outer mutual inclinations\footnote{This is surprising from the perspective of planet eccentricities from radial velocity surveys. Typical cold Jupiters are found to have an eccentricity dispersion $e_0 \sim0.3$. If we assume that $i_0 \sim e_0/2$, as is often found in dynamical experiments, this would imply a large $i_0 \sim 10^\circ$ among cold Jupiters.}. Any larger value of $i_0$ would further strengthen our argument below.

Given the above argument, equation (\ref{eq:psecj}) suggests that there must be a strong correlation between small inner planets and outer planets. Almost all systems with outer planets should also have small inner planets.\footnote{ Whether the reverse is true, i.e., that all small inner planets have large outer planets, has not yet been answered.} Interestingly, such a conclusion was also obtained by \citet{ZhuWu18} through analysis of RV data. They found that almost all cold Jupiters should have inner {\it Kepler} systems. Such a conclusion, now obtained independently from two studies (and extended to planets smaller than Jupiters) suggests that the formation of cold planets is conditional upon the presence of inner planets. 

Turning the above argument around, we may also argue that the mutual inclinations between inner and outer planets, spanning a period range from a few days to a few years, must be low. Even assuming $P({\rm SE}|{\rm CJ}) = 100\%$, the mutual inclination cannot be much greater than $4^\circ$. These are very flat dynamical systems.

Anomalously, one of the cold giant candidates has five inner transiting planets. At face value, this is statistically unlikely given that systems with five inner transiting planets only account for $0.5\%$ (3/589) of all {\it Kepler} systems. In other words, one should expect to have 0.025 of such systems among a sample of five inner systems. This is, however, naturally explained by the above cited result of \citet{Zhu18}: because 
the inclination dispersion of the inner system correlates with its multiplicity, adopting $i_0 = 0.8^\circ$ for $k=5$ boosts the possibility of detecting a single system containing five transiting planets to 0.6.

%%%%%%%%%%%%%%%%%%%%%%%%%%%%%%%%%%%%%%%%%%%%%%%%%%%%%%%%%%

\section{Conclusions} \label{sec:Conclusions}

By employing the automated transit search pipeline of FM16 to an extended sample of target stars, we detect 19 planet candidates, including 12 in common with FM16. Updated stellar radii from {\it Gaia} DR2 allow us to exclude a number of them as false positives, and we are left with 15 likely candidates. We further restrict ourselves to $13$ of those for our analysis. Among these, five have known inner transiting companions.

The FM16 pipeline also yields the detection efficiency and search completeness. Using these, we are able to report a total occurrence rate of $0.70_{-0.20}^{+0.40}$ per Sun-like star for outer planets within a radius range of $0.3 - 1 R_{\rm J}$ and a period range of $2 - 10$ yr. We also find that the radius distribution of cold planets can be expressed as a single power-law of the form $dN/d\log{R_{\rm p}} \propto R_{\rm p}^{\alpha}$, where $\alpha=-1.6^{+1.0}_{-0.9}$.

Our occurrence rate for Jovian planets is compatible with those found by long-term RV monitoring, and our finding that Neptunes are more common than Jupiters is an independent confirmation of the results from microlensing studies. While microlensing has discovered a nearly continuous mass function going from Jupiter- to Neptune-mass planets, we find a continuous size distribution between the two ends, challenging current theories of core accretion.

Finally, we investigate the seemingly surprising fact that five of our {13 long-period planets} orbit stars with known transiting inner companions. This brings us to two conclusions. First, there is a strong correlation between the presence of cold planets, including both Jovian and smaller planets, with the inner system. The formation of cold planets appears to be conditioned upon the presence of small inner planets. This independently confirms the conclusions drawn by \citet{ZhuWu18} and \citet{Bryan19} using RV data. It also provides a valuable clue for planet formation, one that unfortunately remains undeciphered. Second, the mutual inclinations between the inner and outer parts of these planetary systems must be quite small. We find that, across two decades in period, the average inclination should remain below $4^\circ$.

Despite our small sample size, our ability to constrain the prevalence of outer planets and elucidate the relationship between inner and outer planets is a testament to the quality and versatility of the {\it Kepler} data. Looking forward, it is important to note that current and upcoming transit surveys (e.g., {\it TESS} and {\it PLATO}) will have considerably shorter observational baselines. However, the number of planets with periods longer than these baselines will still undoubtedly be numerous \citep{Villanueva19}. Such detections will enable further statistical analysis of the population of long-period planets, better illuminating the architecture of and relationship between planets occupying the inner and outer reaches of their systems.

%%%%%%%%%%%%%%%%%%%%%%%%%%%%%%%%%%%%%%%%%%%%%%%%%%%%%%%%%

\acknowledgements

The authors would like to thank Dan Foreman-Mackey and his coauthors to FM16 for the \texttt{peerless} code, which is well-documented and entirely open-source. The authors also thank Kento Masuda for informing us of an earlier mistake in our draft, and the anonymous referee for improving the clarity of our manuscript. This work has made use of data collected by the {\it Kepler} mission and retrieved from the Mikulski Archive for Space Telescopes (MAST). STScI is operated by the Association of Universities for Research in Astronomy, Inc., under NASA contract NAS5-26555. Funding for the {\it Kepler} mission is provided by the NASA Science Mission directorate. This research has also made use of the NASA Exoplanet Archive and the Exoplanet Follow-up Observation Program website, which are operated by the California Institute of Technology, under contract with the National Aeronautics and Space Administration under the Exoplanet Exploration Program. M.K.H. and Y.W. additionally acknowledge funding from the Natural Sciences and Engineering Research Council (NSERC) of Canada. W.Z. was supported by the Beatrice and Vincent Tremaine Fellowship at CITA.

%%%%%%%%%%%%%%%%%%%%%%%%%%%%%%%%%%%%%%%%%%%%%%%%%%%%%%%%%

\appendix

\section{Modification to the Search and Vetting Procedure of FM16}\label{sec:peerless_mods}

We use \texttt{peerless} to complete a fully automated search for long-period planetary transits in our target light curves. In this appendix, we provide a description of the modifications we make to the search and vetting procedure of FM16.

To determine whether a candidate signal is the result of a planetary transit or simply a systematic effect, FM16 fit five separate models to the potential transit. We elect to discard their box model, which was designed to account for signals that are neither convincingly transit-like nor well-described by the other models. We find that this model has little physical motivation behind it and in practice tends to be too stringent, rejecting planetary transits that would otherwise be accepted at all other levels and/or by eye.

\begin{figure}
	\centering
   \includegraphics[width=0.48\textwidth]{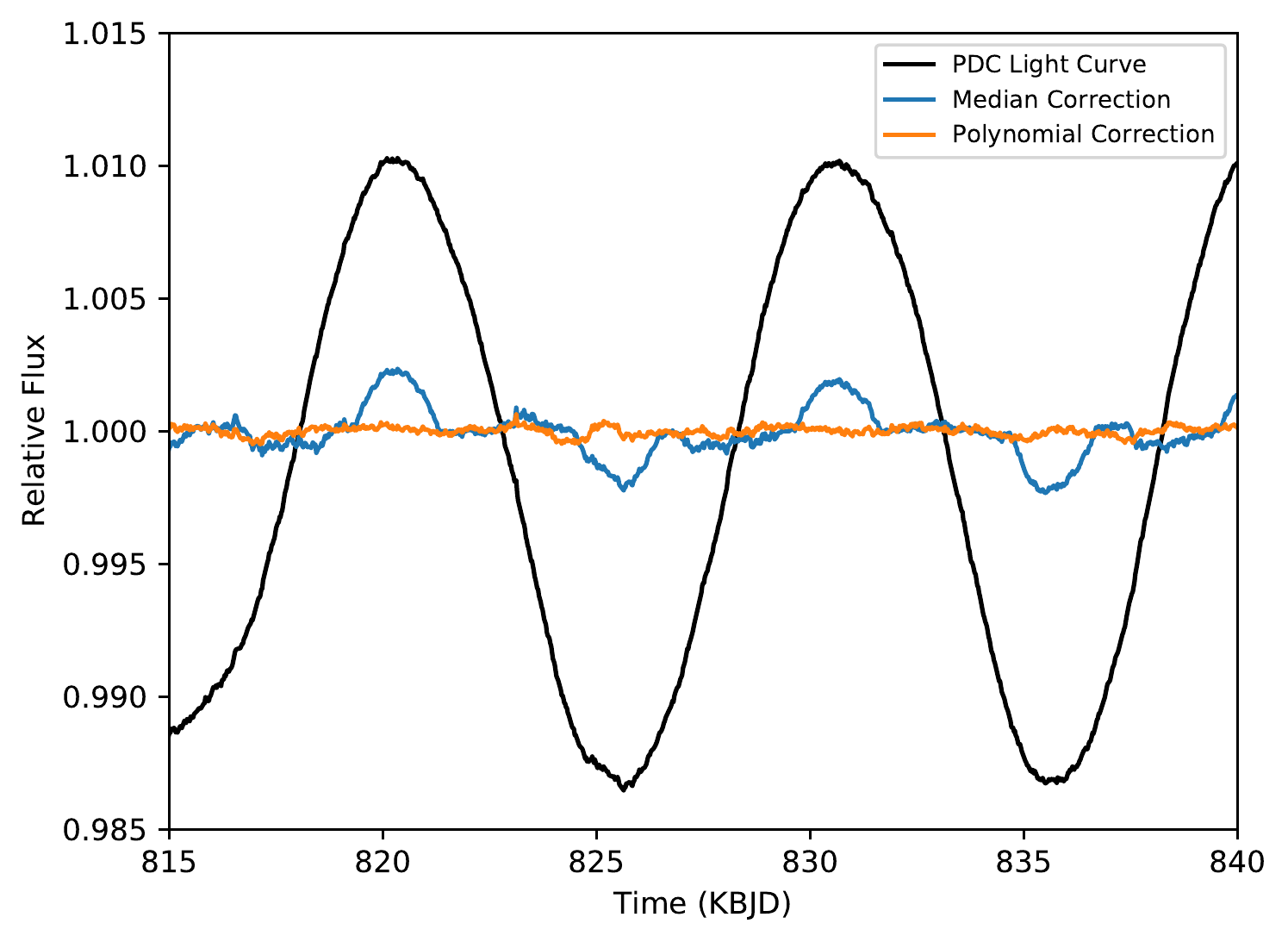}
    \caption{A representative example of a section of a PDC light curve displaying stellar variability (black). The RWM correction is shown in blue, while the second-order polynomial correction is shown in orange. Both corrections employ a half-width of two days.
    }  
    \label{fig:polynomial}
\end{figure}

We make two supplemental changes to the \texttt{peerless} code to eliminate false positive signals caused in part by the running windowed median (RWM) correction. The RWM correction is applied with a two-day half-width to all PDC light curves \citep{Smith12, Stumpe12} to reduce the effects of stellar variability -- but it does not work in all cases. A subset of light curves display large-amplitude stellar variability on timescales of a few to tens of days, which are not excluded from our target sample because each star's CDPP is low on the 7.5 hr timescale we consider for our selection cuts. This stellar variability is not fully removed by the RWM with a two-day half-width, and in some cases, the variability is actually made to look more transit-like after the correction (see Figure \ref{fig:polynomial}). As such, these signals are not rejected by the variability model described in FM16. To eliminate such signals, rather than use the RWM, we fit a second-order polynomial to the PDC light curves, using a running window with an identical two-day half-width. This polynomial correction considerably decreases the noise of such variable light curves, and thus increases the S/N of planetary transits found in those light curves. However, transits are not well-modeled by a polynomial (the change in flux at ingress and egress is far too abrupt). We therefore revert to the median correction at any point in the light curve where the polynomial residuals are more than $3 \sigma$ above their median value.

Unfortunately, this running windowed polynomial correction is very computationally expensive. We therefore choose to only apply it to the light curves of real signals that have already passed all other steps in the vetting process. If any vetted signals are removed by the polynomial correction to the original PDC light curves, we discard the signals as artifacts of the RWM correction applied to stellar variability.

The third modification we make to \texttt{peerless} is to introduce a symmetry criterion to the vetting process for real signals. For each signal, we mirror the data set about the time of mid-transit to produce a `left' transit and a `right' transit. We fit these using the same transit model as FM16, and compare the fitted radii such that

\begin{equation}
    \frac{|R_{\rm p, ~\mathrm{left}} - R_{\rm p, ~\mathrm{right}}|}{R_{\rm p, ~\mathrm{original}}} \leq 0.005 ~.
\end{equation}
If the difference between the `left' and `right' radii (scaled by the fitted radius of the original signal) is larger than 0.5\%, we consider the signal asymmetric and reject it. This limit is chosen to reflect similar methods of asymmetry-based transit rejection, such as that of \citet{Turner16} and references therein.

This symmetry criterion is particularly important for removing signals resulting from sudden drops followed by gradual rises in pixel sensitivity, as well as signals that are produced at the union of two sections of the PDC light curve. In these cases, the step function model (or any other model) is insufficient to describe the change in flux well enough to exceed the Bayesian information criterion of the transit model (see FM16 for details). This symmetry restriction therefore serves as a final automated check of the validity of any identified candidates. We also confirm by eye that none of the signals rejected in this step are convincingly transit-like in appearance.

We further correct the factor of $\pi^{1/3}$ missing from \texttt{peerless} in the calculation of transit probability, which resulted in an underestimation of the occurrence rate by 1.46 in FM16. 

Additionally, though it is not a modification we chose to adopt in this paper, we deem it worth addressing the fact that we experimented with lowering the S/N threshold for detecting transits. Decreasing this threshold from 25 to 20 introduced a plethora of signals that were not convincingly transit-like by eye, yet could not be removed by the vetting process. We therefore chose to retain the S/N threshold used by FM16 so as to avoid these signals and preserve the automated nature of the search and vetting procedure.

\begin{figure*}[!ht]
	\centering
   \includegraphics[width=0.98\textwidth]{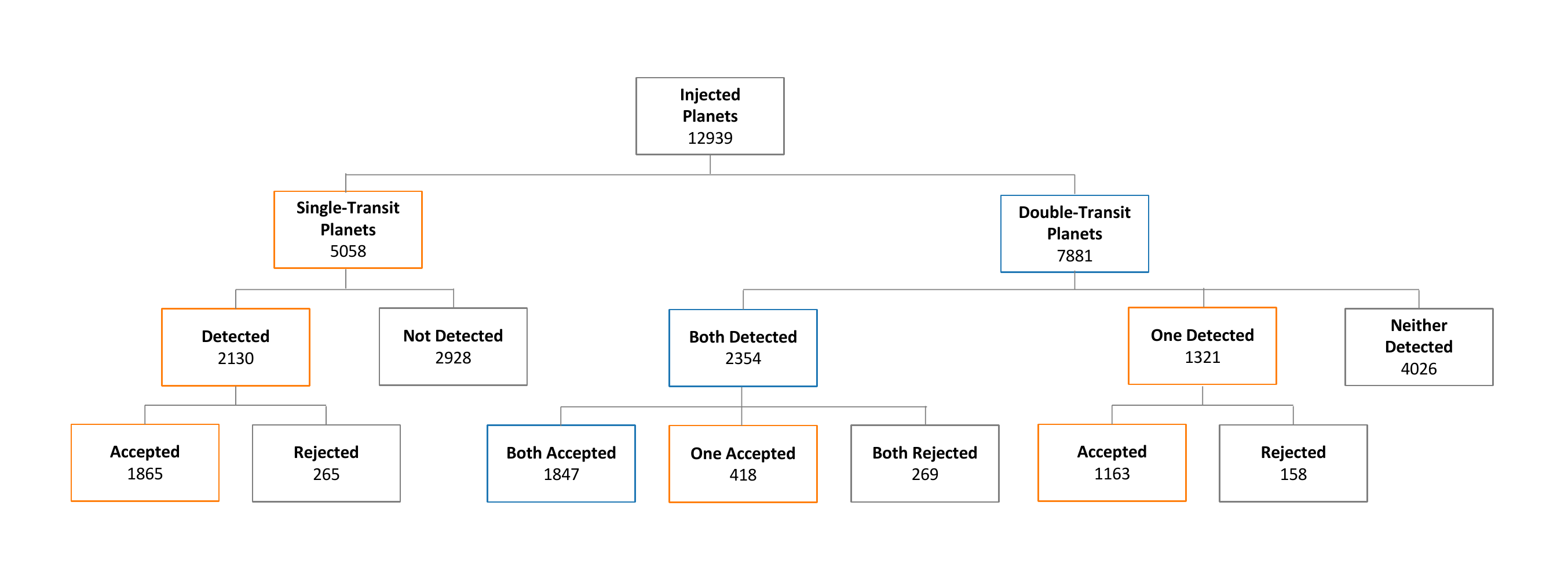}
    \caption{Double- vs. single-transit injection results for $0.03 < R_{\rm p}/R_* < 0.3$ and 2 yr $< P <$ 3.767 yr (corresponding to the first column of the bottom right plot in Figure \ref{fig:completeness}). While some double-transit injections have both transits accepted, nearly as many masquerade as single-transit candidates when only one transit passes all steps in the vetting procedure.
    }
    \label{fig:flowchart}
\end{figure*}

%%%%%%%%%%%%%%%%%%%%%%%%%%%%%%%%%%%%%%%%%%%%%%%%%%%%%%%%%

\section{Double- versus Single-transit Recovery}\label{sec:dvs}

An important caveat of the injection and recovery results presented in Section \ref{sec:Completeness} is that, depending on the injected orbital period and time of mid-transit, a set of simulated transit parameters can produce up to two transits in a single light curve, but only one transit needs to pass all steps of the vetting process to consider the injection recovered. This increases the chance of a double-transit injection being recovered, as it essentially provides twice as many opportunities for the candidate to be accepted. As one might expect, however, a candidate whose orbital period permits it to transit twice within the baseline of observations could masquerade as a candidate with a single-transiting event (i.e., having a much longer period) if the second transit is rejected by the vetting process, takes place during a gap in the observations, or simply is not picked up by the search procedure for any number of reasons. The results of our artificial injections allow us to investigate the prevalence of this phenomenon, assuming a logarithmically flat distribution in orbital period. We may then compare the ratio of double- to single-transiting candidates expected to be detected under this assumption to the ratio observed in our candidate catalog.

First, we note that a candidate can only produce two transits if its orbital period is shorter than the baseline of observations (less than $\sim4$ yr for {\it Kepler}) and if the time of mid-transit is appropriate. Recall also that we only consider orbital periods greater than two years, as this value will be just out of range of transit searches requiring three or more transits within the {\it Kepler} baseline. For instance, consider a transiting planet with an orbital period of three years. Its transit could appear twice in a {\it Kepler} light curve only if its time of mid-transit occurred in the first quarter of the observations. For simplicity, we restrict our consideration to candidates with orbital periods between 2 and 3.8 yr (coinciding with the first column of each plot in Figure \ref{fig:completeness}), but allow the full range of values for all other transit parameters and stellar magnitudes. We also neglect the effect of decreasing transit probability across this relatively small period range for our simple analysis.

With a logarithmically flat period distribution and our chosen period range, the number of injected planets capable of transiting twice in a {\it Kepler} light curve (we will call these double injections) is in a nearly 3:2 ratio with the number of single injections: $\frac{D}{S}|_{\rm inj} = 1.56$. Following the application of our search and vetting procedure to these artificial signals, however, the ratio of {\it accepted} double- versus single-transit events is very different: $\frac{D}{S}|_{\rm acc} = 0.54$. This is because about one-half of all accepted double injections are mistaken for single-transit events. While these injected planets are still considered recovered, their periods are likely to be more crudely estimated (based on their transit duration) compared to double injections where both transits are accepted and a precise period prediction can be made. A breakdown of each step in our double vs. single transit analysis is shown in Figure \ref{fig:flowchart}.

How, then, does this compare to the observed ratio of double- versus single-transit events among our planet candidates? Of the 15 candidates we deem genuine with a period range of $2 - 3.8$ yr and a fractional planet radius of $0.03 \leq R_{\rm p}/R_* \leq 0.3$, we find four double-transit candidates and six single-transit candidates, such that $\frac{D}{S}|_{\rm obs} = 0.67$. This ratio is not totally removed from our injection and recovery results assuming a logarithmically flat period distribution, but it is difficult to confidently compare the two given such a small sample size. Minimally adjusting our chosen range in period or fractional radius substantially changes this $\frac{D}{S}|_{\rm obs}$ ratio while having a much smaller effect on $\frac{D}{S}|_{\rm acc}$ from our injection results.

Therefore, we find it difficult to determine whether a logarithmically flat distribution in period is representative of the population of outer planets in nature. While this assumed distribution is acceptable for the purposes of this work, we recommend further analysis following the inevitable expansion of the long-period planet catalog in the coming years.

%%%%%%%%%%%%%%%%%%%%%%%%%%%%%%%%%%%%%%%%%%%%%%%%%%%%

\section{An Alternative Derivation of the Size Distribution of Outer Planets} \label{sec:size}

The intrinsic size distribution is indicative of the physical compositions of outer planets. The occurrence estimate of Section \ref{sec:occurrence} already points to the higher incidence of smaller planets. Here, we illustrate our resulting size distribution, using a slightly different argument, independent from the search completeness calculated by \texttt{peerless}.

\begin{figure}
	\centering
   \includegraphics[width=0.48\textwidth]{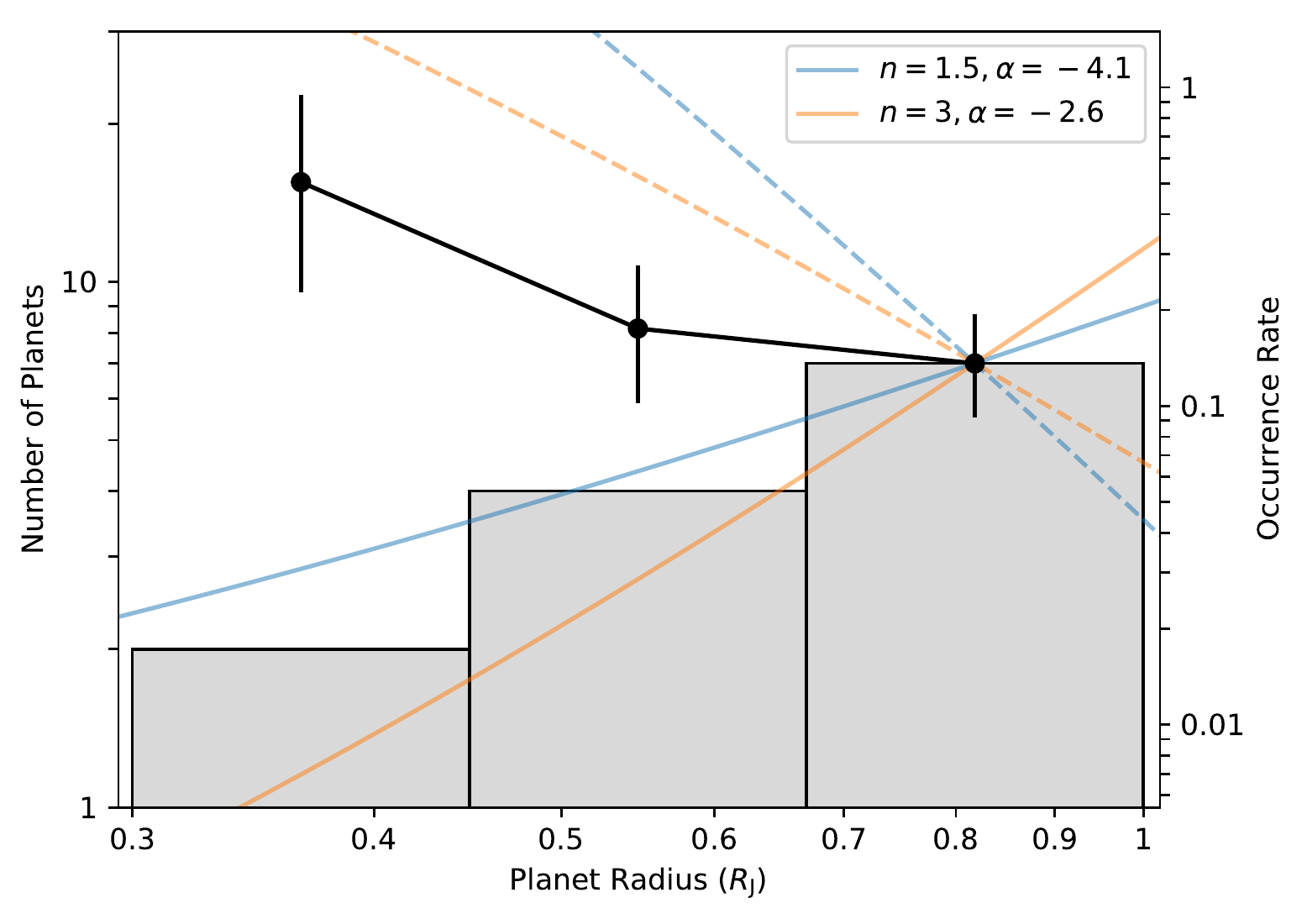}
    \caption{The gray histogram shows the size distribution of our detected planets within $0.3 ~R_{\rm J} < R_{\rm p} < 1 ~R_{\rm J}$ and 2 yr $< P < $ 10 yr, while the colored lines show power laws describing the possible underlying population. 
    The solid colored lines indicate approximate fits to the value of $n$ and the dashed lines their corresponding $\alpha$, where $n$ is the observed size power law and $\alpha$ the intrinsic size power law.
    Each power law is tethered to the histogram at $R_{\rm p} \sim 0.82~ R_{\rm J}$, roughly the radius of Saturn. In black, we plot the calculated occurrence rate and error bars for the full magnitude range from Table \ref{tab:occurrence}.
    }
    \label{fig:occur_hist}
\end{figure}

Adopting the same power law as in equation (\ref{eq:defalpha}), and taking $\beta=0$, we can constrain the value of $\alpha$ using the observed sample, correcting for the fact that smaller planets are more difficult to detect. Planet candidates in our sample are detected when their transit signal surpasses an S/N threshold of 25. While the signal strength is determined by the transit depth, noise in {\it Kepler} light curves is a combination of photon (shot) noise, measurement error (e.g., pointing error and instrument noise), and stellar variability \citep{Koch10}. The {\it Kepler} team encapsulates the total noise of each star into quarterly transit durations of 3, 6 and 12 hr, known as the Combined Differential Photometric Precision (CDPP) \citep[][]{Christiansen12}. For dwarf stars (nongiants) dimmer than 12th magnitude, Table 2 of \citet{Christiansen12} shows that the noise is dominated by photon shot noise and the CDPP value scales with stellar magnitude roughly as $CDPP \propto \sqrt{{ F}}$, where $F$ is the stellar flux.  This fails for brighter stars for which the CDPP is dominated by intrinsic stellar variability (and for which the CDPP is constant). Due to their prevalence, the former population is of relevance for our search. Therefore, the single-transit S/N goes as
\begin{equation}
{\rm S/N} \propto \left({{R_{\rm p}}\over{R_*}}\right)^2 \sqrt{\Delta T \times {F}}~ .
\label{eq:roughSNR}
\end{equation}
Here, we ignore the dependence on transit duration $\Delta T$ and on stellar radius $R_*$. In reality, both affect the S/N to some degree. The minimum planet size at which a transit is detected thus goes as $F^{-1/4}$, meaning smaller planets can only be detected around brighter stars\footnote{Indeed, the smallest planet in our sample (KIC 8505215) is detected around the brightest star in our list, at $K_p = 13.0$. It is also among the smallest stars, making detection easier.}. We denote the minimum flux at which a planet of size $R_{\rm p}$ is detected as $F_{\rm min}(R_{\rm p})$. The stellar sample we use has a flux distribution of the form $dN/d\log F \propto F^{\gamma}$ (Figure \ref{fig:mag-hist}), with $\gamma \approx -1.4$, so the observed size distribution is related to the intrinsic one (equation \ref{eq:defalpha}) as
\begin{eqnarray}
\left({{dN}\over{d\log R_{\rm p}}}\right)_{\rm obs} & = & A R_{\rm p}^\alpha \times {{N_*(F \geq F_{\rm min}(R_{\rm p}))}\over{N_{\rm *,tot}}} \nonumber \\
& \propto & R_{\rm p}^\alpha \times \int_{f_{\rm max}}^{f_{\rm min}} {{dN_*}\over{d\log F}}
\nonumber \\
& \propto & R_{\rm p}^{\alpha - 4 \gamma}~ .
\label{eq:alphaton}
\end{eqnarray}
In other words, because we can only detect small planets around stars that are brighter, a single detection of one such small planet carries a lot of weight.  From Figure \ref{fig:occur_hist}, we find that the observed size distribution scales as $R_{\rm p}^{n}$ where, very roughly, $n \in [1.5,3]$. Therefore, $\alpha = n + 4 \gamma \in [-4.1,-2.6]$. Using a median value of $\alpha \sim -3.5$ would imply that the occurrence rate rises by a factor of $\sim 16$ between $1 ~R_{\rm J}$ and $0.45 ~R_{\rm J}$. In other words, cold Neptunes are much more prevalent than cold Jupiters; a conclusion similar to that of \citet{Suzuki16}.

Figure \ref{fig:occur_hist} also shows the detailed size distribution calculated using detection completeness. Our size distribution above captures the essence of the rising behavior, but is steeper, likely because our simple derivation has ignored effects of stellar variability and variance in stellar radius.

%%%%%%%%%%%%%%%%%%%%%%%%%%%%%%%%%%%%%%%%%%%%%%%%%%%%

\bibliographystyle{apj}
\bibliography{bibliography}

\end{CJK*}
\end{document}

%% file: lcfig.tex
\includegraphics[width=0.24\textwidth]{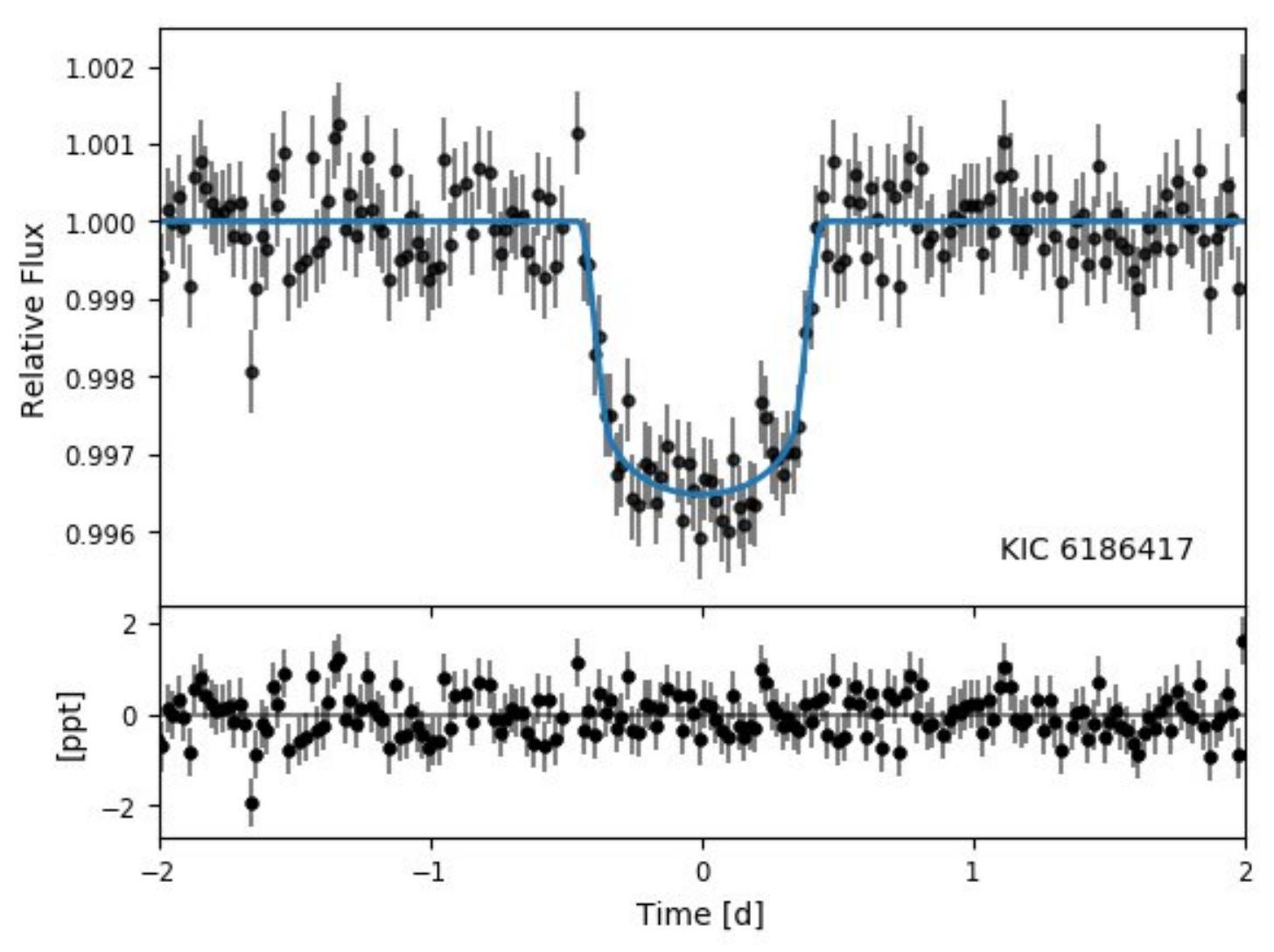}
\includegraphics[width=0.24\textwidth]{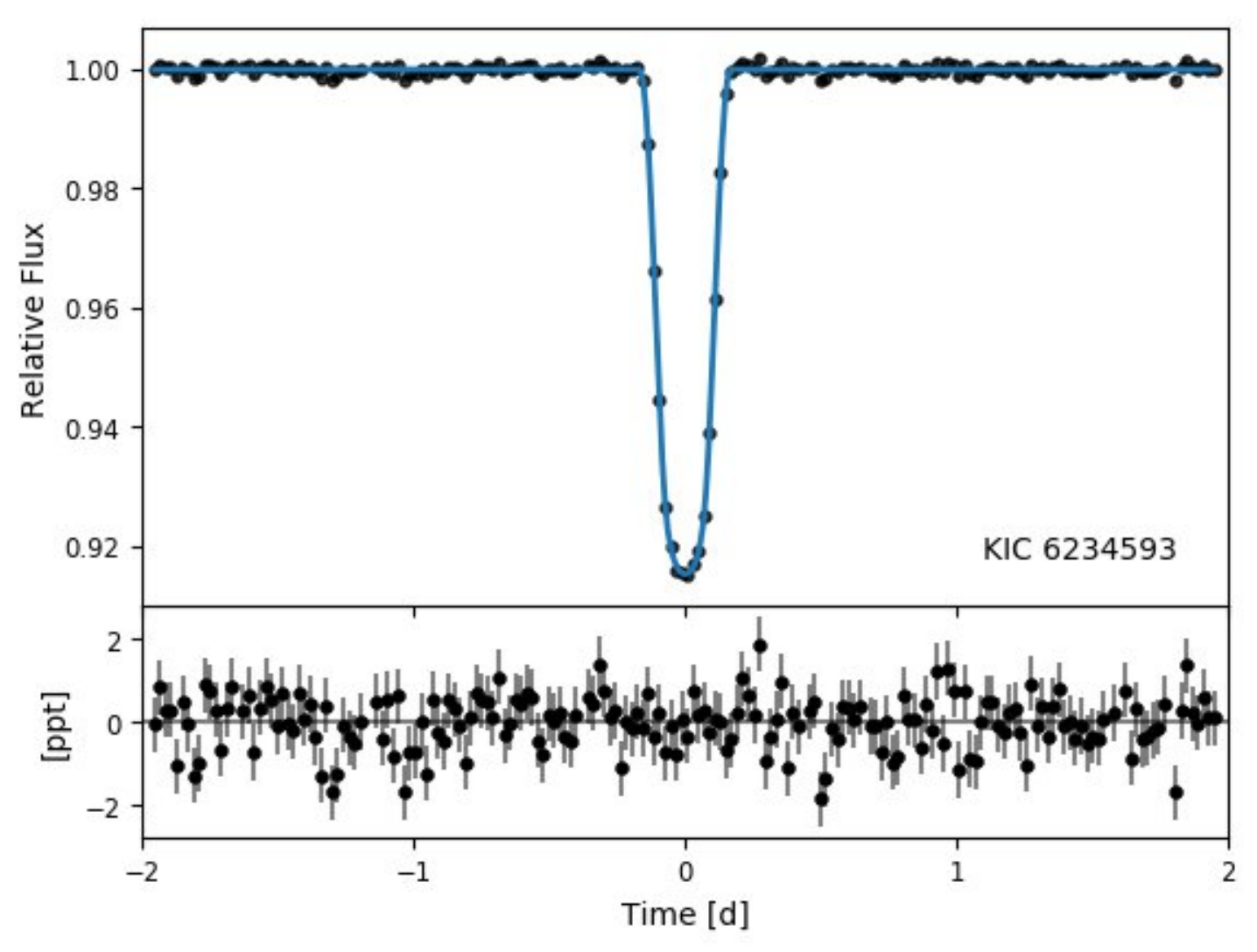}
\includegraphics[width=0.24\textwidth]{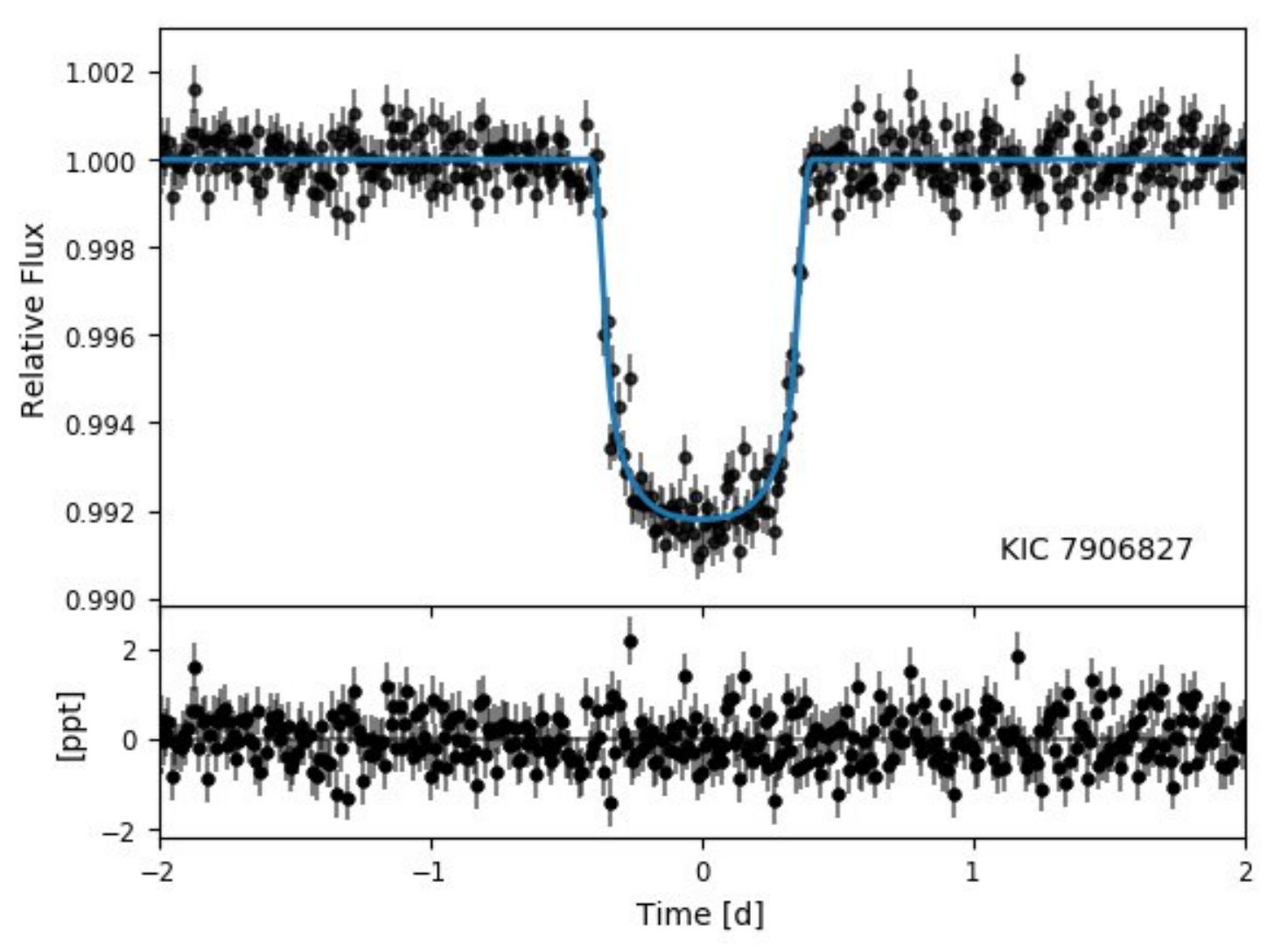}
\includegraphics[width=0.24\textwidth]{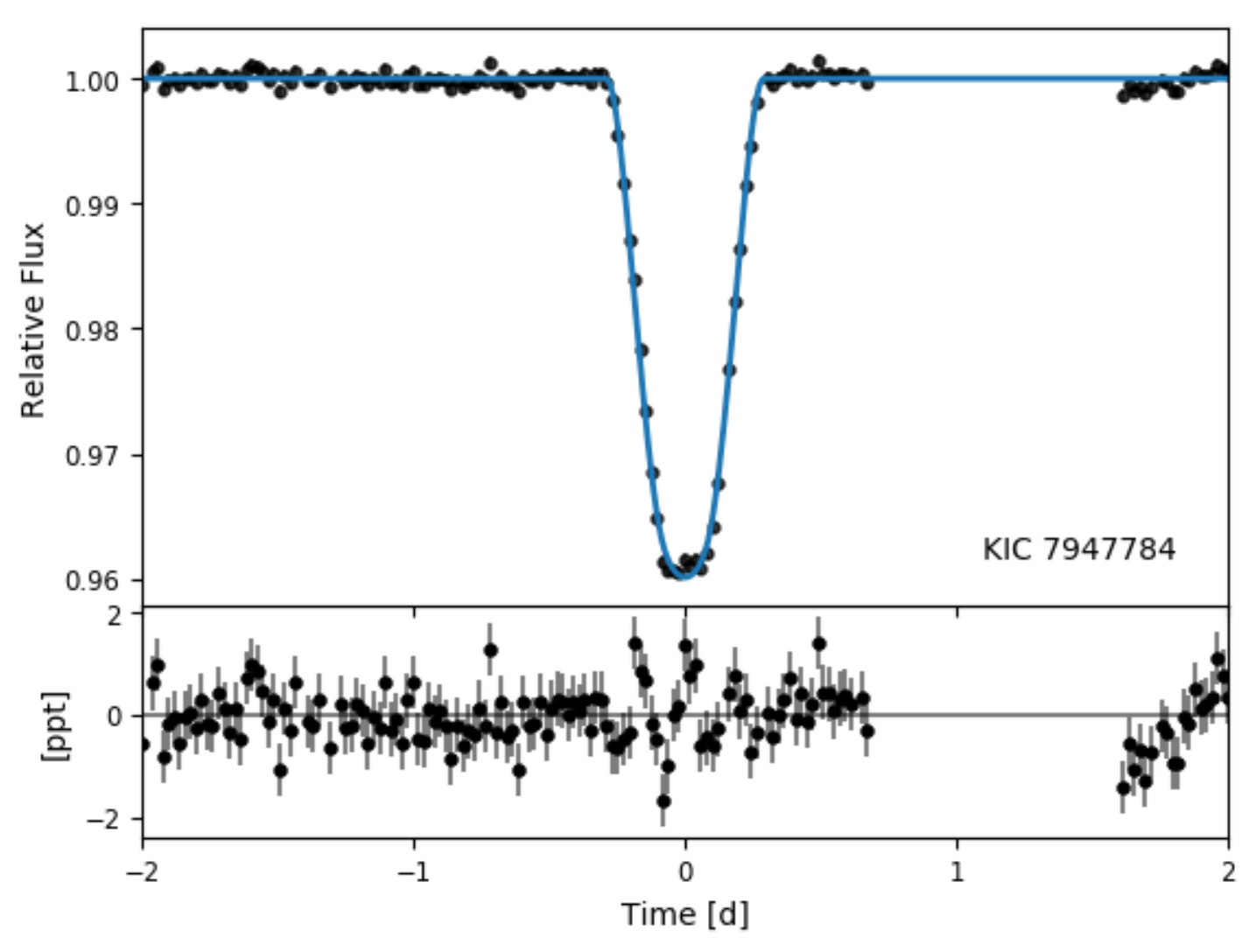}
\includegraphics[width=0.24\textwidth]{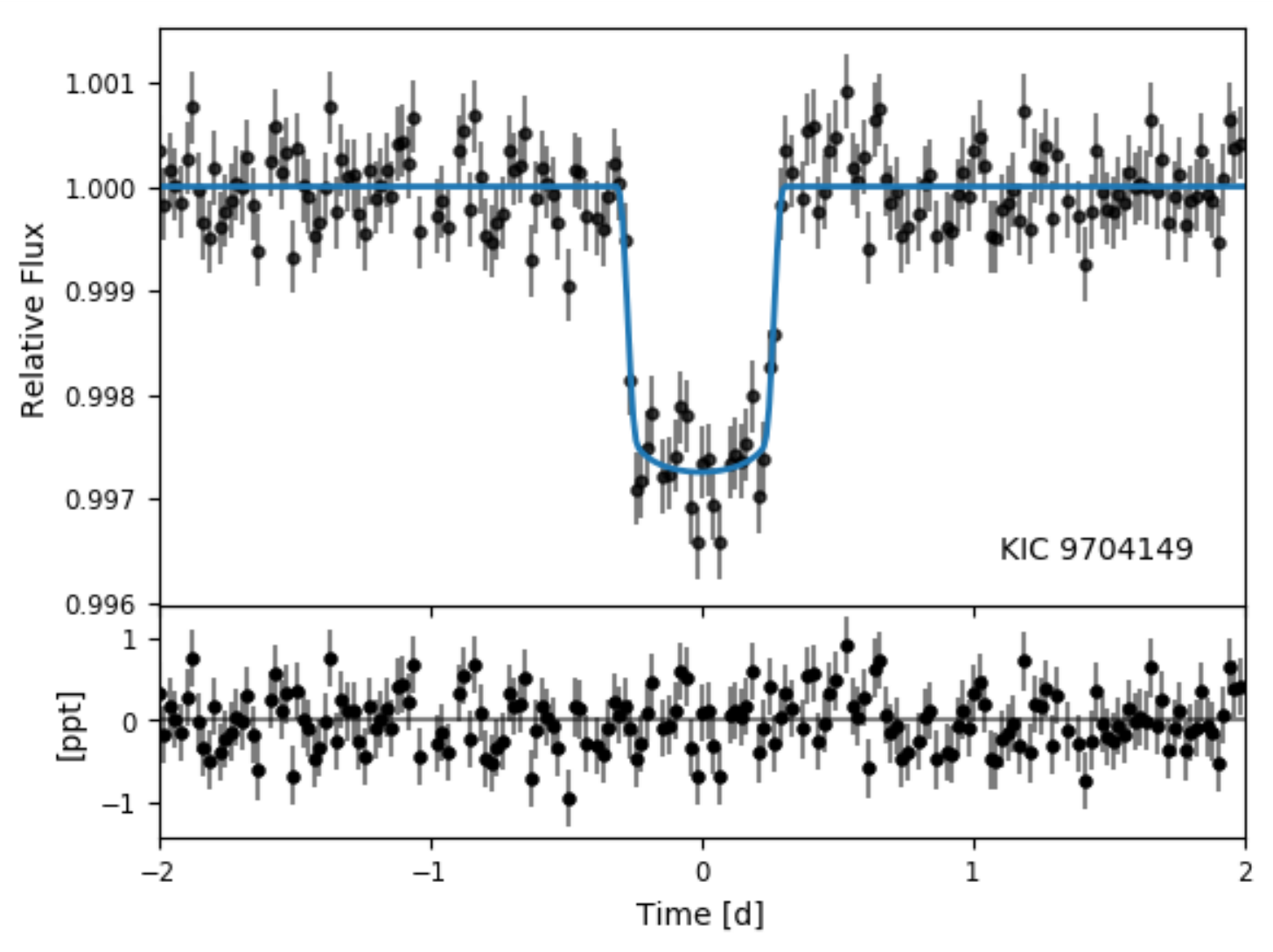}
\includegraphics[width=0.24\textwidth]{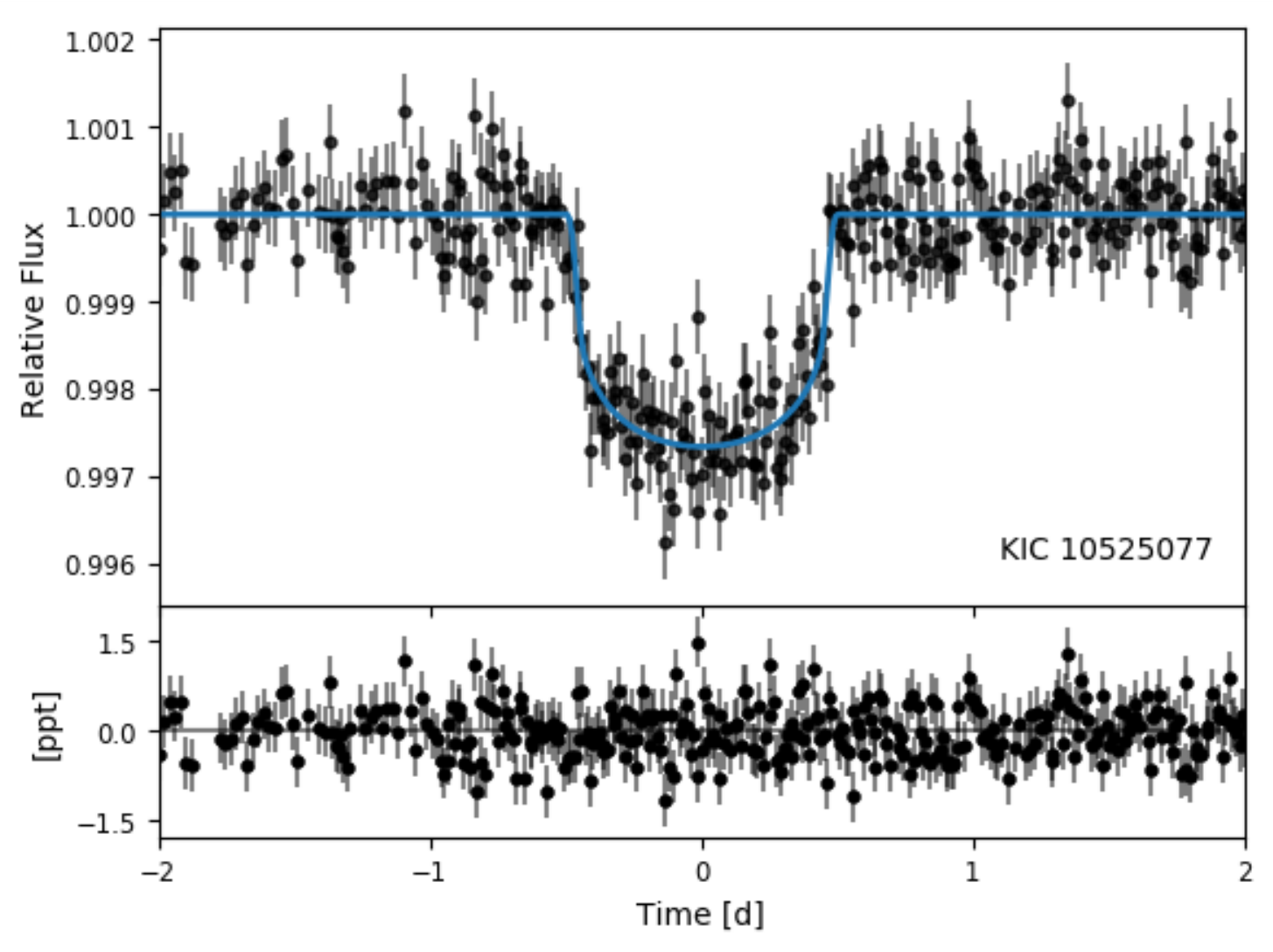}
\includegraphics[width=0.24\textwidth]{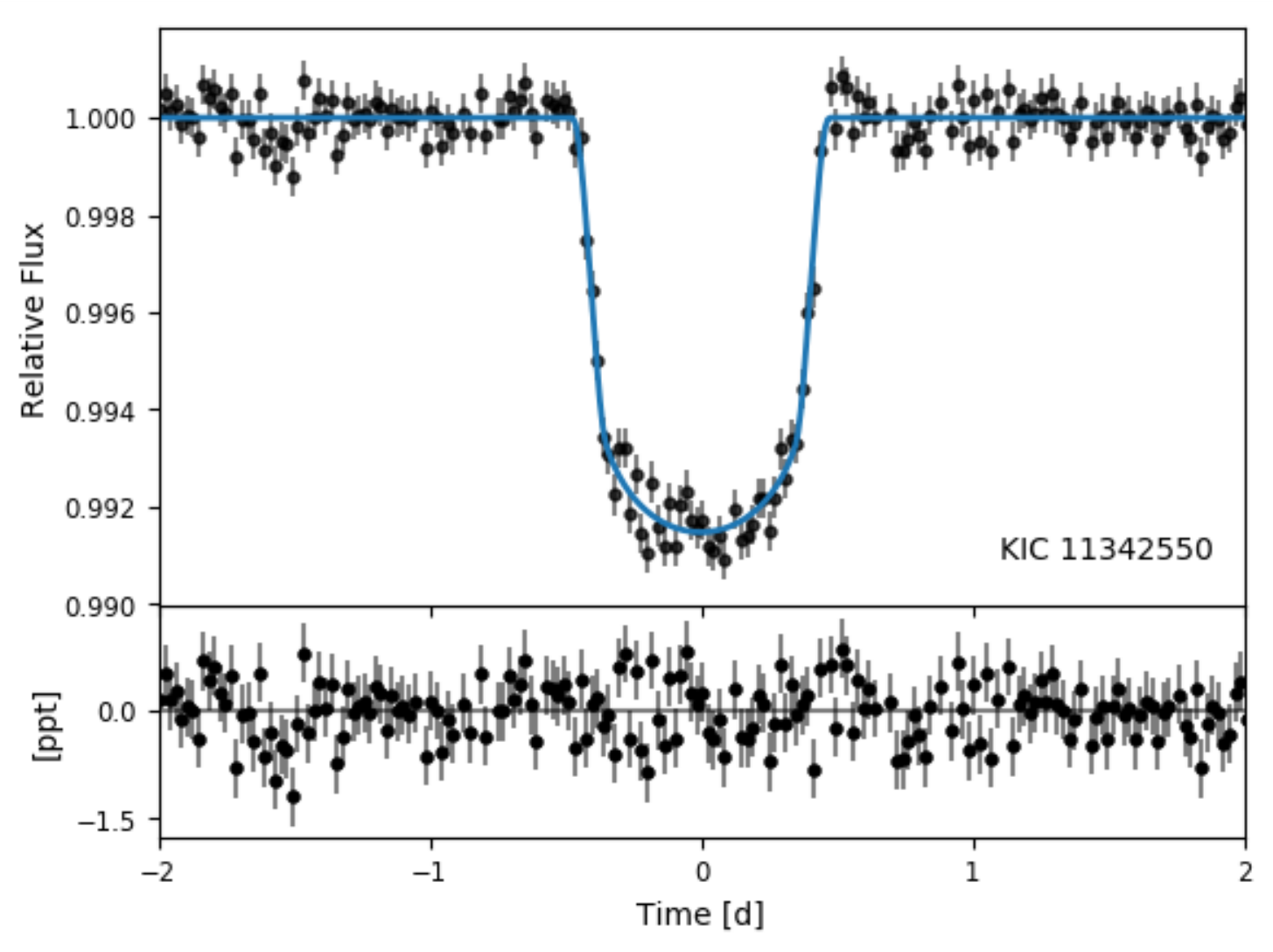}

%% file: param_tab.tex
\tablehead{
    \colhead{KIC ID} &
    \colhead{$R_*$} &
    \colhead{$K_p$} &
    \colhead{$t_0$} & 
    \colhead{Period} &
    \colhead{Radius} &
    \colhead{$b$} &
    \colhead{S/N} &
    \colhead{FAP} &
    \colhead{KOI/Kepler\tablenotemark{a}} &
    \colhead{$n_{\rm{inner}}$}
    \\
    & \colhead{($R_\odot$)} && \colhead{(KBJD)} & \colhead{(days)} & \colhead{($R_{\rm J}$)} &&&&&
}

\startdata

3218908$^{\mathrm{c}}$ & $1.22_{-0.06}^{+0.05}$ & $14.6$ &  $766.6769_{-0.0094}^{+0.0078}$ & $1246.35_{-83.09}^{+150.22}$ & $0.783_{-0.044}^{+0.039}$ & $0.00_{-0.39}^{+0.39}$ & $90.6$ & $0.04$ &  1108/770 & 3\\

3239945$^{\mathrm{d}}$ & $0.74_{-0.03}^{+0.03}$ & $14.0$ & $420.2867_{-0.0004}^{+0.0004}$ & $1071.23_{-0.01}^{+0.01}$ &  $0.904_{-0.038}^{+0.038}$ & $0.14_{-0.17}^{+0.10}$ & $358.6$ & $0.27$ & 490/167 & 3\\

4754460 & $1.17_{-0.09}^{+0.09}$ & $14.9$ & $826.8364_{-0.0044}^{+0.0045}$ & $1772.14_{-326.60}^{+174.75}$ & $0.662_{-0.060}^{+0.058}$ & $0.88_{-0.10}^{+0.03}$ & $30.6$ & $0.05$ & & \\

6551440 & $1.14_{-0.08}^{+0.08}$ & $13.6$ & $1039.0583_{-0.0054}^{+0.0051}$ & $1018.14_{-97.08}^{+194.53}$ & $0.374_{-0.031}^{+0.032}$ & $0.76_{-0.07}^{+0.08}$ & $25.0$ & $0.03$ & & \\

8410697$^{\mathrm{b,d}}$ & $1.05_{-0.07}^{+0.08}$ & $13.4$ & $542.1321_{-0.0018}^{+0.0017}$ & $1047.83_{-0.01}^{+0.01}$ & $0.727_{-0.050}^{+0.057}$ & $0.09_{-0.29}^{+0.28}$ & $73.2$ & $0.05$ & & \\

8505215$^{\mathrm{c}}$ & $0.8_{-0.03}^{+0.04}$ & $13.0$ & $140.0488_{-0.0019}^{+0.0018}$ & $2608.45_{-114.47}^{+195.89}$ & $0.310_{-0.013}^{+0.017}$ & $0.04_{-0.35}^{+0.32}$ & $104.7$ & $0.04$ & 99/none & \\

8800954$^{\mathrm{d}}$ &  $0.83_{-0.03}^{+0.04}$ & $13.4$ & $492.7665_{-0.0010}^{+0.0009}$ & $704.20_{-0.01}^{+0.01}$ & $0.411_{-0.016}^{+0.021}$ & $0.18_{-0.26}^{+0.22}$ & $121.2$ & $0.05$ & 1274 /421 & \\

9306307 & $1.38_{-0.11}^{+0.11}$ & $14.0$ & $1191.3562_{-0.0001}^{+0.0001}$ & $730.83_{-0.24}^{+0.52}$ & $3.296_{-0.266}^{+0.264}$ & $0.64_{-0.02}^{+0.01}$ & $31.8$ & $1.00$ & & \\

10187159$^{\mathrm{c}}$ & $0.78_{-0.05}^{+0.06}$ & $14.4$ & $604.1067_{-0.0025}^{+0.0024}$ & $1006.63_{-47.67}^{+95.27}$ & $0.577_{-0.039}^{+0.048}$ & $0.02_{-0.38}^{+0.35}$ & $29.2$ & $0.09$ & 1870/989 & 1\\

10602068 & $0.84_{-0.06}^{+0.07}$ & $14.9$ & $830.8089_{-0.0002}^{+0.0002}$ & $769.19_{-15.98}^{+15.45}$ & $2.368_{-0.171}^{+0.199}$ & $0.60_{-0.01}^{+0.01}$ & $693.6$ & $1.00$ & & \\

10842718$^{\mathrm{b}}$ & $0.86_{-0.06}^{+0.07}$ & $14.6$ & $226.2347_{-0.0041}^{+0.0040}$ & $8375.64_{-267.18}^{+473.31}$ & $0.580_{-0.041}^{+0.048}$ & $0.01_{-0.29}^{+0.28}$ & $38.2$ & $0.10$ & & \\

11709124$^{\mathrm{c}}$ & $1.07_{-0.05}^{+0.05}$ & $14.5$ & $657.2675_{-0.0018}^{+0.0020}$ & $1183.93_{-88.71}^{+148.27}$ & $0.926_{-0.048}^{+0.050}$ & $0.21_{-0.63}^{+0.32}$ & $155.7$ & $0.06$  & 435/154 & 5\\

& & & & & & & & & & \\
\hline
& & & & & & & & & & \\

6186417 & $1.34_{-0.13}^{+0.15}$ & $15.4$ & $958.7534_{-0.0059}^{+0.0062}$ & $936.06_{-107.32}^{+203.22}$ & $0.733_{-0.081}^{+0.093}$ & $0.05_{-0.57}^{+0.57}$ & $32.6$ & $0.17$ & & \\

6234593 & $0.98_{-0.08}^{+0.09}$ & $15.7$ & $1147.9908_{-0.0003}^{+0.0003}$ & $732.02_{-1.13}^{+2.50}$ & $2.298_{-0.188}^{+0.212}$ & $0.42_{-0.01}^{+0.01}$ & $293.1$ & $1.00$ & & \\

7906827$^{\mathrm{d}}$  & $1.17_{-0.11}^{+0.11}$ & $15.7$ & $772.1934_{-0.0019}^{+0.0020}$ & $737.11_{-0.01}^{+0.01}$ & $0.955_{-0.099}^{+0.093}$ & $0.03_{-0.29}^{+0.27}$ & $106.8$ & $0.02$ & & \\

7947784 & $1.35_{-0.12}^{+0.12}$ & $15.5$ & $905.2546_{-0.0006}^{+0.0006}$ & $741.58_{-8.02}^{+15.24}$ & $2.713_{-0.245}^{+0.243}$ & $0.75_{-0.02}^{+0.02}$ & $266.1$ & $0.98$ & & \\

9704149$^{\mathrm{b}}$ & $0.91_{-0.06}^{+0.07}$ & $15.1$ & $419.7203_{-0.0036}^{+0.0035}$ & $1245.41_{-119.54}^{+312.09}$ & $0.464_{-0.033}^{+0.039}$ & $0.03_{-0.56}^{+0.60}$ & $35.8$ & $0.32$ & & \\

10525077$^{\mathrm{b,d}}$  & $1.15_{-0.09}^{+0.1}$ & $15.4$ & $335.2493_{-0.0065}^{+0.0080}$ & $854.09_{-0.01}^{+0.01}$ & $0.541_{-0.047}^{+0.053}$ & $0.47_{-0.98}^{+0.31}$ & $40.0$ & $0.54$ & 5800/459 & 1 \\ 

11342550$^{\mathrm{c}}$ & $1.04_{-0.06}^{+0.06}$ & $15.3$ & $524.2814_{-0.0019}^{+0.0020}$ & $1632.13_{-111.35}^{+151.10}$ & $0.889_{-0.057}^{+0.058}$ & $0.08_{-0.52}^{+0.42}$ & $169.5$ & $0.03$ & 1421/none & \\

\enddata

%% file: occur_tab.tex
\tablehead{
    \colhead{$R_{\rm p} ~(R_{\rm J})$} & 
    \colhead{$K_p$ Range} & 
    \colhead{Occurrence Rate\tablenotemark{a}} & 
    \colhead{$n_{\rm pl}$}
    }
    
\startdata
$0.30 - 0.45$ &  bright & $0.52_{-0.29}^{+0.46}$ & 2 \\
& mid &  &  \\
& dim & &  \\
& {\bf all} & \bm{$0.50_{-0.28}^{+0.44}$} & \bm{$2$}\\

$0.45 - 0.67$ &  bright & $0.07_{-0.05}^{+0.10}$ & 1 \\
& mid & $0.31_{-0.17}^{+0.28}$ & 2 \\
& dim & $0.33_{-0.23}^{+0.45}$ & 1 \\
& {\bf all} & \bm{$0.17_{-0.07}^{+0.10}$} & \bm{$4$}\\

$0.67 - 1.00$ &  bright & $0.20_{-0.08}^{+0.12}$ & 4 \\
& mid & $0.06_{-0.04}^{+0.09}$ & 1 \\
& dim & $0.13_{-0.07}^{+0.12}$ & 2 \\
& {\bf all} & \bm{$0.14_{-0.04}^{+0.06}$} & \bm{$7$}\\

& & & \\
\hline
& & & \\

$0.30 - 1.00$ & {\bf total\tablenotemark{b}} & 
\bm{$0.70_{-0.20}^{+0.40}$} & \bm{$13$}\\

\enddata

%% file: article.bbl
\begin{thebibliography}{}
\expandafter\ifx\csname natexlab\endcsname\relax\def\natexlab#1{#1}\fi

\bibitem[{{Benz} {et~al.}(2014){Benz}, {Ida}, {Alibert}, {Lin}, \&
  {Mordasini}}]{Benz14}
{Benz}, W., {Ida}, S., {Alibert}, Y., {Lin}, D., \& {Mordasini}, C. 2014, in
  Protostars and Planets VI, ed. H.~{Beuther}, R.~S. {Klessen}, C.~P.
  {Dullemond}, \& T.~{Henning}, 691

\bibitem[{{Berger} {et~al.}(2018){Berger}, {Huber}, {Gaidos}, \& {van
  Saders}}]{Berger18}
{Berger}, T.~A., {Huber}, D., {Gaidos}, E., \& {van Saders}, J.~L. 2018, \apj,
  866, 99

\bibitem[{{Borucki} {et~al.}(2010){Borucki}, {Koch}, {Basri}, {Batalha},
  {Brown}, {Caldwell}, {Caldwell}, {Christensen-Dalsgaard}, {Cochran},
  {DeVore}, {Dunham}, {Dupree}, {Gautier}, {Geary}, {Gilliland}, {Gould},
  {Howell}, {Jenkins}, {Kondo}, {Latham}, {Marcy}, {Meibom}, {Kjeldsen},
  {Lissauer}, {Monet}, {Morrison}, {Sasselov}, {Tarter}, {Boss}, {Brownlee},
  {Owen}, {Buzasi}, {Charbonneau}, {Doyle}, {Fortney}, {Ford}, {Holman},
  {Seager}, {Steffen}, {Welsh}, {Rowe}, {Anderson}, {Buchhave}, {Ciardi},
  {Walkowicz}, {Sherry}, {Horch}, {Isaacson}, {Everett}, {Fischer}, {Torres},
  {Johnson}, {Endl}, {MacQueen}, {Bryson}, {Dotson}, {Haas}, {Kolodziejczak},
  {Van Cleve}, {Chandrasekaran}, {Twicken}, {Quintana}, {Clarke}, {Allen},
  {Li}, {Wu}, {Tenenbaum}, {Verner}, {Bruhweiler}, {Barnes}, \&
  {Prsa}}]{Borucki10}
{Borucki}, W.~J., {Koch}, D., {Basri}, G., {et~al.} 2010, Science, 327, 977

\bibitem[{{Brown} {et~al.}(2011){Brown}, {Latham}, {Everett}, \&
  {Esquerdo}}]{Brown11}
{Brown}, T.~M., {Latham}, D.~W., {Everett}, M.~E., \& {Esquerdo}, G.~A. 2011,
  \aj, 142, 112

\bibitem[{{Bryan} {et~al.}(2019){Bryan}, {Knutson}, {Lee}, {Fulton}, {Batygin},
  {Ngo}, \& {Meshkat}}]{Bryan19}
{Bryan}, M.~L., {Knutson}, H.~A., {Lee}, E.~J., {et~al.} 2019, \aj, 157, 52

\bibitem[{{Bryan} {et~al.}(2016){Bryan}, {Knutson}, {Howard}, {Ngo}, {Batygin},
  {Crepp}, {Fulton}, {Hinkley}, {Isaacson}, {Johnson}, {Marcy}, \&
  {Wright}}]{Bryan16}
{Bryan}, M.~L., {Knutson}, H.~A., {Howard}, A.~W., {et~al.} 2016, \apj, 821, 89

\bibitem[{{Bryson} {et~al.}(2013){Bryson}, {Jenkins}, {Gilliland}, {Twicken},
  {Clarke}, {Rowe}, {Caldwell}, {Batalha}, {Mullally}, {Haas}, \&
  {Tenenbaum}}]{Bryson13}
{Bryson}, S.~T., {Jenkins}, J.~M., {Gilliland}, R.~L., {et~al.} 2013,
  Publications of the Astronomical Society of the Pacific, 125, 889

\bibitem[{{Burke} \& {McCullough}(2014)}]{Burke14}
{Burke}, C.~J., \& {McCullough}, P.~R. 2014, \apj, 792, 79

\bibitem[{{Cassan} {et~al.}(2012){Cassan}, {Kubas}, {Beaulieu}, {Dominik},
  {Horne}, {Greenhill}, {Wambsganss}, {Menzies}, {Williams}, {J{\o}rgensen},
  {Udalski}, {Bennett}, {Albrow}, {Batista}, {Brillant}, {Caldwell}, {Cole},
  {Coutures}, {Cook}, {Dieters}, {Dominis Prester}, {Donatowicz}, {Fouqu{\'e}},
  {Hill}, {Kains}, {Kane}, {Marquette}, {Martin}, {Pollard}, {Sahu}, {Vinter},
  {Warren}, {Watson}, {Zub}, {Sumi}, {Szyma{\'n}ski}, {Kubiak}, {Poleski},
  {Soszynski}, {Ulaczyk}, {Pietrzy{\'n}ski}, \& {Wyrzykowski}}]{Cassan12}
{Cassan}, A., {Kubas}, D., {Beaulieu}, J.-P., {et~al.} 2012, \nat, 481, 167

\bibitem[{{Christiansen} {et~al.}(2012){Christiansen}, {Jenkins}, {Caldwell},
  {Burke}, {Tenenbaum}, {Seader}, {Thompson}, {Barclay}, {Clarke}, {Li},
  {Smith}, {Stumpe}, {Twicken}, \& {Van Cleve}}]{Christiansen12}
{Christiansen}, J.~L., {Jenkins}, J.~M., {Caldwell}, D.~A., {et~al.} 2012,
  Publications of the Astronomical Society of the Pacific, 124, 1279

\bibitem[{{Christiansen} {et~al.}(2015){Christiansen}, {Clarke}, {Burke},
  {Seader}, {Jenkins}, {Twicken}, {Catanzarite}, {Smith}, {Batalha}, {Haas},
  {Thompson}, {Campbell}, {Sabale}, \& {Kamal Uddin}}]{Christiansen15}
{Christiansen}, J.~L., {Clarke}, B.~D., {Burke}, C.~J., {et~al.} 2015, \apj,
  810, 95

\bibitem[{{Cumming} {et~al.}(2008){Cumming}, {Butler}, {Marcy}, {Vogt},
  {Wright}, \& {Fischer}}]{Cumming08}
{Cumming}, A., {Butler}, R.~P., {Marcy}, G.~W., {et~al.} 2008, \pasp, 120, 531

\bibitem[{{Dressing} \& {Charbonneau}(2015)}]{Dressing15}
{Dressing}, C.~D., \& {Charbonneau}, D. 2015, \apj, 807, 45

\bibitem[{{Fabrycky} {et~al.}(2014){Fabrycky}, {Lissauer}, {Ragozzine}, {Rowe},
  {Steffen}, {Agol}, {Barclay}, {Batalha}, {Borucki}, {Ciardi}, {Ford},
  {Gautier}, {Geary}, {Holman}, {Jenkins}, {Li}, {Morehead}, {Morris},
  {Shporer}, {Smith}, {Still}, \& {Van Cleve}}]{Fabrycky14}
{Fabrycky}, D.~C., {Lissauer}, J.~J., {Ragozzine}, D., {et~al.} 2014, \apj,
  790, 146

\bibitem[{{Foreman-Mackey} {et~al.}(2013){Foreman-Mackey}, {Hogg}, {Lang}, \&
  {Goodman}}]{Foreman2013}
{Foreman-Mackey}, D., {Hogg}, D.~W., {Lang}, D., \& {Goodman}, J. 2013, \pasp,
  125, 306

\bibitem[{{Foreman-Mackey} {et~al.}(2016){Foreman-Mackey}, {Morton}, {Hogg},
  {Agol}, \& {Sch{\"o}lkopf}}]{FM16}
{Foreman-Mackey}, D., {Morton}, T.~D., {Hogg}, D.~W., {Agol}, E., \&
  {Sch{\"o}lkopf}, B. 2016, \aj, 152, 206

\bibitem[{{Fortney} {et~al.}(2007){Fortney}, {Marley}, \& {Barnes}}]{Fortney07}
{Fortney}, J.~J., {Marley}, M.~S., \& {Barnes}, J.~W. 2007, \apj, 659, 1661

\bibitem[{{Fressin} {et~al.}(2013){Fressin}, {Torres}, {Charbonneau}, {Bryson},
  {Christiansen}, {Dressing}, {Jenkins}, {Walkowicz}, \& {Batalha}}]{Fressin13}
{Fressin}, F., {Torres}, G., {Charbonneau}, D., {et~al.} 2013, \apj, 766, 81

\bibitem[{{Furlan} {et~al.}(2017){Furlan}, {Ciardi}, {Everett}, {Saylors},
  {Teske}, {Horch}, {Howell}, {van Belle}, {Hirsch}, {Gautier}, {Adams},
  {Barrado}, {Cartier}, {Dressing}, {Dupree}, {Gilliland}, {Lillo-Box},
  {Lucas}, \& {Wang}}]{Furlan17}
{Furlan}, E., {Ciardi}, D.~R., {Everett}, M.~E., {et~al.} 2017, \aj, 153, 71

\bibitem[{{Gaia Collaboration} {et~al.}(2016){Gaia Collaboration}, {Prusti},
  {de Bruijne}, {Brown}, {Vallenari}, {Babusiaux}, {Bailer-Jones}, {Bastian},
  {Biermann}, {Evans}, {Eyer}, {Jansen}, {Jordi}, {Klioner}, {Lammers},
  {Lindegren}, {Luri}, {Mignard}, {Milligan}, {Panem}, {Poinsignon},
  {Pourbaix}, {Randich}, {Sarri}, {Sartoretti}, {Siddiqui}, {Soubiran},
  {Valette}, {van Leeuwen}, {Walton}, {Aerts}, {Arenou}, {Cropper}, {Drimmel},
  {H{\o}g}, {Katz}, {Lattanzi}, {O'Mullane}, {Grebel}, {Holland}, {Huc},
  {Passot}, {Bramante}, {Cacciari}, {Casta{\~n}eda}, {Chaoul}, {Cheek}, {De
  Angeli}, {Fabricius}, {Guerra}, {Hern{\'a}ndez}, {Jean-Antoine-Piccolo},
  {Masana}, {Messineo}, {Mowlavi}, {Nienartowicz}, {Ord{\'o}{\~n}ez- Blanco},
  {Panuzzo}, {Portell}, {Richards}, {Riello}, {Seabroke}, {Tanga},
  {Th{\'e}venin}, {Torra}, {Els}, {Gracia- Abril}, {Comoretto},
  {Garcia-Reinaldos}, {Lock}, {Mercier}, {Altmann}, {Andrae}, {Astraatmadja},
  {Bellas-Velidis}, {Benson}, {Berthier}, {Blomme}, {Busso}, {Carry},
  {Cellino}, {Clementini}, {Cowell}, {Creevey}, {Cuypers}, {Davidson}, {De
  Ridder}, {de Torres}, {Delchambre}, {Dell'Oro}, {Ducourant}, {Fr{\'e}mat},
  {Garc{\'\i}a-Torres}, {Gosset}, {Halbwachs}, {Hambly}, {Harrison}, {Hauser},
  {Hestroffer}, {Hodgkin}, {Huckle}, {Hutton}, {Jasniewicz}, {Jordan},
  {Kontizas}, {Korn}, {Lanzafame}, {Manteiga}, {Moitinho}, {Muinonen},
  {Osinde}, {Pancino}, {Pauwels}, {Petit}, {Recio-Blanco}, {Robin}, {Sarro},
  {Siopis}, {Smith}, {Smith}, {Sozzetti}, {Thuillot}, {van Reeven}, {Viala},
  {Abbas}, {Abreu Aramburu}, {Accart}, {Aguado}, {Allan}, {Allasia},
  {Altavilla}, {{\'A}lvarez}, {Alves}, {Anderson}, {Andrei}, {Anglada Varela},
  {Antiche}, {Antoja}, {Ant{\'o}n}, {Arcay}, {Atzei}, {Ayache}, {Bach},
  {Baker}, {Balaguer-N{\'u}{\~n}ez}, {Barache}, {Barata}, {Barbier}, {Barblan},
  {Baroni}, {Barrado y Navascu{\'e}s}, {Barros}, {Barstow}, {Becciani},
  {Bellazzini}, {Bellei}, {Bello Garc{\'\i}a}, {Belokurov}, {Bendjoya},
  {Berihuete}, {Bianchi}, {Bienaym{\'e}}, {Billebaud}, {Blagorodnova},
  {Blanco-Cuaresma}, {Boch}, {Bombrun}, {Borrachero}, {Bouquillon}, {Bourda},
  {Bouy}, {Bragaglia}, {Breddels}, {Brouillet}, {Br{\"u}semeister},
  {Bucciarelli}, {Budnik}, {Burgess}, {Burgon}, {Burlacu}, {Busonero}, {Buzzi},
  {Caffau}, {Cambras}, {Campbell}, {Cancelliere}, {Cantat-Gaudin}, {Carlucci},
  {Carrasco}, {Castellani}, {Charlot}, {Charnas}, {Charvet}, {Chassat},
  {Chiavassa}, {Clotet}, {Cocozza}, {Collins}, {Collins}, {Costigan}, {Crifo},
  {Cross}, {Crosta}, {Crowley}, {Dafonte}, {Damerdji}, {Dapergolas}, {David},
  {David}, {De Cat}, {de Felice}, {de Laverny}, {De Luise}, {De March}, {de
  Martino}, {de Souza}, {Debosscher}, {del Pozo}, {Delbo}, {Delgado},
  {Delgado}, {di Marco}, {Di Matteo}, {Diakite}, {Distefano}, {Dolding}, {Dos
  Anjos}, {Drazinos}, {Dur{\'a}n}, {Dzigan}, {Ecale}, {Edvardsson}, {Enke},
  {Erdmann}, {Escolar}, {Espina}, {Evans}, {Eynard Bontemps}, {Fabre},
  {Fabrizio}, {Faigler}, {Falc{\~a}o}, {Farr{\`a}s Casas}, {Faye}, {Federici},
  {Fedorets}, {Fern{\'a}ndez-Hern{\'a}ndez}, {Fernique}, {Fienga}, {Figueras},
  {Filippi}, {Findeisen}, {Fonti}, {Fouesneau}, {Fraile}, {Fraser}, {Fuchs},
  {Furnell}, {Gai}, {Galleti}, {Galluccio}, {Garabato}, {Garc{\'\i}a-Sedano},
  {Gar{\'e}}, {Garofalo}, {Garralda}, {Gavras}, {Gerssen}, {Geyer}, {Gilmore},
  {Girona}, {Giuffrida}, {Gomes}, {Gonz{\'a}lez-Marcos},
  {Gonz{\'a}lez-N{\'u}{\~n}ez}, {Gonz{\'a}lez-Vidal}, {Granvik}, {Guerrier},
  {Guillout}, {Guiraud}, {G{\'u}rpide}, {Guti{\'e}rrez-S{\'a}nchez}, {Guy},
  {Haigron}, {Hatzidimitriou}, {Haywood}, {Heiter}, {Helmi}, {Hobbs},
  {Hofmann}, {Holl}, {Holland}, {Hunt}, {Hypki}, {Icardi}, {Irwin}, {Jevardat
  de Fombelle}, {Jofr{\'e}}, {Jonker}, {Jorissen}, {Julbe}, {Karampelas},
  {Kochoska}, {Kohley}, {Kolenberg}, {Kontizas}, {Koposov}, {Kordopatis},
  {Koubsky}, {Kowalczyk}, {Krone-Martins}, {Kudryashova}, {Kull}, {Bachchan},
  {Lacoste-Seris}, {Lanza}, {Lavigne}, {Le Poncin-Lafitte}, {Lebreton},
  {Lebzelter}, {Leccia}, {Leclerc}, {Lecoeur-Taibi}, {Lemaitre}, {Lenhardt},
  {Leroux}, {Liao}, {Licata}, {Lindstr{\o}m}, {Lister}, {Livanou}, {Lobel},
  {L{\"o}ffler}, {L{\'o}pez}, {Lopez-Lozano}, {Lorenz}, {Loureiro},
  {MacDonald}, {Magalh{\~a}es Fernandes}, {Managau}, {Mann}, {Mantelet},
  {Marchal}, {Marchant}, {Marconi}, {Marie}, {Marinoni}, {Marrese},
  {Marschalk{\'o}}, {Marshall}, {Mart{\'\i}n-Fleitas}, {Martino}, {Mary},
  {Matijevi{\v{c}}}, {Mazeh}, {McMillan}, {Messina}, {Mestre}, {Michalik},
  {Millar}, {Miranda}, {Molina}, {Molinaro}, {Molinaro}, {Moln{\'a}r},
  {Moniez}, {Montegriffo}, {Monteiro}, {Mor}, {Mora}, {Morbidelli}, {Morel},
  {Morgenthaler}, {Morley}, {Morris}, {Mulone}, {Muraveva}, {Musella},
  {Narbonne}, {Nelemans}, {Nicastro}, {Noval}, {Ord{\'e}novic},
  {Ordieres-Mer{\'e}}, {Osborne}, {Pagani}, {Pagano}, {Pailler}, {Palacin},
  {Palaversa}, {Parsons}, {Paulsen}, {Pecoraro}, {Pedrosa}, {Pentik{\"a}inen},
  {Pereira}, {Pichon}, {Piersimoni}, {Pineau}, {Plachy}, {Plum}, {Poujoulet},
  {Pr{\v{s}}a}, {Pulone}, {Ragaini}, {Rago}, {Rambaux}, {Ramos-Lerate},
  {Ranalli}, {Rauw}, {Read}, {Regibo}, {Renk}, {Reyl{\'e}}, {Ribeiro},
  {Rimoldini}, {Ripepi}, {Riva}, {Rixon}, {Roelens}, {Romero-G{\'o}mez},
  {Rowell}, {Royer}, {Rudolph}, {Ruiz-Dern}, {Sadowski}, {Sagrist{\`a}
  Sell{\'e}s}, {Sahlmann}, {Salgado}, {Salguero}, {Sarasso}, {Savietto},
  {Schnorhk}, {Schultheis}, {Sciacca}, {Segol}, {Segovia}, {Segransan},
  {Serpell}, {Shih}, {Smareglia}, {Smart}, {Smith}, {Solano}, {Solitro},
  {Sordo}, {Soria Nieto}, {Souchay}, {Spagna}, {Spoto}, {Stampa}, {Steele},
  {Steidelm{\"u}ller}, {Stephenson}, {Stoev}, {Suess}, {S{\"u}veges}, {Surdej},
  {Szabados}, {Szegedi-Elek}, {Tapiador}, {Taris}, {Tauran}, {Taylor},
  {Teixeira}, {Terrett}, {Tingley}, {Trager}, {Turon}, {Ulla}, {Utrilla},
  {Valentini}, {van Elteren}, {Van Hemelryck}, {van Leeuwen}, {Varadi},
  {Vecchiato}, {Veljanoski}, {Via}, {Vicente}, {Vogt}, {Voss}, {Votruba},
  {Voutsinas}, {Walmsley}, {Weiler}, {Weingrill}, {Werner}, {Wevers},
  {Whitehead}, {Wyrzykowski}, {Yoldas}, {{\v{Z}}erjal}, {Zucker}, {Zurbach},
  {Zwitter}, {Alecu}, {Allen}, {Allende Prieto}, {Amorim},
  {Anglada-Escud{\'e}}, {Arsenijevic}, {Azaz}, {Balm}, {Beck}, {Bernstein},
  {Bigot}, {Bijaoui}, {Blasco}, {Bonfigli}, {Bono}, {Boudreault}, {Bressan},
  {Brown}, {Brunet}, {Bunclark}, {Buonanno}, {Butkevich}, {Carret}, {Carrion},
  {Chemin}, {Ch{\'e}reau}, {Corcione}, {Darmigny}, {de Boer}, {de Teodoro}, {de
  Zeeuw}, {Delle Luche}, {Domingues}, {Dubath}, {Fodor}, {Fr{\'e}zouls},
  {Fries}, {Fustes}, {Fyfe}, {Gallardo}, {Gallegos}, {Gardiol}, {Gebran},
  {Gomboc}, {G{\'o}mez}, {Grux}, {Gueguen}, {Heyrovsky}, {Hoar}, {Iannicola},
  {Isasi Parache}, {Janotto}, {Joliet}, {Jonckheere}, {Keil}, {Kim},
  {Klagyivik}, {Klar}, {Knude}, {Kochukhov}, {Kolka}, {Kos}, {Kutka}, {Lainey},
  {LeBouquin}, {Liu}, {Loreggia}, {Makarov}, {Marseille}, {Martayan},
  {Martinez-Rubi}, {Massart}, {Meynadier}, {Mignot}, {Munari}, {Nguyen},
  {Nordlander}, {Ocvirk}, {O'Flaherty}, {Olias Sanz}, {Ortiz}, {Osorio},
  {Oszkiewicz}, {Ouzounis}, {Palmer}, {Park}, {Pasquato}, {Peltzer}, {Peralta},
  {P{\'e}turaud}, {Pieniluoma}, {Pigozzi}, {Poels}, {Prat}, {Prod'homme},
  {Raison}, {Rebordao}, {Risquez}, {Rocca-Volmerange}, {Rosen}, {Ruiz-Fuertes},
  {Russo}, {Sembay}, {Serraller Vizcaino}, {Short}, {Siebert}, {Silva},
  {Sinachopoulos}, {Slezak}, {Soffel}, {Sosnowska}, {Strai{\v{z}}ys}, {ter
  Linden}, {Terrell}, {Theil}, {Tiede}, {Troisi}, {Tsalmantza}, {Tur},
  {Vaccari}, {Vachier}, {Valles}, {Van Hamme}, {Veltz}, {Virtanen}, {Wallut},
  {Wichmann}, {Wilkinson}, {Ziaeepour}, \& {Zschocke}}]{Gaia16}
{Gaia Collaboration}, {Prusti}, T., {de Bruijne}, J.~H.~J., {et~al.} 2016,
  \aap, 595, A1

\bibitem[{{Gaia Collaboration} {et~al.}(2018){Gaia Collaboration}, {Brown},
  {Vallenari}, {Prusti}, {de Bruijne}, {Babusiaux}, {Bailer-Jones}, {Biermann},
  {Evans}, {Eyer}, {Jansen}, {Jordi}, {Klioner}, {Lammers}, {Lindegren},
  {Luri}, {Mignard}, {Panem}, {Pourbaix}, {Randich}, {Sartoretti}, {Siddiqui},
  {Soubiran}, {van Leeuwen}, {Walton}, {Arenou}, {Bastian}, {Cropper},
  {Drimmel}, {Katz}, {Lattanzi}, {Bakker}, {Cacciari}, {Casta{\~n}eda},
  {Chaoul}, {Cheek}, {De Angeli}, {Fabricius}, {Guerra}, {Holl}, {Masana},
  {Messineo}, {Mowlavi}, {Nienartowicz}, {Panuzzo}, {Portell}, {Riello},
  {Seabroke}, {Tanga}, {Th{\'e}venin}, {Gracia-Abril}, {Comoretto},
  {Garcia-Reinaldos}, {Teyssier}, {Altmann}, {Andrae}, {Audard},
  {Bellas-Velidis}, {Benson}, {Berthier}, {Blomme}, {Burgess}, {Busso},
  {Carry}, {Cellino}, {Clementini}, {Clotet}, {Creevey}, {Davidson}, {De
  Ridder}, {Delchambre}, {Dell'Oro}, {Ducourant},
  {Fern{\'a}ndez-Hern{\'a}ndez}, {Fouesneau}, {Fr{\'e}mat}, {Galluccio},
  {Garc{\'\i}a-Torres}, {Gonz{\'a}lez-N{\'u}{\~n}ez}, {Gonz{\'a}lez-Vidal},
  {Gosset}, {Guy}, {Halbwachs}, {Hambly}, {Harrison}, {Hern{\'a}ndez},
  {Hestroffer}, {Hodgkin}, {Hutton}, {Jasniewicz}, {Jean-Antoine-Piccolo},
  {Jordan}, {Korn}, {Krone-Martins}, {Lanzafame}, {Lebzelter}, {L{\"o}ffler},
  {Manteiga}, {Marrese}, {Mart{\'\i}n-Fleitas}, {Moitinho}, {Mora}, {Muinonen},
  {Osinde}, {Pancino}, {Pauwels}, {Petit}, {Recio-Blanco}, {Richards},
  {Rimoldini}, {Robin}, {Sarro}, {Siopis}, {Smith}, {Sozzetti}, {S{\"u}veges},
  {Torra}, {van Reeven}, {Abbas}, {Abreu Aramburu}, {Accart}, {Aerts},
  {Altavilla}, {{\'A}lvarez}, {Alvarez}, {Alves}, {Anderson}, {Andrei},
  {Anglada Varela}, {Antiche}, {Antoja}, {Arcay}, {Astraatmadja}, {Bach},
  {Baker}, {Balaguer-N{\'u}{\~n}ez}, {Balm}, {Barache}, {Barata}, {Barbato},
  {Barblan}, {Barklem}, {Barrado}, {Barros}, {Barstow}, {Bartholom{\'e}
  Mu{\~n}oz}, {Bassilana}, {Becciani}, {Bellazzini}, {Berihuete}, {Bertone},
  {Bianchi}, {Bienaym{\'e}}, {Blanco-Cuaresma}, {Boch}, {Boeche}, {Bombrun},
  {Borrachero}, {Bossini}, {Bouquillon}, {Bourda}, {Bragaglia}, {Bramante},
  {Breddels}, {Bressan}, {Brouillet}, {Br{\"u}semeister}, {Brugaletta},
  {Bucciarelli}, {Burlacu}, {Busonero}, {Butkevich}, {Buzzi}, {Caffau},
  {Cancelliere}, {Cannizzaro}, {Cantat-Gaudin}, {Carballo}, {Carlucci},
  {Carrasco}, {Casamiquela}, {Castellani}, {Castro-Ginard}, {Charlot},
  {Chemin}, {Chiavassa}, {Cocozza}, {Costigan}, {Cowell}, {Crifo}, {Crosta},
  {Crowley}, {Cuypers}, {Dafonte}, {Damerdji}, {Dapergolas}, {David}, {David},
  {de Laverny}, {De Luise}, {De March}, {de Martino}, {de Souza}, {de Torres},
  {Debosscher}, {del Pozo}, {Delbo}, {Delgado}, {Delgado}, {Di Matteo},
  {Diakite}, {Diener}, {Distefano}, {Dolding}, {Drazinos}, {Dur{\'a}n},
  {Edvardsson}, {Enke}, {Eriksson}, {Esquej}, {Eynard Bontemps}, {Fabre},
  {Fabrizio}, {Faigler}, {Falc{\~a}o}, {Farr{\`a}s Casas}, {Federici},
  {Fedorets}, {Fernique}, {Figueras}, {Filippi}, {Findeisen}, {Fonti},
  {Fraile}, {Fraser}, {Fr{\'e}zouls}, {Gai}, {Galleti}, {Garabato},
  {Garc{\'\i}a-Sedano}, {Garofalo}, {Garralda}, {Gavel}, {Gavras}, {Gerssen},
  {Geyer}, {Giacobbe}, {Gilmore}, {Girona}, {Giuffrida}, {Glass}, {Gomes},
  {Granvik}, {Gueguen}, {Guerrier}, {Guiraud}, {Guti{\'e}rrez-S{\'a}nchez},
  {Haigron}, {Hatzidimitriou}, {Hauser}, {Haywood}, {Heiter}, {Helmi}, {Heu},
  {Hilger}, {Hobbs}, {Hofmann}, {Holland}, {Huckle}, {Hypki}, {Icardi},
  {Jan{\ss}en}, {Jevardat de Fombelle}, {Jonker}, {Juh{\'a}sz}, {Julbe},
  {Karampelas}, {Kewley}, {Klar}, {Kochoska}, {Kohley}, {Kolenberg},
  {Kontizas}, {Kontizas}, {Koposov}, {Kordopatis}, {Kostrzewa-Rutkowska},
  {Koubsky}, {Lambert}, {Lanza}, {Lasne}, {Lavigne}, {Le Fustec}, {Le
  Poncin-Lafitte}, {Lebreton}, {Leccia}, {Leclerc}, {Lecoeur-Taibi},
  {Lenhardt}, {Leroux}, {Liao}, {Licata}, {Lindstr{\o}m}, {Lister}, {Livanou},
  {Lobel}, {L{\'o}pez}, {Managau}, {Mann}, {Mantelet}, {Marchal}, {Marchant},
  {Marconi}, {Marinoni}, {Marschalk{\'o}}, {Marshall}, {Martino}, {Marton},
  {Mary}, {Massari}, {Matijevi{\v{c}}}, {Mazeh}, {McMillan}, {Messina},
  {Michalik}, {Millar}, {Molina}, {Molinaro}, {Moln{\'a}r}, {Montegriffo},
  {Mor}, {Morbidelli}, {Morel}, {Morris}, {Mulone}, {Muraveva}, {Musella},
  {Nelemans}, {Nicastro}, {Noval}, {O'Mullane}, {Ord{\'e}novic},
  {Ord{\'o}{\~n}ez-Blanco}, {Osborne}, {Pagani}, {Pagano}, {Pailler},
  {Palacin}, {Palaversa}, {Panahi}, {Pawlak}, {Piersimoni}, {Pineau}, {Plachy},
  {Plum}, {Poggio}, {Poujoulet}, {Pr{\v{s}}a}, {Pulone}, {Racero}, {Ragaini},
  {Rambaux}, {Ramos-Lerate}, {Regibo}, {Reyl{\'e}}, {Riclet}, {Ripepi}, {Riva},
  {Rivard}, {Rixon}, {Roegiers}, {Roelens}, {Romero-G{\'o}mez}, {Rowell},
  {Royer}, {Ruiz-Dern}, {Sadowski}, {Sagrist{\`a} Sell{\'e}s}, {Sahlmann},
  {Salgado}, {Salguero}, {Sanna}, {Santana-Ros}, {Sarasso}, {Savietto},
  {Schultheis}, {Sciacca}, {Segol}, {Segovia}, {S{\'e}gransan}, {Shih},
  {Siltala}, {Silva}, {Smart}, {Smith}, {Solano}, {Solitro}, {Sordo}, {Soria
  Nieto}, {Souchay}, {Spagna}, {Spoto}, {Stampa}, {Steele},
  {Steidelm{\"u}ller}, {Stephenson}, {Stoev}, {Suess}, {Surdej}, {Szabados},
  {Szegedi-Elek}, {Tapiador}, {Taris}, {Tauran}, {Taylor}, {Teixeira},
  {Terrett}, {Teyssand ier}, {Thuillot}, {Titarenko}, {Torra Clotet}, {Turon},
  {Ulla}, {Utrilla}, {Uzzi}, {Vaillant}, {Valentini}, {Valette}, {van Elteren},
  {Van Hemelryck}, {van Leeuwen}, {Vaschetto}, {Vecchiato}, {Veljanoski},
  {Viala}, {Vicente}, {Vogt}, {von Essen}, {Voss}, {Votruba}, {Voutsinas},
  {Walmsley}, {Weiler}, {Wertz}, {Wevers}, {Wyrzykowski}, {Yoldas},
  {{\v{Z}}erjal}, {Ziaeepour}, {Zorec}, {Zschocke}, {Zucker}, {Zurbach}, \&
  {Zwitter}}]{Gaia18b}
{Gaia Collaboration}, {Brown}, A.~G.~A., {Vallenari}, A., {et~al.} 2018, \aap,
  616, A1

\bibitem[{{Gould} {et~al.}(2010){Gould}, {Dong}, {Gaudi}, {Udalski}, {Bond},
  {Greenhill}, {Street}, {Dominik}, {Sumi}, {Szyma{\'n}ski}, {Han}, {Allen},
  {Bolt}, {Bos}, {Christie}, {DePoy}, {Drummond}, {Eastman}, {Gal-Yam},
  {Higgins}, {Janczak}, {Kaspi}, {Koz{\l}owski}, {Lee}, {Mallia}, {Maury},
  {Maoz}, {McCormick}, {Monard}, {Moorhouse}, {Morgan}, {Natusch}, {Ofek},
  {Park}, {Pogge}, {Polishook}, {Santallo}, {Shporer}, {Spector}, {Thornley},
  {Yee}, {{\ensuremath{\mu}}FUN Collaboration}, {Kubiak}, {Pietrzy{\'n}ski},
  {Soszy{\'n}ski}, {Szewczyk}, {Wyrzykowski}, {Ulaczyk}, {Poleski}, {OGLE
  Collaboration}, {Abe}, {Bennett}, {Botzler}, {Douchin}, {Freeman}, {Fukui},
  {Furusawa}, {Hearnshaw}, {Hosaka}, {Itow}, {Kamiya}, {Kilmartin}, {Korpela},
  {Lin}, {Ling}, {Makita}, {Masuda}, {Matsubara}, {Miyake}, {Muraki}, {Nagaya},
  {Nishimoto}, {Ohnishi}, {Okumura}, {Perrott}, {Philpott}, {Rattenbury},
  {Saito}, {Sako}, {Sullivan}, {Sweatman}, {Tristram}, {von Seggern}, {Yock},
  {MOA Collaboration}, {Albrow}, {Batista}, {Beaulieu}, {Brillant}, {Caldwell},
  {Calitz}, {Cassan}, {Cole}, {Cook}, {Coutures}, {Dieters}, {Dominis Prester},
  {Donatowicz}, {Fouqu{\'e}}, {Hill}, {Hoffman}, {Jablonski}, {Kane}, {Kains},
  {Kubas}, {Marquette}, {Martin}, {Martioli}, {Meintjes}, {Menzies},
  {Pedretti}, {Pollard}, {Sahu}, {Vinter}, {Wambsganss}, {Watson}, {Williams},
  {Zub}, {PLANET Collaboration}, {Allan}, {Bode}, {Bramich}, {Burgdorf},
  {Clay}, {Fraser}, {Hawkins}, {Horne}, {Kerins}, {Lister}, {Mottram},
  {Saunders}, {Snodgrass}, {Steele}, {Tsapras}, {RoboNet Collaboration},
  {J{\o}rgensen}, {Anguita}, {Bozza}, {Calchi Novati}, {Harps{\o}e}, {Hinse},
  {Hundertmark}, {Kj{\ae}rgaard}, {Liebig}, {Mancini}, {Masi}, {Mathiasen},
  {Rahvar}, {Ricci}, {Scarpetta}, {Southworth}, {Surdej}, {Th{\"o}ne}, \&
  {MiNDSTEp Consortium}}]{Gould10}
{Gould}, A., {Dong}, S., {Gaudi}, B.~S., {et~al.} 2010, \apj, 720, 1073

\bibitem[{{Grether} \& {Lineweaver}(2006)}]{Grether06}
{Grether}, D., \& {Lineweaver}, C.~H. 2006, \apj, 640, 1051

\bibitem[{{Hirsch} {et~al.}(2017){Hirsch}, {Ciardi}, {Howard}, {Everett},
  {Furlan}, {Saylors}, {Horch}, {Howell}, {Teske}, \& {Marcy}}]{Hirsch17}
{Hirsch}, L.~A., {Ciardi}, D.~R., {Howard}, A.~W., {et~al.} 2017, \aj, 153, 117

\bibitem[{{Huber} {et~al.}(2014){Huber}, {Silva Aguirre}, {Matthews},
  {Pinsonneault}, {Gaidos}, {Garc{\'{\i}}a}, {Hekker}, {Mathur}, {Mosser},
  {Torres}, {Bastien}, {Basu}, {Bedding}, {Chaplin}, {Demory}, {Fleming},
  {Guo}, {Mann}, {Rowe}, {Serenelli}, {Smith}, \& {Stello}}]{Huber14}
{Huber}, D., {Silva Aguirre}, V., {Matthews}, J.~M., {et~al.} 2014, \apjs, 211,
  2

\bibitem[{{Kipping}(2013)}]{Kipping13b}
{Kipping}, D.~M. 2013, \mnras, 434, L51

\bibitem[{{Kipping} {et~al.}(2016){Kipping}, {Torres}, {Henze}, {Teachey},
  {Isaacson}, {Petigura}, {Marcy}, {Buchhave}, {Chen}, {Bryson}, \&
  {Sandford}}]{Kipping16}
{Kipping}, D.~M., {Torres}, G., {Henze}, C., {et~al.} 2016, \apj, 820, 112

\bibitem[{{Koch} {et~al.}(2010){Koch}, {Borucki}, {Basri}, {Batalha}, {Brown},
  {Caldwell}, {Christensen-Dalsgaard}, {Cochran}, {DeVore}, {Dunham},
  {Gautier}, {Geary}, {Gilliland}, {Gould}, {Jenkins}, {Kondo}, {Latham},
  {Lissauer}, {Marcy}, {Monet}, {Sasselov}, {Boss}, {Brownlee}, {Caldwell},
  {Dupree}, {Howell}, {Kjeldsen}, {Meibom}, {Morrison}, {Owen}, {Reitsema},
  {Tarter}, {Bryson}, {Dotson}, {Gazis}, {Haas}, {Kolodziejczak}, {Rowe}, {Van
  Cleve}, {Allen}, {Chandrasekaran}, {Clarke}, {Li}, {Quintana}, {Tenenbaum},
  {Twicken}, \& {Wu}}]{Koch10}
{Koch}, D.~G., {Borucki}, W.~J., {Basri}, G., {et~al.} 2010, \apj, 713, L79

\bibitem[{{Kreidberg}(2015)}]{batman}
{Kreidberg}, L. 2015, \pasp, 127, 1161

\bibitem[{{Law} {et~al.}(2014){Law}, {Morton}, {Baranec}, {Riddle},
  {Ravichandran}, {Ziegler}, {Johnson}, {Tendulkar}, {Bui}, {Burse}, {Das},
  {Dekany}, {Kulkarni}, {Punnadi}, \& {Ramaprakash}}]{Law14}
{Law}, N.~M., {Morton}, T., {Baranec}, C., {et~al.} 2014, \apj, 791, 35

\bibitem[{{Lissauer} {et~al.}(2012){Lissauer}, {Marcy}, {Rowe}, {Bryson},
  {Adams}, {Buchhave}, {Ciardi}, {Cochran}, {Fabrycky}, {Ford}, {Fressin},
  {Geary}, {Gilliland}, {Holman}, {Howell}, {Jenkins}, {Kinemuchi}, {Koch},
  {Morehead}, {Ragozzine}, {Seader}, {Tanenbaum}, {Torres}, \&
  {Twicken}}]{Lissauer12}
{Lissauer}, J.~J., {Marcy}, G.~W., {Rowe}, J.~F., {et~al.} 2012, \apj, 750, 112

\bibitem[{{Mayor} {et~al.}(2011){Mayor}, {Marmier}, {Lovis}, {Udry},
  {S{\'e}gransan}, {Pepe}, {Benz}, {Bertaux}, {Bouchy}, {Dumusque}, {Lo Curto},
  {Mordasini}, {Queloz}, \& {Santos}}]{Mayor11}
{Mayor}, M., {Marmier}, M., {Lovis}, C., {et~al.} 2011, ArXiv e-prints,
  arXiv:1109.2497

\bibitem[{{Morton}(2015)}]{isocrones}
{Morton}, T.~D. 2015, {isochrones: Stellar model grid package}, Astrophysics
  Source Code Library, ascl:1503.010

\bibitem[{{Morton} {et~al.}(2016){Morton}, {Bryson}, {Coughlin}, {Rowe},
  {Ravichandran}, {Petigura}, {Haas}, \& {Batalha}}]{Morton16}
{Morton}, T.~D., {Bryson}, S.~T., {Coughlin}, J.~L., {et~al.} 2016, \apj, 822,
  86

\bibitem[{{Murphy} {et~al.}(2018){Murphy}, {Moe}, {Kurtz}, {Bedding},
  {Shibahashi}, \& {Boffin}}]{Murphy18}
{Murphy}, S.~J., {Moe}, M., {Kurtz}, D.~W., {et~al.} 2018, \mnras, 474, 4322

\bibitem[{{Petigura} {et~al.}(2013){Petigura}, {Howard}, \&
  {Marcy}}]{Petigura13}
{Petigura}, E.~A., {Howard}, A.~W., \& {Marcy}, G.~W. 2013, Proceedings of the
  National Academy of Science, 110, 19273

\bibitem[{{Raghavan} {et~al.}(2010){Raghavan}, {McAlister}, {Henry}, {Latham},
  {Marcy}, {Mason}, {Gies}, {White}, \& {ten Brummelaar}}]{Raghavan10}
{Raghavan}, D., {McAlister}, H.~A., {Henry}, T.~J., {et~al.} 2010, \apjs, 190,
  1

\bibitem[{{Rauer} {et~al.}(2014){Rauer}, {Catala}, {Aerts}, {Appourchaux},
  {Benz}, {Brandeker}, {Christensen-Dalsgaard}, {Deleuil}, {Gizon}, {Goupil},
  {G{\"u}del}, {Janot-Pacheco}, {Mas-Hesse}, {Pagano}, {Piotto}, {Pollacco},
  {Santos}, {Smith}, {Su{\'a}rez}, {Szab{\'o}}, {Udry}, {Adibekyan}, {Alibert},
  {Almenara}, {Amaro-Seoane}, {Eiff}, {Asplund}, {Antonello}, {Barnes},
  {Baudin}, {Belkacem}, {Bergemann}, {Bihain}, {Birch}, {Bonfils}, {Boisse},
  {Bonomo}, {Borsa}, {Brand{\~a}o}, {Brocato}, {Brun}, {Burleigh}, {Burston},
  {Cabrera}, {Cassisi}, {Chaplin}, {Charpinet}, {Chiappini}, {Church},
  {Csizmadia}, {Cunha}, {Damasso}, {Davies}, {Deeg}, {D{\'{\i}}az}, {Dreizler},
  {Dreyer}, {Eggenberger}, {Ehrenreich}, {Eigm{\"u}ller}, {Erikson}, {Farmer},
  {Feltzing}, {de Oliveira Fialho}, {Figueira}, {Forveille}, {Fridlund},
  {Garc{\'{\i}}a}, {Giommi}, {Giuffrida}, {Godolt}, {Gomes da Silva},
  {Granzer}, {Grenfell}, {Grotsch-Noels}, {G{\"u}nther}, {Haswell}, {Hatzes},
  {H{\'e}brard}, {Hekker}, {Helled}, {Heng}, {Jenkins}, {Johansen},
  {Khodachenko}, {Kislyakova}, {Kley}, {Kolb}, {Krivova}, {Kupka}, {Lammer},
  {Lanza}, {Lebreton}, {Magrin}, {Marcos-Arenal}, {Marrese}, {Marques},
  {Martins}, {Mathis}, {Mathur}, {Messina}, {Miglio}, {Montalban}, {Montalto},
  {Monteiro}, {Moradi}, {Moravveji}, {Mordasini}, {Morel}, {Mortier},
  {Nascimbeni}, {Nelson}, {Nielsen}, {Noack}, {Norton}, {Ofir}, {Oshagh},
  {Ouazzani}, {P{\'a}pics}, {Parro}, {Petit}, {Plez}, {Poretti}, {Quirrenbach},
  {Ragazzoni}, {Raimondo}, {Rainer}, {Reese}, {Redmer}, {Reffert},
  {Rojas-Ayala}, {Roxburgh}, {Salmon}, {Santerne}, {Schneider}, {Schou},
  {Schuh}, {Schunker}, {Silva-Valio}, {Silvotti}, {Skillen}, {Snellen}, {Sohl},
  {Sousa}, {Sozzetti}, {Stello}, {Strassmeier}, {{\v S}vanda}, {Szab{\'o}},
  {Tkachenko}, {Valencia}, {Van Grootel}, {Vauclair}, {Ventura}, {Wagner},
  {Walton}, {Weingrill}, {Werner}, {Wheatley}, \& {Zwintz}}]{Rauer14}
{Rauer}, H., {Catala}, C., {Aerts}, C., {et~al.} 2014, Experimental Astronomy,
  38, 249

\bibitem[{{Ricker} {et~al.}(2014){Ricker}, {Winn}, {Vanderspek}, {Latham},
  {Bakos}, {Bean}, {Berta-Thompson}, {Brown}, {Buchhave}, {Butler}, {Butler},
  {Chaplin}, {Charbonneau}, {Christensen-Dalsgaard}, {Clampin}, {Deming},
  {Doty}, {De Lee}, {Dressing}, {Dunham}, {Endl}, {Fressin}, {Ge}, {Henning},
  {Holman}, {Howard}, {Ida}, {Jenkins}, {Jernigan}, {Johnson}, {Kaltenegger},
  {Kawai}, {Kjeldsen}, {Laughlin}, {Levine}, {Lin}, {Lissauer}, {MacQueen},
  {Marcy}, {McCullough}, {Morton}, {Narita}, {Paegert}, {Palle}, {Pepe},
  {Pepper}, {Quirrenbach}, {Rinehart}, {Sasselov}, {Sato}, {Seager},
  {Sozzetti}, {Stassun}, {Sullivan}, {Szentgyorgyi}, {Torres}, {Udry}, \&
  {Villasenor}}]{Ricker14}
{Ricker}, G.~R., {Winn}, J.~N., {Vanderspek}, R., {et~al.} 2014, in \procspie,
  Vol. 9143, Space Telescopes and Instrumentation, 914320

\bibitem[{{Santerne} {et~al.}(2012){Santerne}, {D{\'{\i}}az}, {Moutou},
  {Bouchy}, {H{\'e}brard}, {Almenara}, {Bonomo}, {Deleuil}, \&
  {Santos}}]{Santerne12}
{Santerne}, A., {D{\'{\i}}az}, R.~F., {Moutou}, C., {et~al.} 2012, \aap, 545,
  A76

\bibitem[{{Seager} \& {Mall{\'e}n-Ornelas}(2003)}]{Seager03}
{Seager}, S., \& {Mall{\'e}n-Ornelas}, G. 2003, \apj, 585, 1038

\bibitem[{{Smith} {et~al.}(2012){Smith}, {Stumpe}, {Van Cleve}, {Jenkins},
  {Barclay}, {Fanelli}, {Girouard}, {Kolodziejczak}, {McCauliff}, {Morris}, \&
  {Twicken}}]{Smith12}
{Smith}, J.~C., {Stumpe}, M.~C., {Van Cleve}, J.~E., {et~al.} 2012,
  Publications of the Astronomical Society of the Pacific, 124, 1000

\bibitem[{{Steffen} {et~al.}(2010){Steffen}, {Batalha}, {Borucki}, {Buchhave},
  {Caldwell}, {Cochran}, {Endl}, {Fabrycky}, {Fressin}, {Ford}, {Fortney},
  {Haas}, {Holman}, {Howell}, {Isaacson}, {Jenkins}, {Koch}, {Latham},
  {Lissauer}, {Moorhead}, {Morehead}, {Marcy}, {MacQueen}, {Quinn},
  {Ragozzine}, {Rowe}, {Sasselov}, {Seager}, {Torres}, \& {Welsh}}]{Steffen10}
{Steffen}, J.~H., {Batalha}, N.~M., {Borucki}, W.~J., {et~al.} 2010, \apj, 725,
  1226

\bibitem[{{Stumpe} {et~al.}(2012){Stumpe}, {Smith}, {Van Cleve}, {Twicken},
  {Barclay}, {Fanelli}, {Girouard}, {Jenkins}, {Kolodziejczak}, {McCauliff}, \&
  {Morris}}]{Stumpe12}
{Stumpe}, M.~C., {Smith}, J.~C., {Van Cleve}, J.~E., {et~al.} 2012,
  Publications of the Astronomical Society of the Pacific, 124, 985

\bibitem[{{Suzuki} {et~al.}(2016){Suzuki}, {Bennett}, {Sumi}, {Bond}, {Rogers},
  {Abe}, {Asakura}, {Bhattacharya}, {Donachie}, {Freeman}, {Fukui}, {Hirao},
  {Itow}, {Koshimoto}, {Li}, {Ling}, {Masuda}, {Matsubara}, {Muraki},
  {Nagakane}, {Onishi}, {Oyokawa}, {Rattenbury}, {Saito}, {Sharan}, {Shibai},
  {Sullivan}, {Tristram}, {Yonehara}, \& {MOA Collaboration}}]{Suzuki16}
{Suzuki}, D., {Bennett}, D.~P., {Sumi}, T., {et~al.} 2016, \apj, 833, 145

\bibitem[{{Suzuki} {et~al.}(2018){Suzuki}, {Bennett}, {Ida}, {Mordasini},
  {Bhattacharya}, {Bond}, {Donachie}, {Fukui}, {Hirao}, {Koshimoto},
  {Miyazaki}, {Nagakane}, {Ranc}, {Rattenbury}, {Sumi}, {Alibert}, \&
  {Lin}}]{Suzuki18}
{Suzuki}, D., {Bennett}, D.~P., {Ida}, S., {et~al.} 2018, \apj, 869, L34

\bibitem[{{Thompson} {et~al.}(2018){Thompson}, {Coughlin}, {Hoffman},
  {Mullally}, {Christiansen}, {Burke}, {Bryson}, {Batalha}, {Haas},
  {Catanzarite}, {Rowe}, {Barentsen}, {Caldwell}, {Clarke}, {Jenkins}, {Li},
  {Latham}, {Lissauer}, {Mathur}, {Morris}, {Seader}, {Smith}, {Klaus},
  {Twicken}, {Van Cleve}, {Wohler}, {Akeson}, {Ciardi}, {Cochran}, {Henze},
  {Howell}, {Huber}, {Pr{\v s}a}, {Ram{\'{\i}}rez}, {Morton}, {Barclay},
  {Campbell}, {Chaplin}, {Charbonneau}, {Christensen-Dalsgaard}, {Dotson},
  {Doyle}, {Dunham}, {Dupree}, {Ford}, {Geary}, {Girouard}, {Isaacson},
  {Kjeldsen}, {Quintana}, {Ragozzine}, {Shabram}, {Shporer}, {Silva Aguirre},
  {Steffen}, {Still}, {Tenenbaum}, {Welsh}, {Wolfgang}, {Zamudio}, {Koch}, \&
  {Borucki}}]{Thompson18}
{Thompson}, S.~E., {Coughlin}, J.~L., {Hoffman}, K., {et~al.} 2018, \apjs, 235,
  38

\bibitem[{{Turner} {et~al.}(2016){Turner}, {Pearson}, {Biddle}, {Smart},
  {Zellem}, {Teske}, {Hardegree-Ullman}, {Griffith}, {Leiter}, {Cates},
  {Nieberding}, {Smith}, {Thompson}, {Hofmann}, {Berube}, {Nguyen}, {Small},
  {Guvenen}, {Richardson}, {McGraw}, {Raphael}, {Crawford}, {Robertson},
  {Tombleson}, {Carleton}, {Towner}, {Walker-LaFollette}, {Hume}, {Watson},
  {Jones}, {Lichtenberger}, {Hoglund}, {Cook}, {Crossen}, {Jorgensen},
  {Romine}, {Thompson}, {Villegas}, {Wilson}, {Sanford}, {Taylor}, \&
  {Henz}}]{Turner16}
{Turner}, J.~D., {Pearson}, K.~A., {Biddle}, L.~I., {et~al.} 2016, \mnras, 459,
  789

\bibitem[{{Uehara} {et~al.}(2016){Uehara}, {Kawahara}, {Masuda}, {Yamada}, \&
  {Aizawa}}]{Uehara16}
{Uehara}, S., {Kawahara}, H., {Masuda}, K., {Yamada}, S., \& {Aizawa}, M. 2016,
  \apj, 822, 2

\bibitem[{{Villanueva} {et~al.}(2019){Villanueva}, {Dragomir}, \&
  {Gaudi}}]{Villanueva19}
{Villanueva}, Steven, J., {Dragomir}, D., \& {Gaudi}, B.~S. 2019, \aj, 157, 84

\bibitem[{{Wang} {et~al.}(2015){Wang}, {Fischer}, {Barclay}, {Picard}, {Ma},
  {Bowler}, {Schmitt}, {Boyajian}, {Jek}, {LaCourse}, {Baranec}, {Riddle},
  {Law}, {Lintott}, {Schawinski}, {Simister}, {Gr{\'e}goire}, {Babin}, {Poile},
  {Jacobs}, {Jebson}, {Omohundro}, {Schwengeler}, {Sejpka}, {Terentev},
  {Gagliano}, {Paakkonen}, {Otnes Berge}, {Winarski}, {Green}, {Schmitt},
  {Kristiansen}, \& {Hoekstra}}]{Wang15}
{Wang}, J., {Fischer}, D.~A., {Barclay}, T., {et~al.} 2015, \apj, 815, 127

\bibitem[{{Winn}(2010)}]{Winn10}
{Winn}, J.~N. 2010, {Exoplanet Transits and Occultations}, 55--77

\bibitem[{{Yee} \& {Gaudi}(2008)}]{Yee08}
{Yee}, J.~C., \& {Gaudi}, B.~S. 2008, \apj, 688, 616

\bibitem[{{Zhu} {et~al.}(2018){Zhu}, {Petrovich}, {Wu}, {Dong}, \&
  {Xie}}]{Zhu18}
{Zhu}, W., {Petrovich}, C., {Wu}, Y., {Dong}, S., \& {Xie}, J. 2018, \apj, 860,
  101

\bibitem[{{Zhu} \& {Wu}(2018)}]{ZhuWu18}
{Zhu}, W., \& {Wu}, Y. 2018, \aj, 156, 92

\bibitem[{{Ziegler} {et~al.}(2018){Ziegler}, {Law}, {Baranec}, {Howard},
  {Morton}, {Riddle}, {Duev}, {Salama}, {Jensen-Clem}, \&
  {Kulkarni}}]{Ziegler18V}
{Ziegler}, C., {Law}, N.~M., {Baranec}, C., {et~al.} 2018, \aj, 156, 83

\end{thebibliography}
